\documentclass[letter,11pt]{article}
\usepackage{jheppub}
\usepackage{multirow}

\def\t{{\bar t}}
\def\Mtt{m_{t\bar t}}
\def\PTtt{p_{T,t\bar t}}
\def\PTt{p_{T,t}}
\def\PTtbar{p_{T,{\bar t}}}
\def\PTavt{p_{T,t/{\bar t}}}
\def\Yavt{y_{t/{\bar t}}}
\def\Ytt{y_{t\bar t}}
\def\GeV{\, \rm GeV}
\def\TeV{\, \rm TeV}

\title{\boldmath Dynamical scales for multi-TeV top-pair production at the LHC}

\author[a]{Micha\l{}  Czakon,}
\author[b]{David Heymes}
\author[b]{and Alexander Mitov}

\affiliation[a]{Institut f\"ur Theoretische Teilchenphysik und Kosmologie,
RWTH Aachen University, D-52056 Aachen, Germany}
\affiliation[b]{Cavendish Laboratory, University of Cambridge, Cambridge CB3 0HE, UK}

\note{Preprint numbers: Cavendish-HEP-16/08, TTK-16-21}

\abstract{We calculate all major differential distributions with stable top-quarks at the LHC. The calculation covers the multi-TeV range that will be explored during LHC Run II and beyond. Our results are in the form of high-quality binned distributions. We offer predictions based on three different parton distribution function (pdf) sets. In the near future we will make our results available also in the more flexible {\tt fastNLO} format that allows fast re-computation with any other pdf set. 
In order to be able to extend our calculation into the multi-TeV range we have had to derive a set of dynamic scales. Such scales are selected based on the principle of fastest perturbative convergence applied to the differential and inclusive cross-section. Many observations from our study are likely to be applicable and useful to other precision processes at the LHC. 
With scale uncertainty now under good control, pdfs arise as the leading source of uncertainty for TeV top production. Based on our findings, true precision in the boosted regime will likely only be possible after new and improved pdf sets appear. We expect that LHC top-quark data will play an important role in this process.}

\begin{document} 
\maketitle
\flushbottom

\section{Introduction}\label{sec:intro}

The recent derivation of the fully differential next-to-next-to-leading order (NNLO) correction to top quark-pair production at the LHC \cite{Czakon:2015owf} and at the Tevatron \cite{Czakon:2014xsa,Czakon:2016ckf} naturally raises the question: what precision can be expected in top-quark pair production at the LHC across observables and in the widest achievable kinematical ranges? To address this question, it is instructive to first recall the situation with the total inclusive cross-section which is well-understood in (resummed) NNLO QCD \cite{Baernreuther:2012ws,Czakon:2012zr,Czakon:2012pz,Czakon:2013goa}.

Upon the inclusion of the NNLO QCD correction, $\sigma_{\rm tot}$ can be predicted with an accuracy of about 5\%. A number of independent sources contribute to this total error, the most important ones being missing higher order terms (beyond NNLO), pdf error and parametric $m_t$ and $\alpha_S$ uncertainties. Significantly, all these sources of error are comparable in magnitude which indicates that further reduction in the error of top-pair production at the LHC would be a significant challenge even in the long run. The next level of uncertainty contributors to $\sigma_{\rm tot}$ are at the level of about 1\% and include EW corrections, finite top width and various non-perturbative effects. 

This uncertainty breakdown for $\sigma_{\rm tot}$ is a good indicator for the sources of uncertainty to be expected in top-pair differential distributions. It is important to recognise, however, that the various sources of uncertainty mentioned in the context of $\sigma_{\rm tot}$ could vary wildly across kinematics. For example, the electroweak (EW) corrections are expected to become on par with the NNLO QCD scale variation in the TeV range \cite{Beenakker:1993yr,Kuhn:2005it,Bernreuther:2005is,Kuhn:2006vh,Bernreuther:2006vg,Bernreuther:2008md,Manohar:2012rs,Kuhn:2013zoa,Campbell:2015vua,Hollik:2007sw,Bernreuther:2010ny,Pagani:2016caq}. Finite top width effects are typically suppressed by powers of $\Gamma_t/m_t$ but can be much larger in special kinematic regions \cite{Denner:2010jp,Bevilacqua:2010qb,Denner:2012yc,Papanastasiou:2013dta,Frederix:2013gra,Cascioli:2013wga,Bevilacqua:2015qha,Frederix:2016rdc}. Non-factorisable effects in inclusive observables are typically suppressed by powers of $1/m_t$ but could be much larger, for example, in presence of jet vetoes if $p_{\rm T,veto} \ll m_t$, in which case they are suppressed only as $1/p_{\rm T,veto}$ \cite{Mitov:2012gt}. 

In this paper we take the first step towards the systematic study of theoretical uncertainties in {\it precision} fully-differential top-pair production at the LHC with stable top quarks. Specifically, we focus our discussion on NNLO QCD scale uncertainty which, at present, is a main source of theoretical error. The framework of our discussion is as follows: 
\begin{enumerate}
\item We consider the variation of factorisation and renormalisation scales as a proxy for missing higher order terms. The scale variation procedure we use is not {\it ad hoc}; its applicability to the total inclusive cross-section has been validated.
\item As a prerequisite to scale variation, one needs to specify a default central scale $\mu_0$. The main goal of this paper is to identify the functional form of $\mu_0$. We choose such a scale based on the criterium of {\it perturbative convergence}. In doing so we account for LO, NLO and NNLO corrections as well as, where available, NNLO plus soft-gluon resummation.
\item We assume that the sought default scale $\mu_0$ is the same for both the renormalisation and factorisation scales, i.e. $\mu_{R,0}=\mu_{F,0}=\mu_0$. Scale variation, however, is done independently for $\mu_F$ and $\mu_R$ \cite{Cacciari:2008zb}:
\begin{equation}
\mu_{F,R}\in(\mu_0/2,2\mu_0) ~~~{\rm with} ~~~ 0.5\leq \mu_R/\mu_F \leq 2 \; .
\label{eq:scalevar}
\end{equation}
\item A dynamic scale is, {\it a priori}, better than a fixed scale. However, the spread among various dynamic scales can be comparable in size to scale variation and therefore a {\it sensible} choice among possible dynamic scales has to be made.
\end{enumerate}

Perturbative convergence is an indicator of the reliability of perturbative predictions. Ever since the early days of heavy flavour NLO calculations \cite{Nason:1989zy,Beenakker:1990maa} running scales -- motivated by physical arguments -- have been used. Clearly, different scale choices affect the rate of convergence through higher-order terms they introduce. Since scales are unphysical, one may promote perturbative convergence to a principle and try to derive the ``correct" scale with it. In this work we only invoke the principle of fastest perturbative convergence in a weak sense
\footnote{Partly, in order to avoid subtleties related to the existence and uniqueness of such hypothetical scale.};
we speak of the {\it criterium of faster perturbative convergence} which we define as follows (related past work is reviewed in sec.~\ref{sec:pastresults}): between two scales, the one that offers faster convergence is better. Clearly, the scale $\mu_0$ will depend on the set of considered functional forms.

We motivate and explain our choices for scale $\mu_0$ in section~\ref{sec:choosescale}, but before going into this, we would like to make the following comment. While the scale choices we identify in this paper are sensible and satisfy the above criteria we do not imply that even ``better" dynamic scales cannot be derived in the future. In particular, such scale modifications may be needed to reflect improved future understanding of the large $p_T$ behaviour of top production due to resummation of large collinear logs $\sim \ln(p_T/m_t)$ as well as the validity of the five-flavour number scheme that is exclusively used in the description of top production at present (see refs.~\cite{Forte:2010ta,Ball:2011mu,Han:2014nja} for related work). As quality LHC data at large $p_T$ starts to appear and these two theoretical issues get scrutinised, the functional form for the scale $\mu_0$ may potentially need to be revisited. We, however, find it unlikely that such potential future scales will lead to significant deviations in observables compared to the scales derived in this work.

The paper is organised as follows: in section~\ref{sec:pastresults} we offer a brief overview of past results on scale setting relevant for our discussion. In section \ref{sec:choosescale} we analyse the total inclusive cross-section and differential distributions for LHC 8 TeV and derive the functional forms for ``best" scales $\mu_0$. As it turns out, two scales are needed: one for the $p_T$ distribution and one for all other distributions. In section~\ref{sec:pdf} we study the sensitivity of NNLO differential distributions and demonstrate that our ``best" scales $\mu_0$ are stable with respect to the choice of pdf. In section~\ref{sec:pheno} we present our best predictions for all stable top differential distributions in NNLO QCD for LHC 8 and 13 TeV. Prospects for further improvements are discussed in the conclusions. All results are made available in electronic form with the Arxiv submission of this paper.

\section{Overview of past work related to scale setting}\label{sec:pastresults}

Interpreting scale variation as theoretical uncertainty due to missing higher order terms has long history. Within such an approach factorisation and renormalisation scales are typically varied up and down by factors of two and one-half around a judiciously chosen default value. Such default scale, often called {\it central scale}, is specific to each process and observable. Clearly, the choices for both the central scale and the variation around it are arbitrary. Nevertheless, as a result of three decades of higher-order calculations for high-energy colliders, a {\it common} choice of scale variation (2,1/2) has emerged. Such variation procedure, which is common across processes and observables, is very useful in practice because it allows to easily interpret and compare theoretical errors derived for different, even unrelated, processes. One can justify the amount of scale variation around a central value a posteriori, by comparing predictions for central scales computed at different orders in perturbation theory. A scale variation procedure is deemed good if the error estimate at certain perturbative order contains the central value of the next higher order. Such procedure requires at least NLO calculations. If NNLO results are available then such checks can be even quantitative.

In top-pair production the scale variation procedure eq.~(\ref{eq:scalevar}) based on restricted independent variation of the factorisation and renormalisation scales has been shown to work very well through NNLO for the total inclusive cross-section \cite{Czakon:2013xaa}. We expect that it will also work well for differential distributions, at least in the bulk low-$p_T$ region, and we also extend this variation procedure to the whole kinematic range for all kinematic variables.
\footnote{We note that such a procedure has not been validated in extreme high-$p_T$ kinematics, where we also expect it to work, possibly after resummation and other relevant procedures have been carried out.}

The choice for the central scale is, however, much less clear and often alternative choices are made in different calculations for the same observable. We hope that with the advent of NNLO collider phenomenology such choices will be more and more scrutinised in the future. We also hope that the present work will serve as an example in this regard. While we cannot give an exhaustive collection of scales used in collider physics, in the following we will review some past work which has some relevance for our present work in top-pair production.

A number of dynamic scales has been used in the past in top-pair production at hadron colliders. In refs.~\cite{Denner:2012yc,Cascioli:2013wga} a geometric average scale (see eq.~(\ref{eq:mu-ET}) below) has been used for both $t\t$ and single top production. $H'_T$-based scales are also used \cite{Frederix:2013gra}, where $H'_T$ includes all final state partons as in eq.~(\ref{eq:mu-HTprime}) below. Scales based on $m_T$ (\ref{eq:mu-mT}) have been used since the early days of NLO calculations \cite{Nason:1989zy,Beenakker:1990maa,Mangano:1991jk,Frixione:1995fj}, as well as, more recently, $\Mtt$-based scales \cite{Ahrens:2010zv,Ferroglia:2013zwa,Ferroglia:2015ivv,Pecjak:2016nee}.

Similar functional forms for the factorisation and renormalisation scales have been used and discussed in other collider processes. For example,  for $W+jets$ production $H'_T/2$ scale has been used at NLO \cite{Berger:2010zx}, while at NNLO a modified version of $H'_T$ was used in ref.~\cite{Boughezal:2016yfp}. A detailed study of dynamic scales in $W+3jets$ was performed in ref.~\cite{Melnikov:2009wh} where scales based on the MLM and CKKW procedures \cite{Alwall:2007fs,Catani:2001cc} were found to offer small corrections across different kinematics, in variance with the case of the $W$-boson transverse mass. Related discussion for $V+jets$ can be found in ref.~\cite{Bauer:2009km}. An often made choice in inclusive jet production is $p_{T}$ or $p_{T,{\rm max}}$ \cite{Chatrchyan:2012bja,Aad:2014vwa,Carrazza:2014hra} while for dijet mass distributions one typically has $p_{T,{\rm ave}}$ and $p_{T,{\rm max}}e^{0.3y^*}$ \cite{Chatrchyan:2012bja,Aad:2013tea}. A recent summary of existing LHC jet measurements can be found in ref.~\cite{Francavilla:2015yxa}. 

Past approaches to scale setting include the {\it Method of Effective Charges} \cite{Grunberg:1980ja,Grunberg:1982fw,Grunberg:1989xf} (sometimes referred to as {\it Fastest Apparent Convergence} \cite{Stevenson:1981vj,Kubo:1982gd}; see also Ref.~14 in \cite{Grunberg:1982fw}), the {\it Principle of Minimal Sensitivity} \cite{Stevenson:1981vj,Stevenson:1986cu}; the {\it Complete Renormalization Group Improvement} approach \cite{Maxwell:2000mm} which provides a factorisation scale based on an alternative collinear factorisation scheme \cite{Maltoni:2007tc}, extending earlier work on factorisation scale setting in Higgs production \cite{Boos:2003yi,Maltoni:2003pn}. Finally, the Brodsky-Lepage-Mackenzie scale setting approach \cite{Brodsky:1982gc} (and its further refinement known as {\it Principle of Maximum Conformality}) \cite{Brodsky:2011ig,Brodsky:2011ta,Brodsky:2012rj,Mojaza:2012mf,Brodsky:2013vpa,Ma:2015dxa} is based on the idea of restoring the conformal symmetry of the QCD Lagrangian in observables. The BLM/PMC approach specifies a value for the renormalisation, but not factorisation, scale. 

Our approach is closest, yet not identical, to the criterion of {\it Fastest Apparent Convergence}. This criterion derives from the {\it Method of Effective Charges} and sets the renormalisation scale at such a (process-dependent) value that the NLO correction for a particular observable vanishes. The {\it Method of Effective Charges} is more general; its application is process-dependent and sets to zero all terms in the perturbative expansion beyond the leading order. The conditions one imposes are such that the truncated perturbative expansion for an observable is renormalisation scheme independent to any finite order. In effect, this method replaces the fixed order expansion in the usual $\overline{\rm MS}$ coupling evaluated at scale $\mu_R$ with a Born-level effective coupling defined in a new, process-dependent renormalisation scheme. As a by product of this procedure the value of the renormalisation constant gets fixed, too. Our approach is similar to the above in that it tries to minimise the size of higher order corrections, but not necessarily set them to zero.  

In this work we choose to follow the usual approach to scale setting due to its {\it broadly-established} applicability from fully inclusive observables to exclusive multi-particle final states. In particular, here we only consider scales which are common to all orders in the strong coupling expansion. For this reason, in the present work we do not study the implications of the BLM/PMC procedures. Recent comparison of predictions based on the BLM/PMC and the usual scale setting approaches can be found in ref.~\cite{Czakon:2016ckf}.

Alternative approaches for estimating theory errors have been proposed in refs.~\cite{Cacciari:2011ze,Bagnaschi:2014wea,David:2013gaa}.

\section{Choosing the scale $\mu_0$}\label{sec:choosescale}

In order to identify the most appropriate dynamical scale for use in top-pair production at the LHC, we perform a number of fully differential calculations based on the following set of functional forms:
\begin{eqnarray}
\mu_0&\sim& m_t \;, \label{eq:mu-mtop}\\
\mu_0&\sim& m_T = \sqrt{m_t^2+p_T^2} \;, \label{eq:mu-mT}\\
\mu_0&\sim& H_T = \sqrt{m_t^2+\PTt^2} + \sqrt{m_t^2+\PTtbar^2} \;, \label{eq:mu-HT}\\
\mu_0&\sim& H'_T = \sqrt{m_t^2+\PTt^2} + \sqrt{m_t^2+\PTtbar^2} + \sum_{i}p_{T,i} \;, \label{eq:mu-HTprime}\\
\mu_0&\sim& E_T = \sqrt{\sqrt{m_t^2+\PTt^2}\sqrt{m_t^2+\PTtbar^2}} \;, \label{eq:mu-ET}\\
\mu_0&\sim& H_{T,\rm int} = \sqrt{(m_t/2)^2+\PTt^2} + \sqrt{(m_t/2)^2+\PTtbar^2} \;, \label{eq:mu-HTint}\\
\mu_0&\sim& \Mtt \label{eq:mu-Mtt} \;,
\label{eq:scales-def}
\end{eqnarray}
where the momentum $p_T$ entering the definition of $m_T$ in eq.~(\ref{eq:mu-mT}) is either that of the top or the antitop, depending on the distribution. The sum in the definition of $H'_T$ runs over all massless partons present in the final state (at NNLO there could be up to two partons). Finally, an important part of the process of choosing the functional form of $\mu_0$ involves the fixing of the proportionality constant, signified by the $\sim$ sign in the above equations. While for brevity we focus our presentation on LHC 8 TeV, we have also verified that our conclusions remain unchanged at LHC 13 TeV. Unless explicitly specified, throughout this work we combine partonic cross-sections with pdf of the same order (LO with LO, NLO with NLO, etc). Resummed NNLO partonic cross-sections are convoluted with NNLO pdf. The strong coupling constant $\alpha_S$ is evaluated through the LHAPDF interface \cite{Buckley:2014ana} as appropriate for the corresponding pdf set. Throughout this paper scale variation in differential distributions is performed by independently varying $\mu_F$ and $\mu_R$ (as defined in sec.~\ref{sec:intro}). Only in sec.~\ref{sec:stot} -- in the context of the total inclusive cross-section -- we use simultaneous $\mu_F=\mu_R$ scale variation.

\subsection{Total cross-section}\label{sec:stot}

We begin our investigation with the total inclusive cross-section based on the standard choice $\mu_0=m_t$ and computed with two pdf sets: MSTW2008 \cite{Martin:2009iq} and NNPDF3.0 \cite{Ball:2014uwa}. The total cross-section is computed with the help of the program {\tt Top++} \cite{Czakon:2011xx}. Besides the LO, NLO and NNLO QCD corrections we also include soft-gluon resummation through NNLL accuracy where available (i.e. for the total cross-section computed with a fixed scale $\mu_0\sim m_t$).

\begin{figure}[t]
\hskip -4mm
\includegraphics[width=0.52\textwidth]{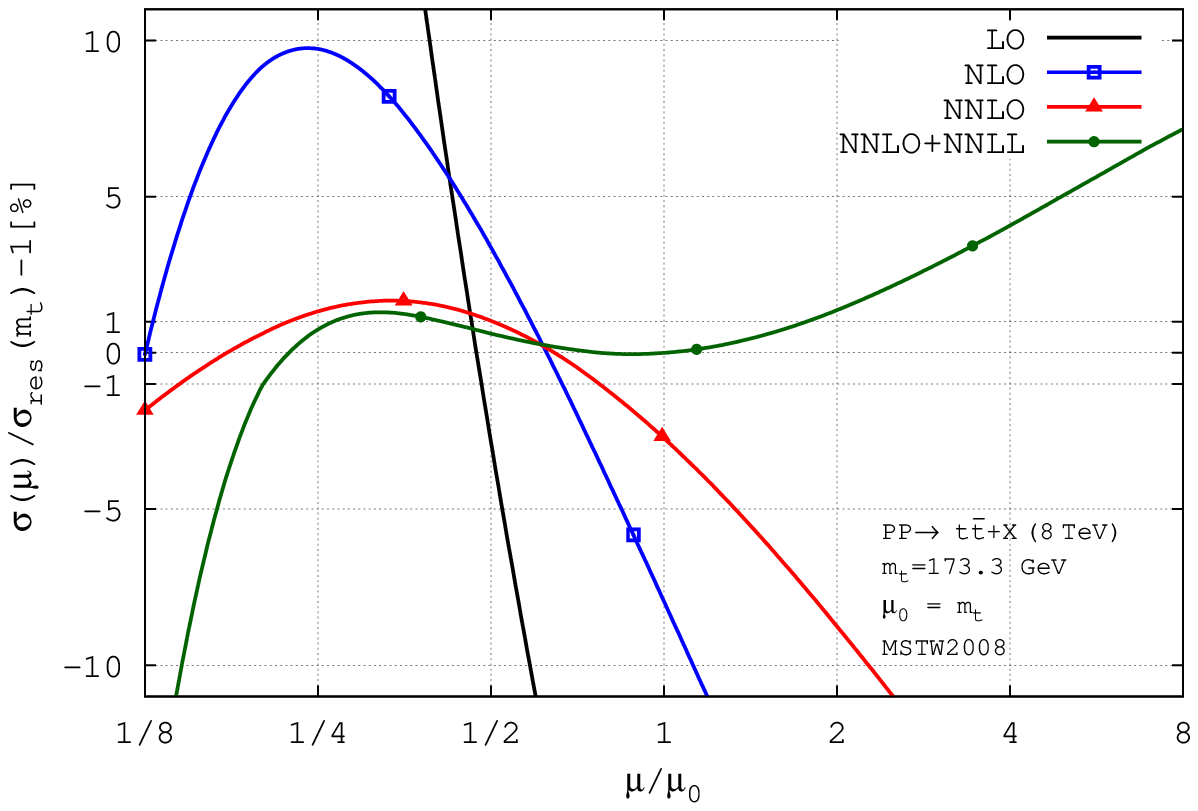}
\includegraphics[width=0.52\textwidth]{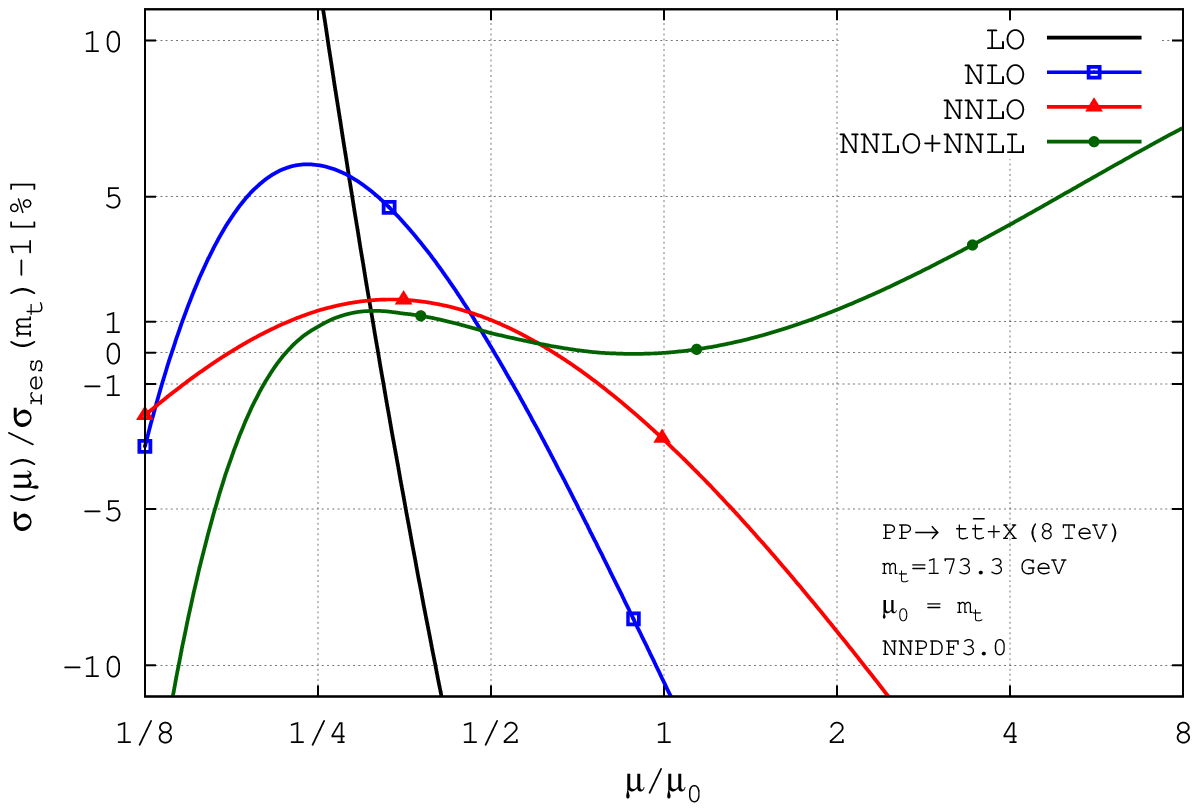}
\caption{\label{fig:stot-mt} Total cross-section at LO, NLO, NNLO and NNLO+NNLL QCD evaluated with a fixed scale $\mu_F=\mu_R=m_t$ with two different pdf sets: MSTW2008 (left) and NNPDF3.0 (right). Each plot is normalised to the NNLO+NNLL cross-section evaluated with the corresponding pdf set at scale $\mu_0=m_t$. The symbols on some of the lines are meant to help distinguish the various lines.}
\end{figure}

Two important observations can be made from fig.~\ref{fig:stot-mt} and they turn out to be central for this work: first, the scale for which {\it perturbative convergence} is maximised is slightly above $m_t/2$, i.e. that scale is significantly lower than the standard one $\mu_0=m_t$. Second, the value of the fixed order NNLO cross-section evaluated at the scale of fastest convergence is only about 0.5\% higher than the NNLO+NNLL resummed one evaluated at the usual scale $\mu_0=m_t$, i.e. the two values essentially agree (recall that 0.5\% difference is only a small fraction of the scale uncertainty of the resummed result).

The numerical agreement between the fixed order result evaluated at a lower scale and the usual resummed result is significant. First, in practical terms, such an agreement allows the use of fixed order results without the need to worry about the numerical impact of soft-gluon resummation
\footnote{We have not investigated the possible validity or breakdown of such a conclusion outside the context of fully inclusive top-pair production at the LHC.}.
The fact that the fixed order result at a smaller scale is larger than the standard resummed prediction (albeit by a tiny amount) is also consistent with what one might expect about yet uncalculated higher-order effects based purely on the behaviour of the known LO, NLO and NNLO corrections to top-pair production, as well as soft-gluon resummation, where one observes reasonably fast convergence of so-far always positive higher order corrections.

Perhaps not surprisingly, given the large uncertainty at LO (as evident from its large slope and from the difference between the two pdf sets), the LO correction is not a reliable input to the above analysis. The difference between the two pdf sets decreases fast with higher orders and is completely negligible at NNLO and at NNLO+NNLL. It thus appears that the point of fastest convergence is not very different for the two pdf sets and the values of the NNLO cross-section one derives from the two pdf sets are within less than 1\% from each other. We also notice that for scales smaller than the one of fastest convergence the hierarchy of perturbative corrections gets completely inverted, i.e. the LO is largest and the inclusion of higher orders decreases the total cross-section. 

With this observation in mind it is interesting to contrast our findings based on the principle of fastest convergence with the principle of minimal sensitivity which has often been invoked in the past. Had we followed the latter principle we would have found NLO correction which is very large compared to the standard NNLO resummed result. The minimal sensitivity scale for which the NLO curve plateaus is particularly low, around $m_t/4$. Furthermore, we notice a significant shift when going from NLO to NNLO both in terms of minimum sensitivity scale and in terms of the values the cross-section takes at these two scales.

The picture emerging from fig.~\ref{fig:stot-mt} has a direct analogue in inclusive Higgs production at the LHC. Following the recent work \cite{Anastasiou:2016cez} on inclusive Higgs production in NNNLO QCD we observe the almost one-to-one behaviour between the top inclusive cross-section at order ${\rm N}^n{\rm LO}$ and the total Higgs cross-section 
\footnote{We only consider the $gg\to h$ channel in the limit of large $m_t$.}
at order ${\rm N}^{n+1}{\rm LO}$ for $n=0,1,2$ as a function of the scale $\mu$. Importantly, the analogy extends also to the resummed NNLO cross-section, especially the rise of the NNLL resummed cross-section for larger values of $\mu$. We have checked, but do not show it in fig.~\ref{fig:stot-mt}, that the inclusion of soft gluon resummation with lower logarithmic accuracy (NLL and LL) does not lead to such a rise for larger values of $\mu$. Similar behaviour is seen also in the case of the Higgs cross-section. From this comparison we can conclude that both inclusive top-pair and Higgs production cross-sections exhibit fastest perturbative convergence at scales lower than the usual ones: $m_t$ for top production and $m_h/2$ for Higgs production (note that in both cases these scales are half the mass of the Born-level final state). On the other hand, the fast rise of the resummed cross-section at larger values of $\mu$ indicates that the perturbative series is not converging well there and therefore such large scales should be avoided.

In the following we verify the above conclusion by considering the full set of scales (\ref{eq:mu-mtop}-\ref{eq:mu-Mtt}). We consider the LO, NLO and NNLO cross-sections but no soft-gluon resummation. 
\begin{figure}[t]
\hskip -4mm
\includegraphics[width=0.52\textwidth]{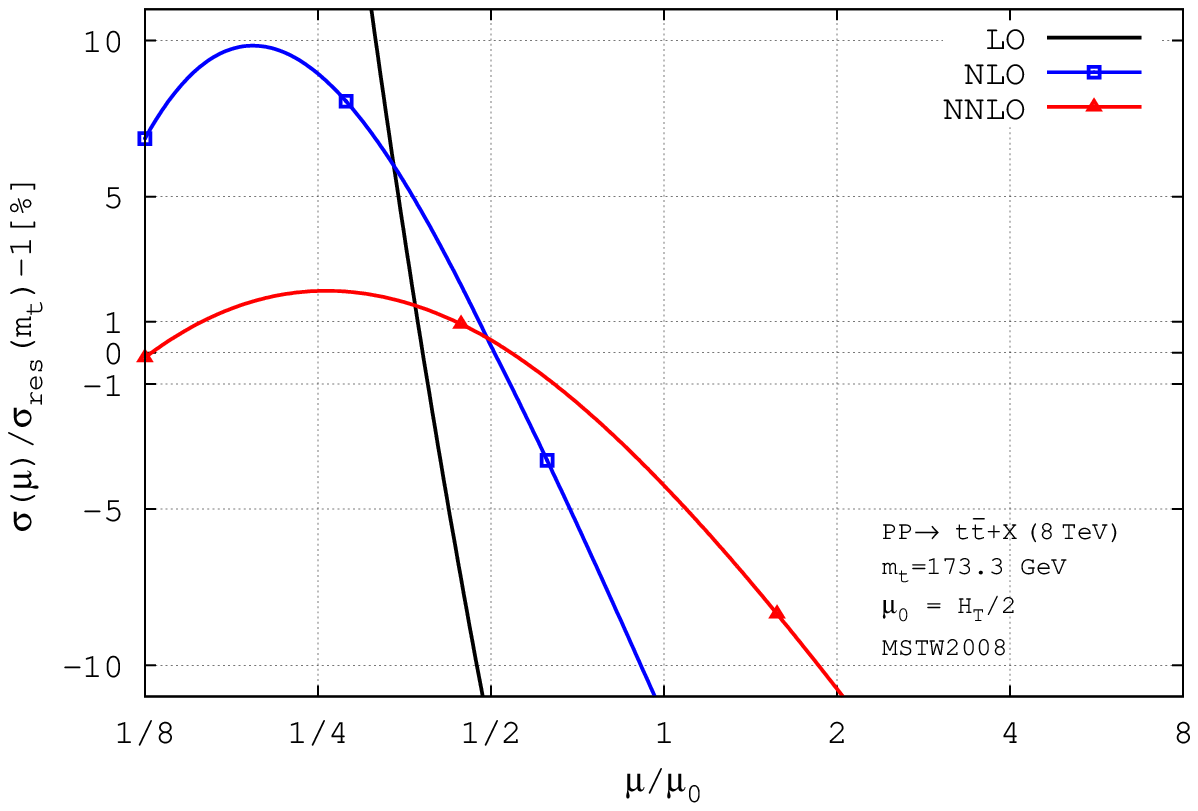}
\includegraphics[width=0.52\textwidth]{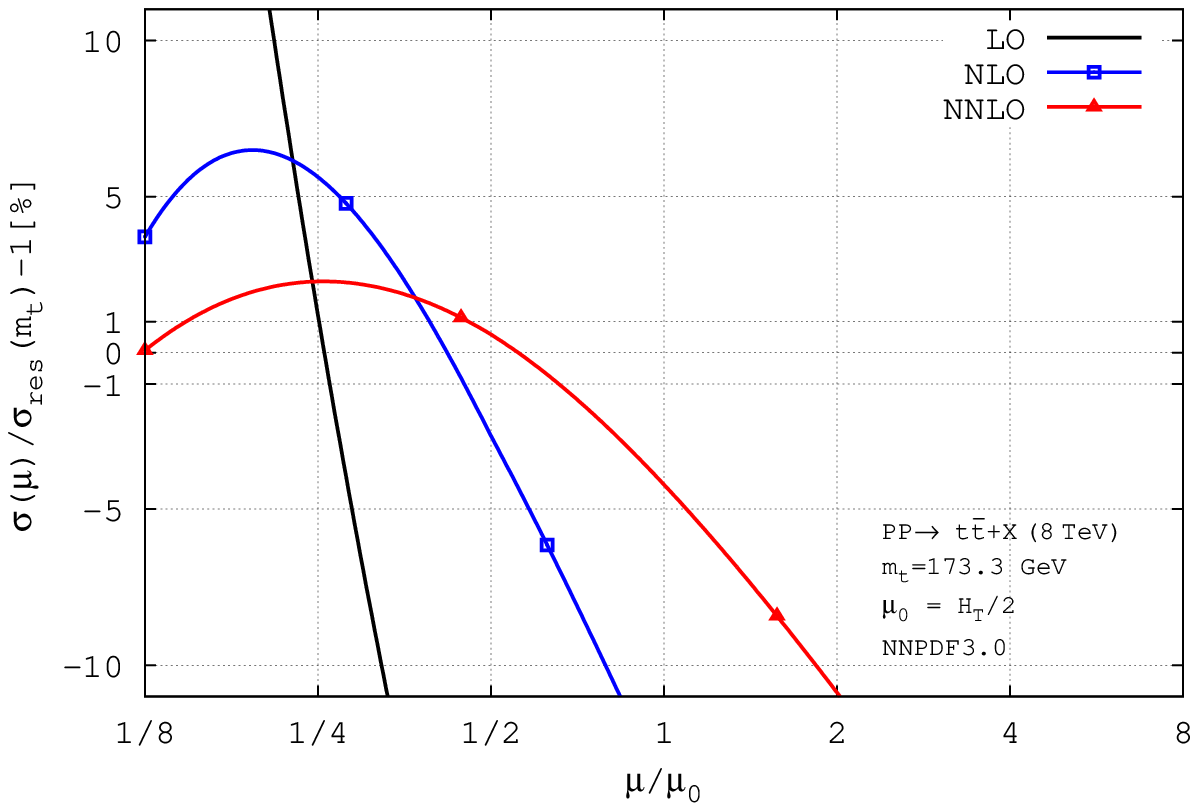}
\caption{\label{fig:stot-HT2} Total cross-section at LO, NLO and NNLO evaluated with a dynamic scale $\mu_F=\mu_R=H_T/2$ (defined in eq.~(\ref{eq:mu-HT})) with two different pdf sets: MSTW2008 (left) and NNPDF3.0 (right). Each plot is normalised as in fig.~\ref{fig:stot-mt}, i.e. to the NNLO+NNLL cross-section evaluated with the corresponding pdf set at scale $\mu_0=m_t$. The symbols on some of the lines are meant to help distinguish the various lines.}
\end{figure}

We first study the most natural choice for a dynamic scale in inclusive top production, namely, $\mu_0=H_T/2$. In fig.~\ref{fig:stot-HT2} we present the $\mu=\mu_F=\mu_R$ dependence of the total cross-section evaluated with this scale. We observe that the behaviour of the cross-section as a function of the scale $\mu$ is rather similar to the one with a fixed scale. The only noticeable difference between the two figures is the shift towards smaller scales, i.e. while the scale of fastest convergence was slightly above $1/2$ of the nominal value ($m_t$ in that case) now it is almost exactly at $1/2$ of the nominal value $H_T/2$. Moreover, the value of the NNLO cross-section at such a scale is only $0.5\%$ larger than the resummed NNLO+NNLL cross-section at scale $m_t$, for both pdf sets studied here. From this we conclude that the optimal choice for a dynamic scale, and one that reproduces well the known total cross-section, is:
\begin{equation}
\mu_0 = {H_T\over 4} \,.
\label{eq:mu0HT4}
\end{equation}

The fact that the optimal value of the dynamic scale is slightly below the value for the fixed scale is easy to understand. At low $\PTt$ -- which is the region that generates the bulk of the total cross-section -- the scale in eq.~(\ref{eq:mu0HT4}) behaves as $m_t/2+O(\PTt^2)$. Upon integration over $\PTt$ the terms $O(\PTt^2)$ generate additional contribution which effectively increases the value of the scale or, in other words, an effective static scale has value larger than $m_t/2$ due to the running scale effects. In this sense we view the scale $m_t$ not as the ``best" scale at which to evaluate the total cross-section, but as the best average value of the running scale which reproduces the total cross-section. The value for the fastest convergence scale of about $0.7m_t$ observed in fig.~\ref{fig:stot-mt} is consistent with this observation.

\begin{figure}[t]
\hskip -4mm
\includegraphics[width=0.52\textwidth]{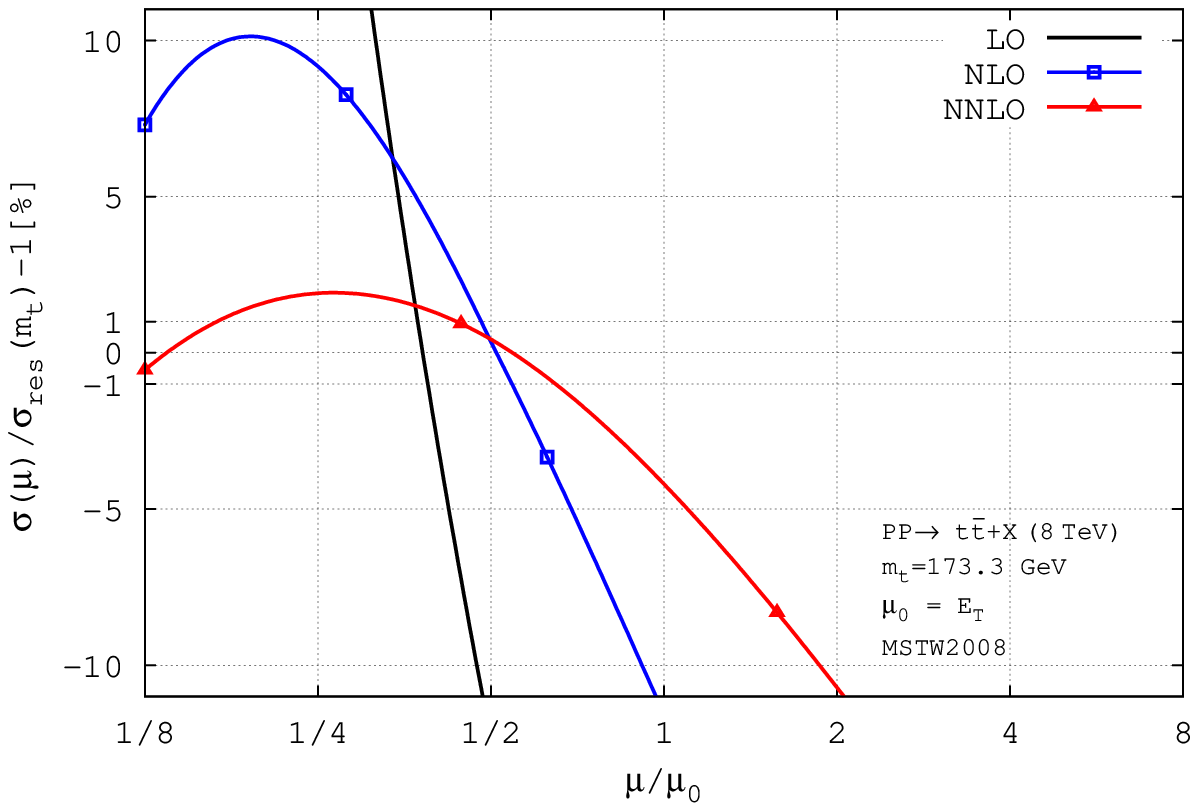}
\includegraphics[width=0.52\textwidth]{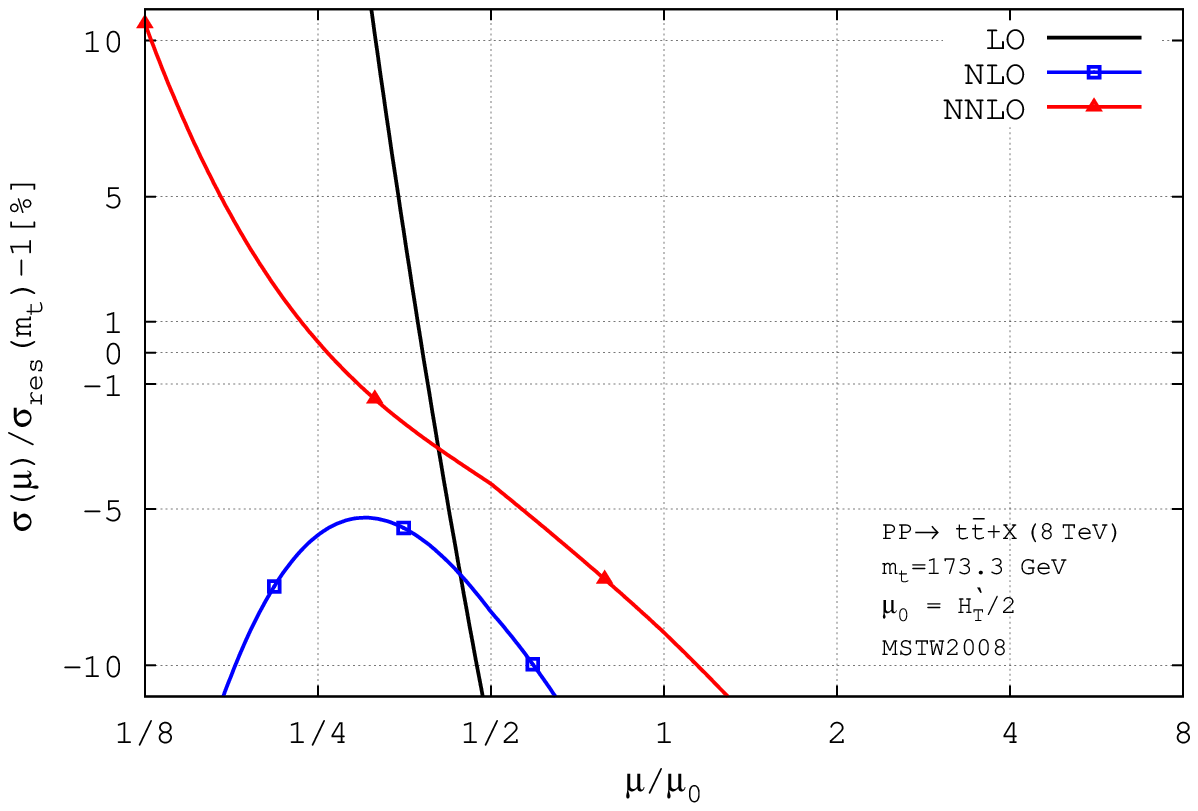}
\caption{\label{fig:stot-ETHprimeT} As in fig.~\ref{fig:stot-HT2} but for scale $E_T$ (\ref{eq:mu-ET}) (left) and $H'_T$ (\ref{eq:mu-HTprime}) (right). Both use pdf set MSTW2008. The symbols on some of the lines are meant to help distinguish the various lines.}
\end{figure}

There are several alternative definitions of the scale $H_T$ that have been considered in the literature. One of them is eq.~(\ref{eq:mu-ET}) which we denote as $E_T$; it differs from $H_T$ by taking the geometric as opposed to arithmetic average of the $t$ and $\bar t$ transverse masses. From fig.~\ref{fig:stot-ETHprimeT} (left) we conclude that the numerical difference between the two scales is immaterial. Another alternative definition (\ref{eq:mu-HTprime}), denoted here as $H'_T$, involves the sum of the transverse masses of all final state partons. In fig.~\ref{fig:stot-ETHprimeT} (right) we see that the behaviour of this scale is very different from $H_T$, especially at NNLO. 
Indeed, the NLO and NNLO curves do not even cross and the NNLO curve has monotonic behaviour over the whole interval $1/8 \leq \mu/\mu_0 \leq8$. We have not studied in depth this peculiar behaviour but point out that such a scale is much more sensitive to singular emissions (real and virtual). For this reason, a definition that relies on clustering the emitted partons into jets may alleviate such behaviour.
\footnote{We thank Bryan Webber for a helpful discussion on this point.}
Anticipating our findings for the scale $\mu_0$ in differential distributions, in this work we find strong support for the idea that a good dynamical scale should, among others, resemble as much as possible the born-level observable for the process of interest. It seems to us this conclusion may also have implications for processes outside top physics, or at a minimum, may warrant similar investigations in other processes. 

\begin{figure}[t]
\hskip -4mm
\includegraphics[width=0.52\textwidth]{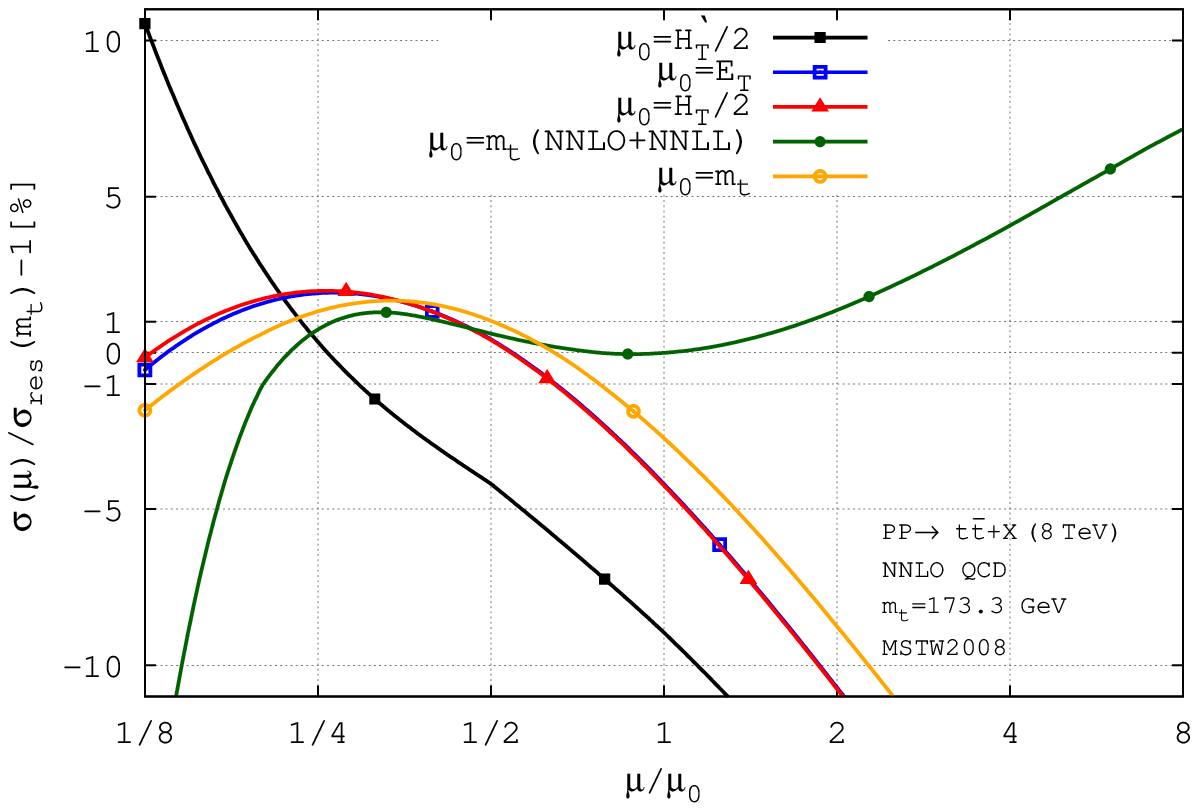}
\includegraphics[width=0.52\textwidth]{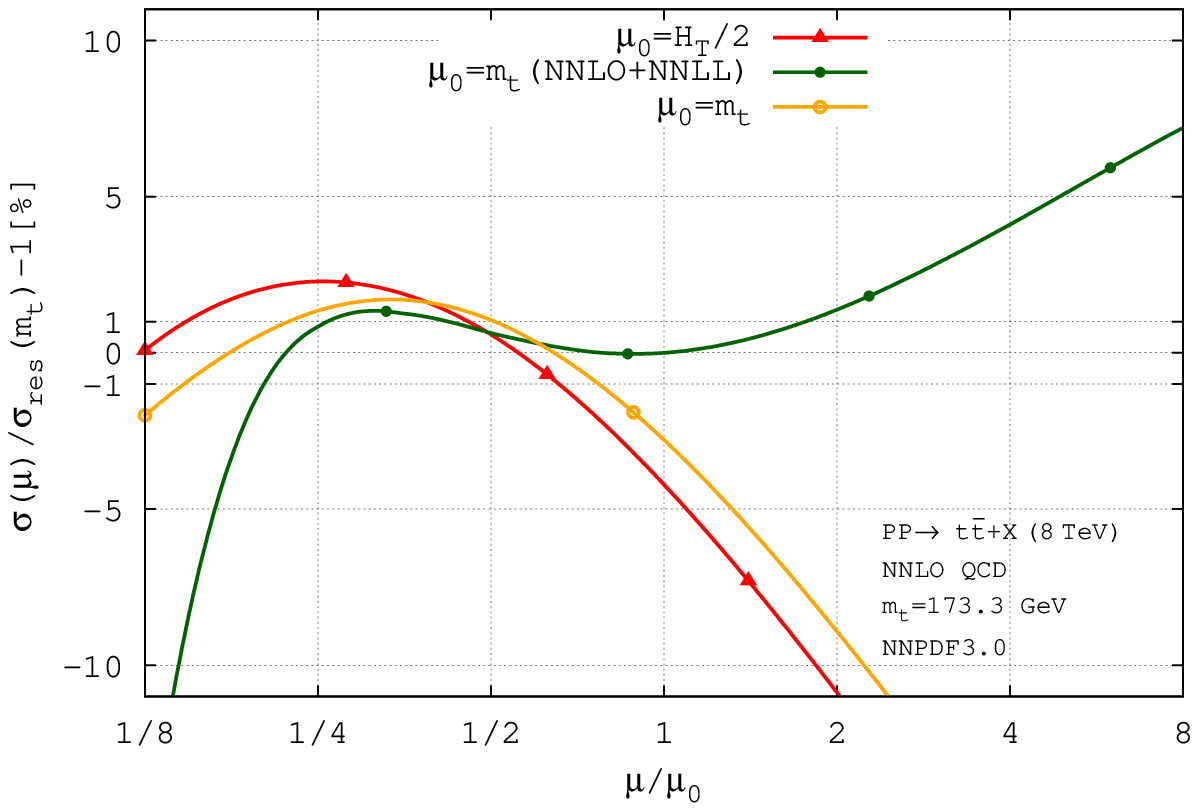}
\caption{\label{fig:stot-compare} Comparison of the total cross-section at NNLO evaluated with different dynamic scales and with two different pdf sets: MSTW2008NNLO (left) and NNPDF3.0NNLO (right). The symbols on some of the lines are meant to help distinguish the various lines.}
\end{figure}
To summarise our discussion of scale-setting for the total cross-section in fig.~\ref{fig:stot-compare} we compare all scales used so far in NNLO QCD (and NNLO+NNLL where available) and for both pdf sets. From this figure it is easy to see that at this order of perturbation theory the predictions are rather stable with respect to the choice of pdf set (at least for the pdf sets we have studied) and that the choice of a scale ensuring fastest convergence is a rather clear cut. Moreover, such scale returns value for $\sigma_{\rm tot}$ which is in nearly perfect agreement with the so-far default value for $\sigma_{\rm tot}$ evaluated with NNLO+NNLL at the scale $\mu=m_t$. From this figure it is also evident that for the fastest convergence scale eq.~(\ref{eq:mu0HT4}), the scale behaviour of the total cross-section is very regular and monotonic around the value $\mu/\mu_0=1/2$.

\subsection{Differential distributions}\label{sec:sdif}

In determining the functional form of the scale $\mu_0$ one is constrained by the following limiting cases: at $p_T\to 0$ we have $\mu_0\approx c_0 m_t$, while for very large $p_T$ we have $\mu_0\approx c_\infty p_T$. The two constants $c_{0}$ and $c_{\infty}$ are a priori unknown as is the scale's functional form that interpolates between these two limits. The limit $p_T\to 0$ is, however, strongly correlated with the total cross-section. We will thus use the scale derived in section \ref{sec:stot} in the context of the total inclusive cross-section, to fix the constant $c_0$. From eq.~(\ref{eq:mu0HT4}) we have $c_0=1/2$. 

\begin{figure}[h]
\includegraphics[width=0.50\textwidth]{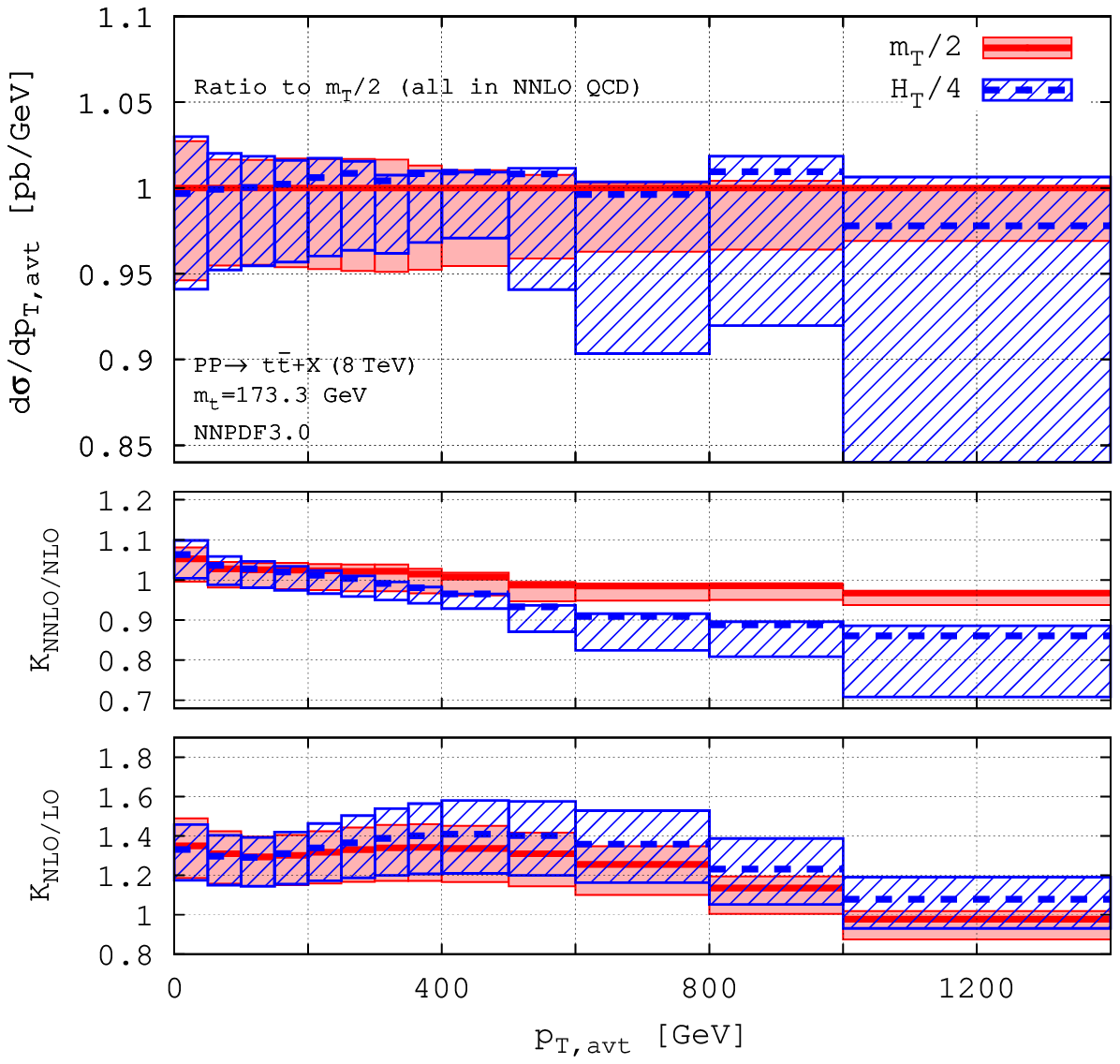}
\includegraphics[width=0.50\textwidth]{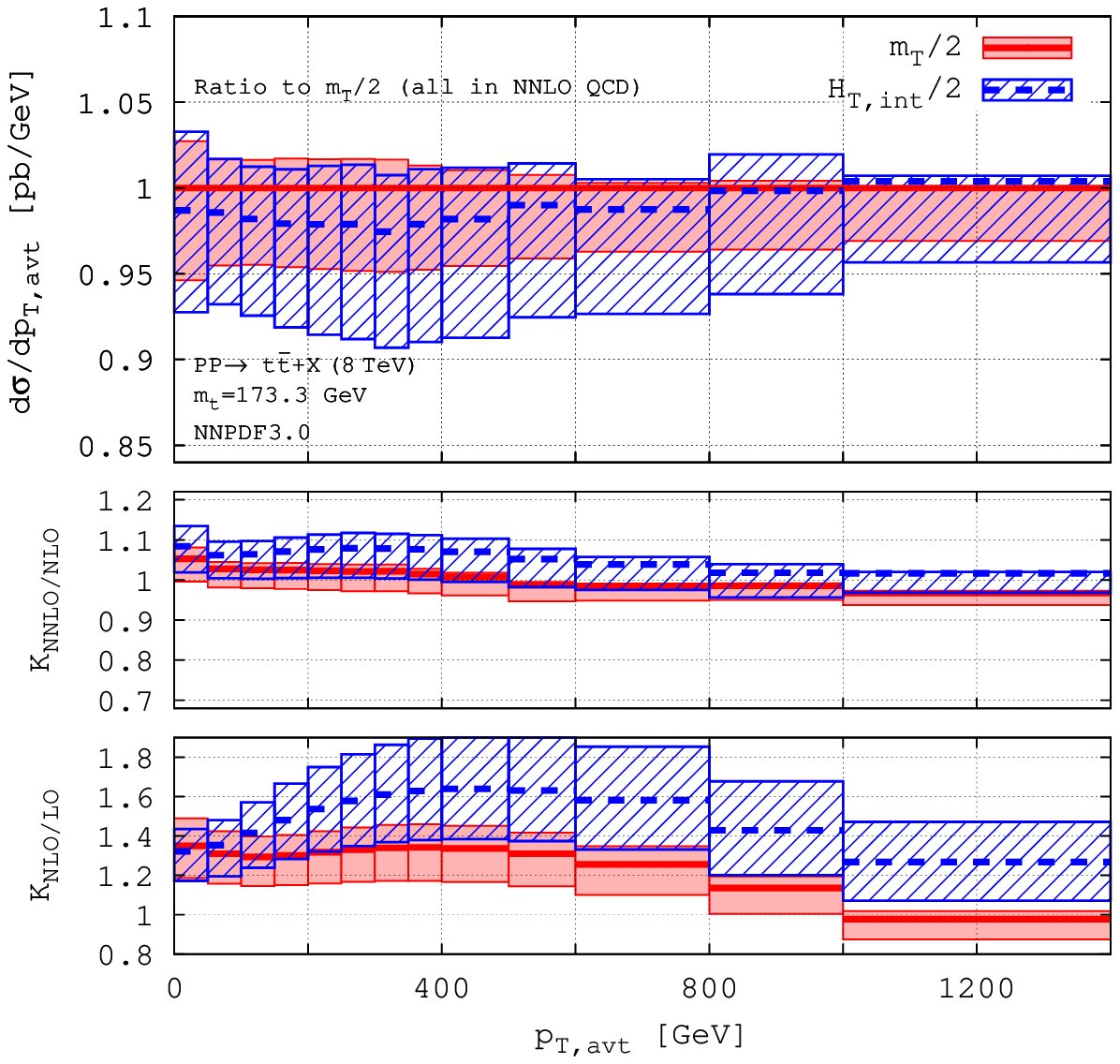}
\vskip 4mm
\includegraphics[width=0.50\textwidth]{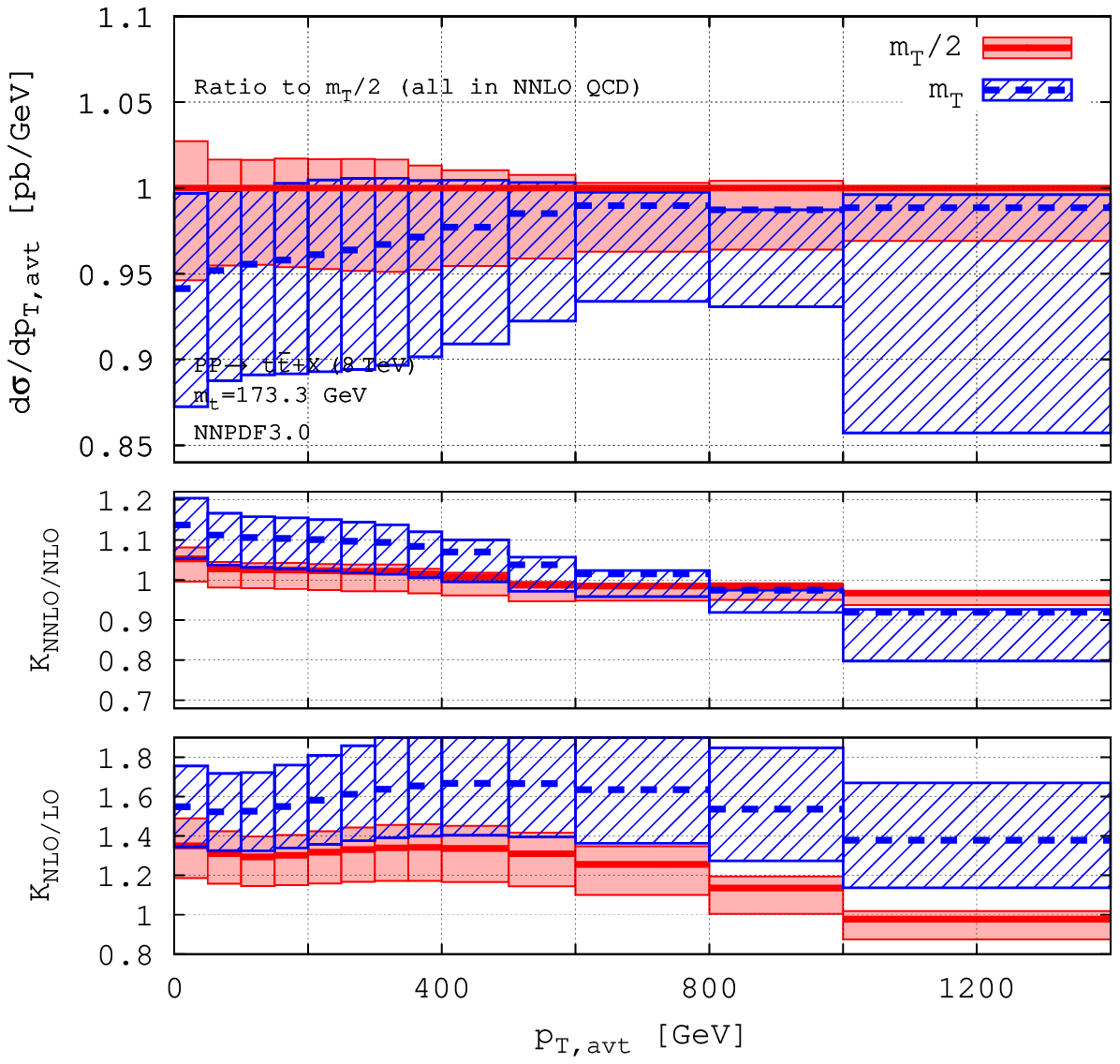}
\includegraphics[width=0.50\textwidth]{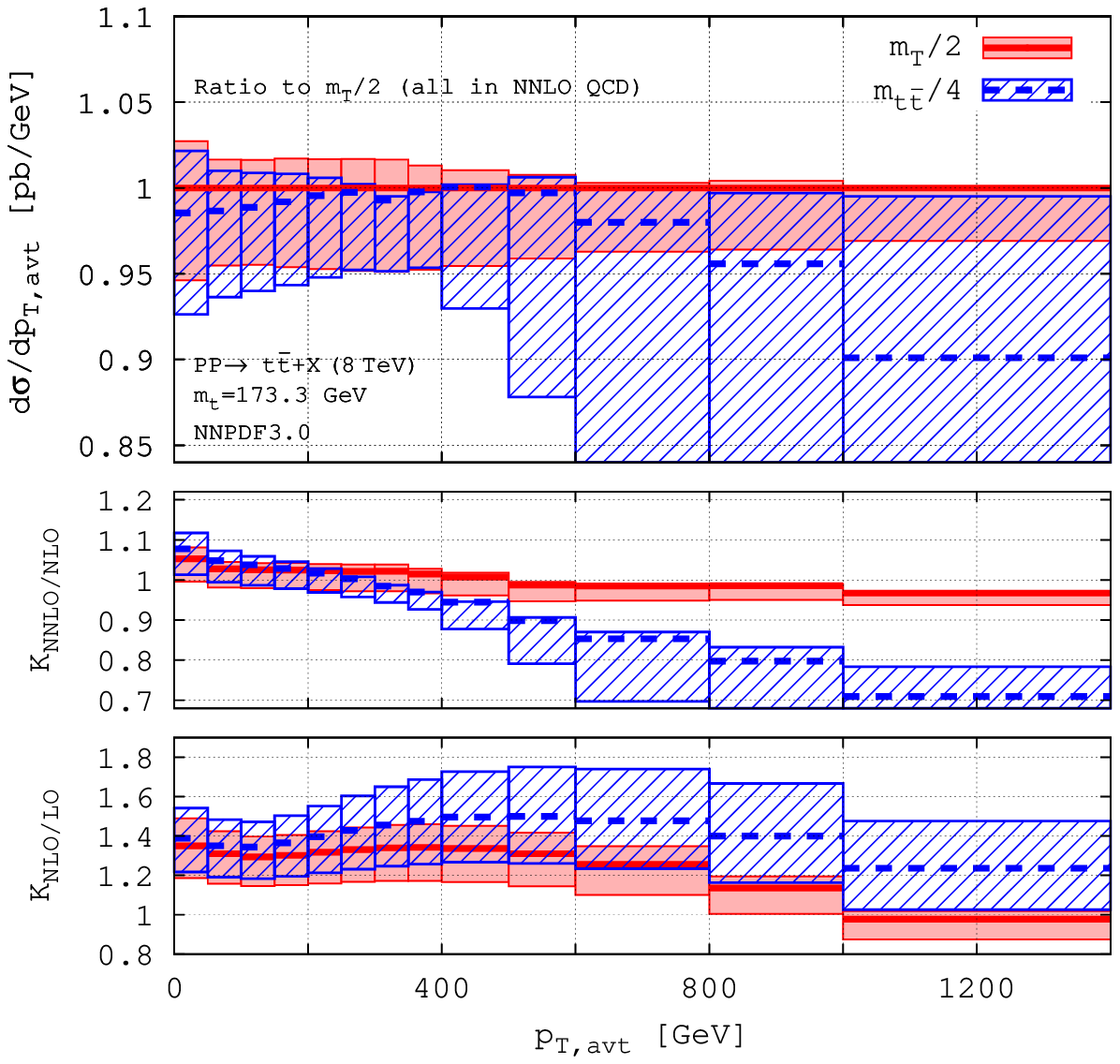}
\caption{\label{fig:diff-PTav-ratios} Comparison of the average top/antitop $p_T$ differential cross-section at NNLO evaluated with five different dynamic scales. All plots show ratios with respect to the default scale $m_T/2$ (\ref{eq:bestscale}): $H_T/4$ (top left), $H_{T,\rm int}/2$ (top right), $m_T$ (bottom left) and $\Mtt/4$ (bottom right). Error bands are from scale variation only.}
\end{figure}

The scale $\mu_0=H_T/4$ (\ref{eq:mu0HT4}) implies that $c_\infty=1/2$. One may wonder, however, if the constants $c_\infty$ and $c_0$ should necessarily be equal. Indeed, the typical value used in the past for the former constant is $c_\infty=1$. 
\footnote{We point out that the scale $H_{T,\rm int}/2$ has been introduced specifically in order to allow interpolation between $c_0=1/2$ and $c_\infty=1$.}
Since $\sigma_{\rm tot}$ is not sensitive to the large-$\PTt$ limit, one will need to investigate differential distributions and we turn to them in the following.

We would like to stress that since the limit of large $p_T$ has not yet been experimentally constrained, in this study we cannot rely on data. For this reason, our only guiding principle will be the principle of fastest perturbative convergence. As it turns out, this principle is actually quite powerful and quite clear picture of a ``good" scale emerges from our analysis. We will allow for scales with different large-$p_T$ behaviour and will nevertheless conclude that the best scale is $\mu_0=H_T/4$. We will also find that for the $\PTt$ distribution (as well as for the $\PTavt$ of the average top/antitop) the best scale will be not $H_T/4$ but $\mu_0=m_T/2$ as defined in eq.~\ref{eq:mu-mT}. Both scales $H_T/4$ and $m_T/2$ have the same asymptotic behaviour in the limits $\PTt\to 0$ and $\PTt\to\infty$ thus arriving at the following ``best" scale
\begin{equation}
\mu_0 = \left\{ 
  \begin{array}{l l}
    {m_T\over 2} & ~ {\rm for:}\quad \PTt,~\PTtbar~{\rm and}~\PTavt\, , \\
    & \\
    {H_T\over 4} & ~ {\rm for:}\quad {\rm all~other~distributions}\, .
  \end{array} \right.
  \label{eq:bestscale}
\end{equation}

Eq.~(\ref{eq:bestscale}) above is the main result of this work. In the following we present its justification by the way of analysing differential distributions. We also compare three different pdf sets: NNPDF3.0 \cite{Ball:2014uwa}, CT14 \cite{Dulat:2015mca} and MMHT2014 \cite{Harland-Lang:2014zoa}.

\begin{figure}[h]
\includegraphics[width=0.50\textwidth]{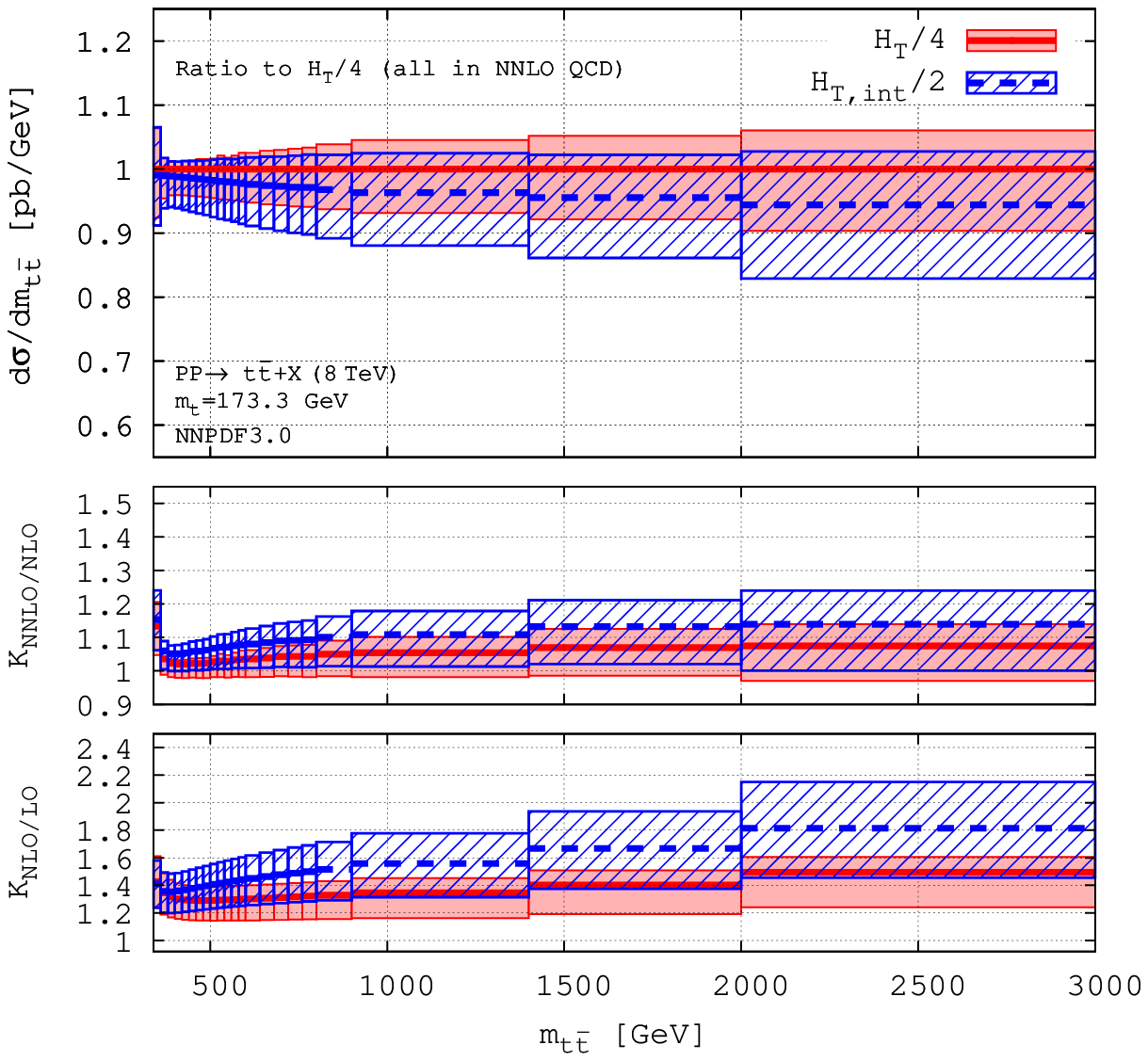}
\includegraphics[width=0.50\textwidth]{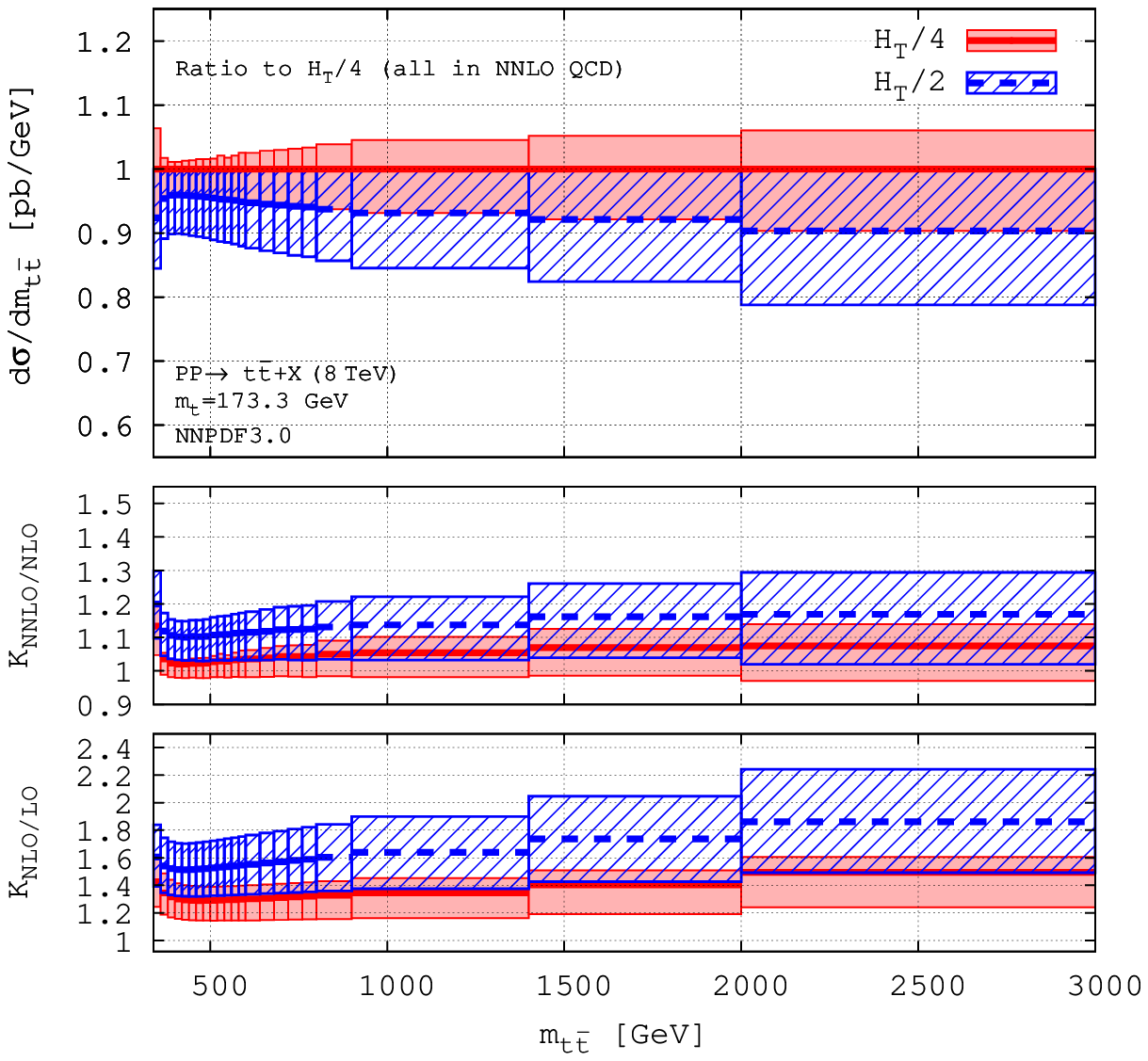}
\vskip 4mm
\includegraphics[width=0.50\textwidth]{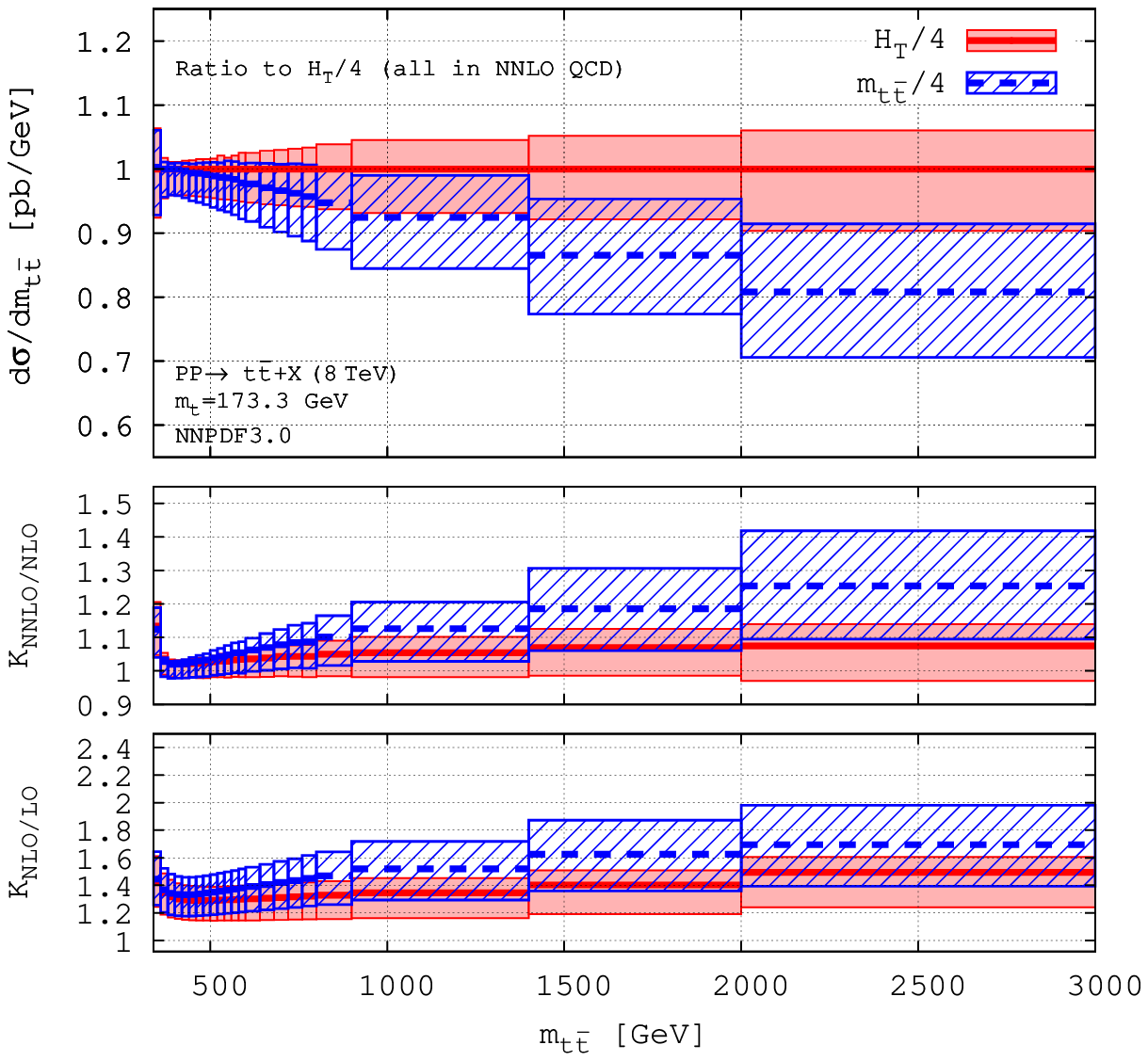}
\includegraphics[width=0.50\textwidth]{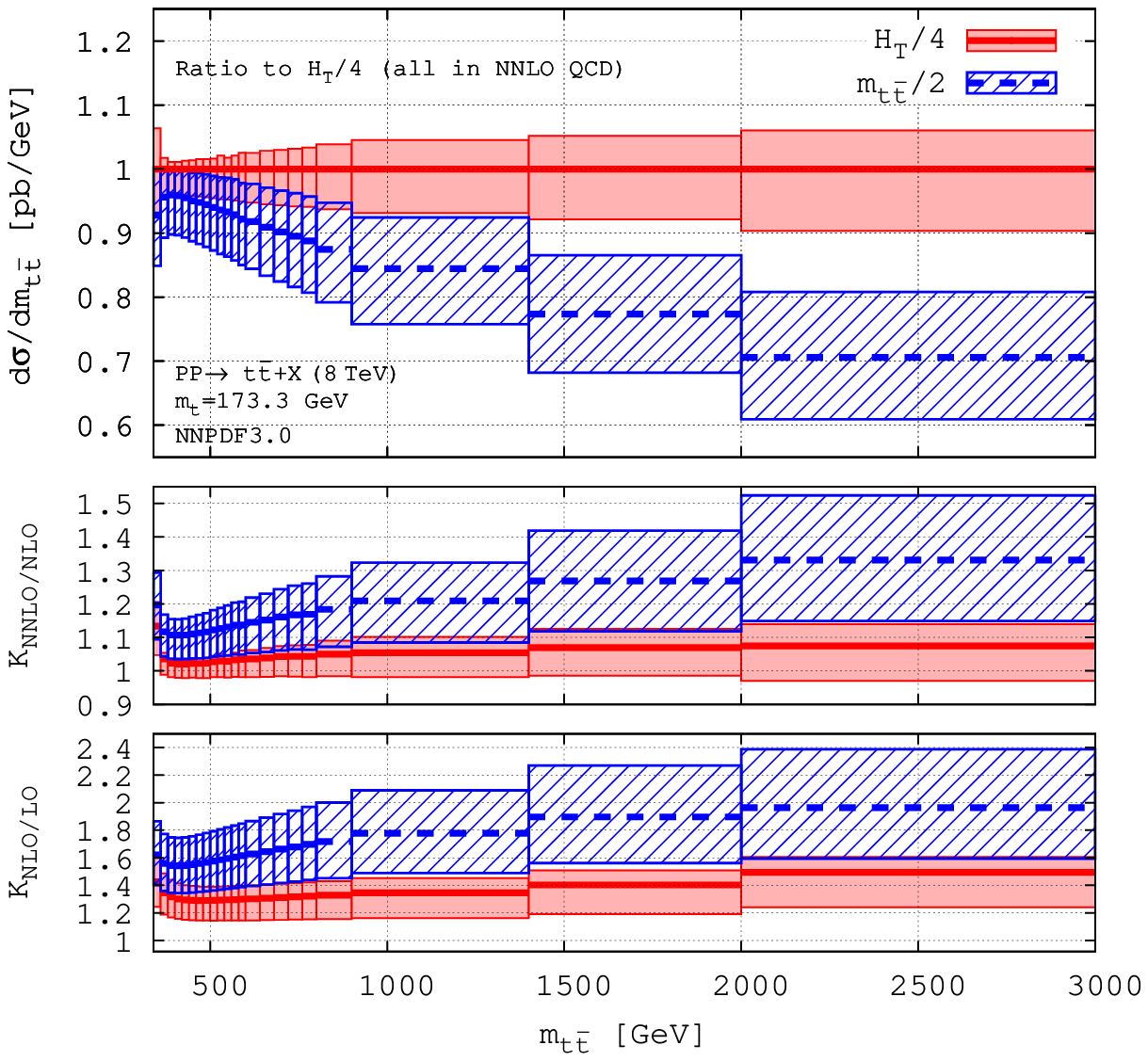}
\caption{\label{fig:diff-Mtt-ratios} Comparison of the $\Mtt$ differential cross-section at NNLO evaluated with five different dynamic scales. All plots show ratios with respect to the default scale $H_T/4$ (\ref{eq:bestscale}): $H_{T,\rm int}/2$ (top left), $H_{T}/2$ (top right), $\Mtt/4$ (bottom left) and $\Mtt/2$ (bottom right). Error bands are from scale variation only.}
\end{figure}

In fig.~\ref{fig:diff-PTav-ratios} we compare predictions for $\PTavt$ computed with five different dynamic scales: $m_T/2,~m_T,~H_T/4,~H_{T,\rm int}/2$ and $\Mtt/4$. We observe that the scale $m_T/2$ consistently leads to K-factors that are closest to unity, i.e. it fits best the requirement for fastest perturbative convergence in the full kinematic range. A K-factor between orders $a$ and $b$, $a\geq b$, is defined:
\begin{equation}
{\rm K}_{{\rm N}^a{\rm LO}/{\rm N}^b{\rm LO}}(\mu) = {d\sigma_{{\rm N}^a{\rm LO}}(\mu)\over d\sigma_{{\rm N}^b{\rm LO}}(\mu_{\rm central})} \,.
\end{equation}
We also notice that the scale $m_T/2$ leads to cross-section with the smallest scale variation. It is worth noting that the difference between the central values for the NNLO $p_T$ distribution based on the scales $m_T/2$ and $H_{T}/4$ never exceeds 2\% for $\PTavt < 1 \TeV$, i.e. the effect of the scale choice at NNLO is rather limited. 

Similarly, in fig.~\ref{fig:diff-Mtt-ratios} we compare predictions for $\Mtt$ also computed with five different dynamic scales: $H_T/4,~H_T/2,~H_{T,\rm int}/2,~\Mtt/2$ and $\Mtt/4$. We observe that the scale $H_T/4$ consistently leads to K-factors that are closest to unity, i.e. it fits best the requirement for fastest perturbative convergence. We also notice that this scale leads to cross-section with the smallest scale variation.

The comparison in fig.~\ref{fig:diff-Mtt-ratios} demonstrates that $\Mtt$-based scales lead to poor perturbative convergence. Even for an $\Mtt$-based scale that is as small as $\Mtt/4$ the deviation between the absolute predictions is large and exceeds the size of the scale error. Such scales have been used in the past \cite{Ahrens:2010zv,Ferroglia:2013zwa} as well as recently in the resummation-based work \cite{Ferroglia:2015ivv,Pecjak:2016nee}. Our findings seem to indicate that the large corrections found in refs.~\cite{Ferroglia:2015ivv,Pecjak:2016nee} are actually due to the particular scale choice. It will be interesting to check if a different scale choice (like, for example, $H_T/4$) will lead to much smaller resummation corrections.

\section{Pdf related issues}\label{sec:pdf}

\begin{figure}[t]
\includegraphics[width=0.50\textwidth]{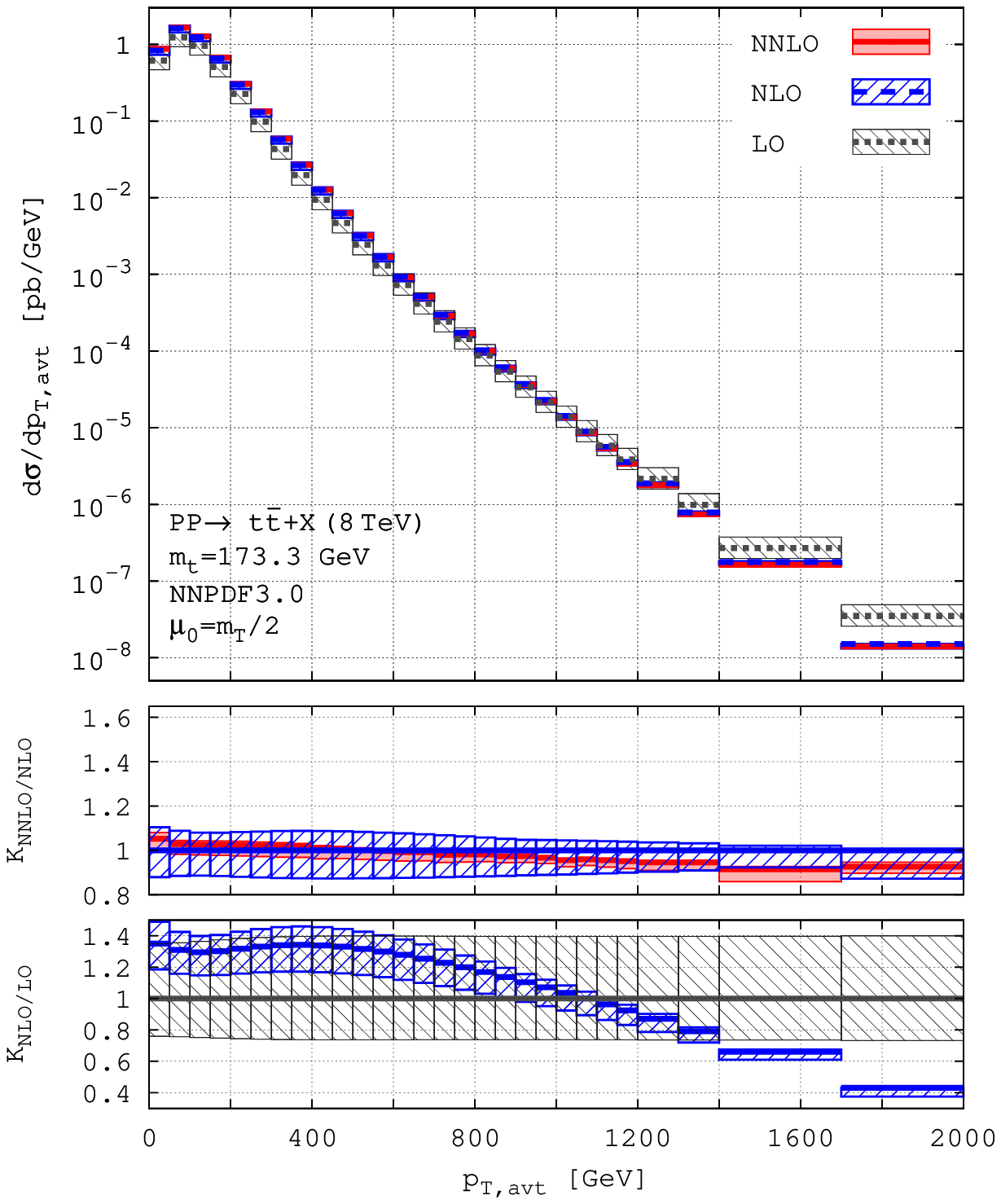}
\includegraphics[width=0.50\textwidth]{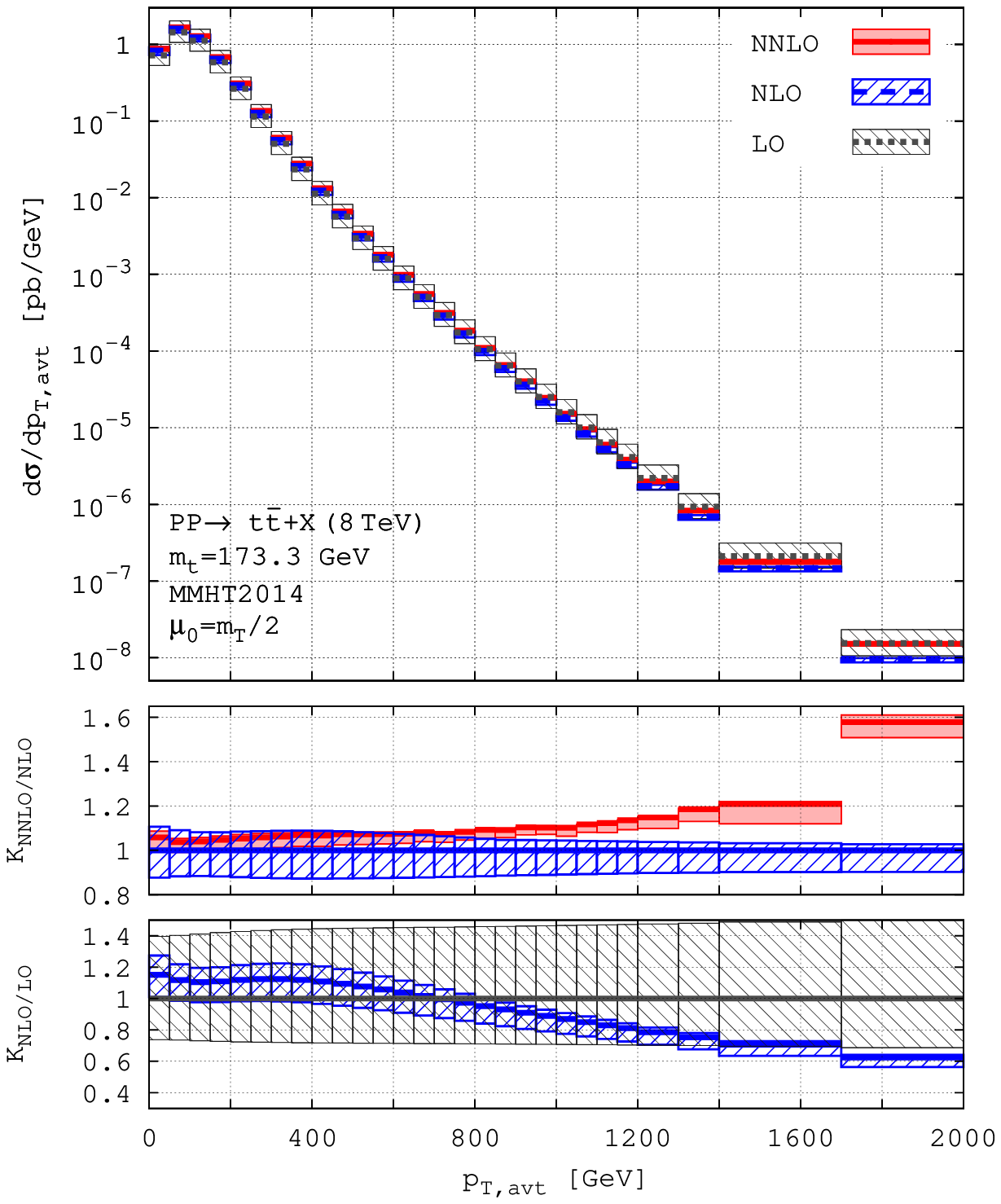}
\includegraphics[width=0.50\textwidth]{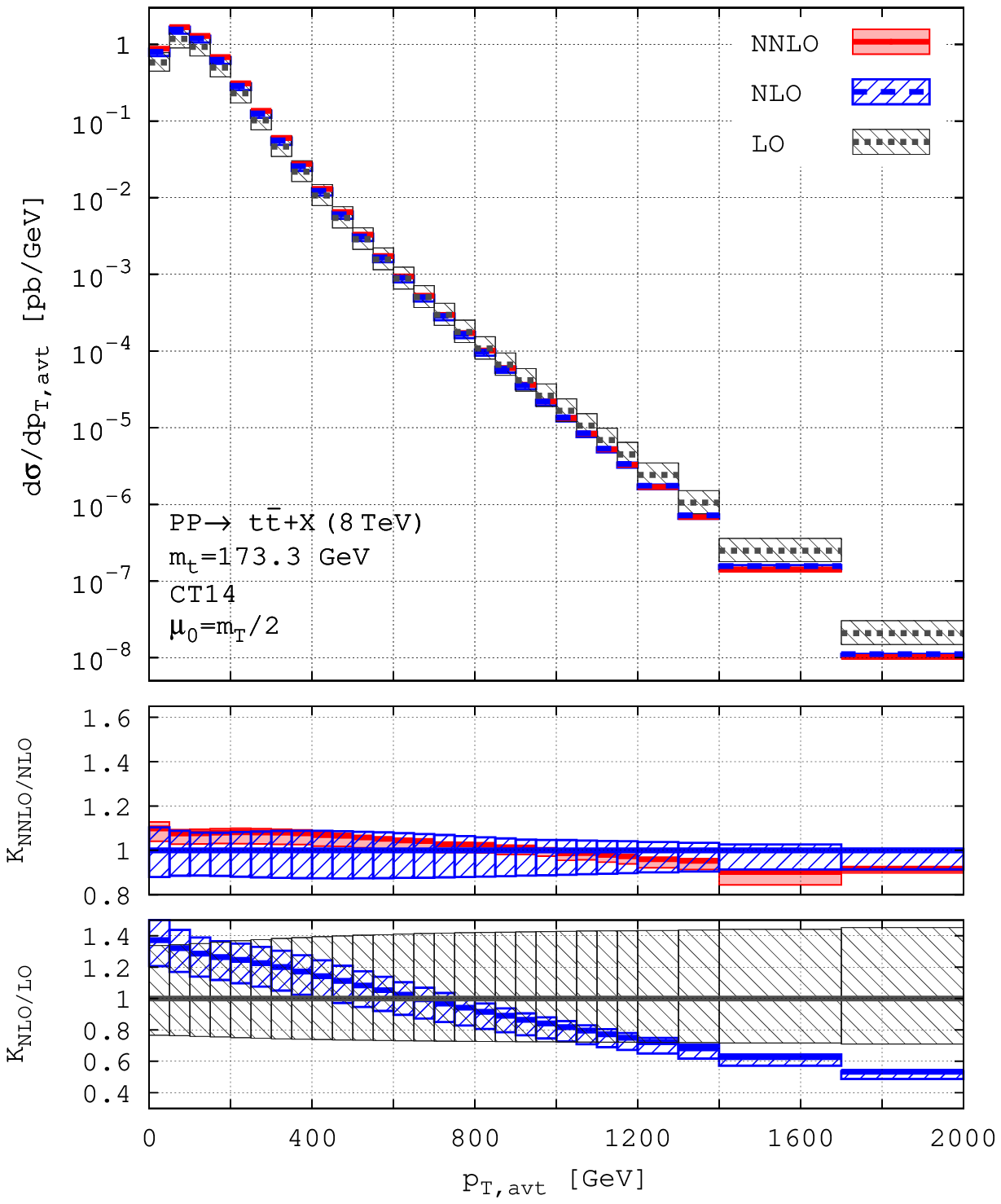}
\includegraphics[width=0.50\textwidth]{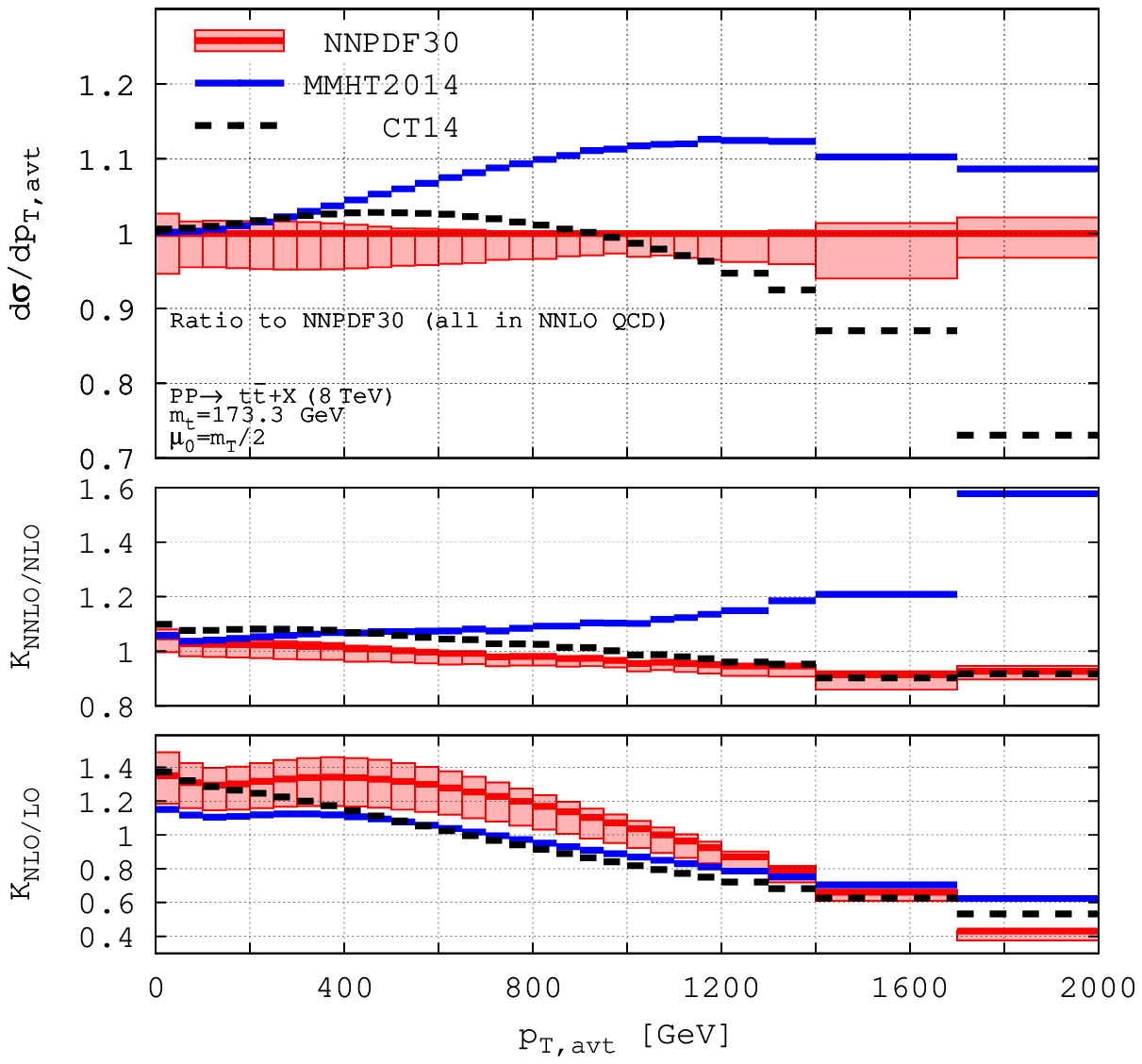}
\caption{\label{fig:diff-PTavt-8TeV-abs} $\PTavt$ distribution for LHC 8 TeV computed with three pdf sets: NNPDF 3.0 (top left), MMHT2014 (top right) and CT14 (bottom left). The ratios of these distributions with respect to NNPDF3.0 are also shown (bottom right). Error bands are from scale variation only.}
\end{figure}
\begin{figure}[t]
\includegraphics[width=0.50\textwidth]{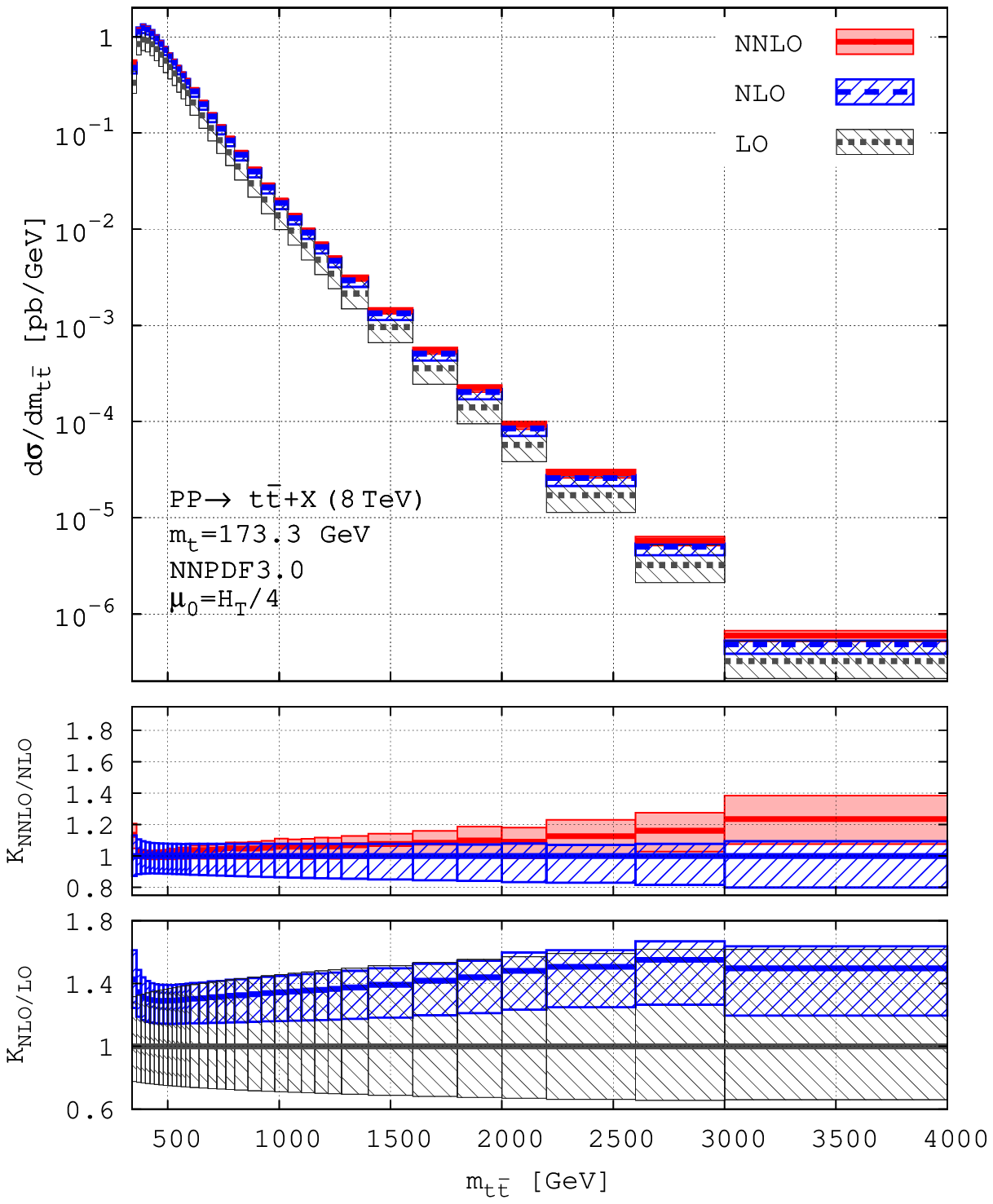}
\includegraphics[width=0.50\textwidth]{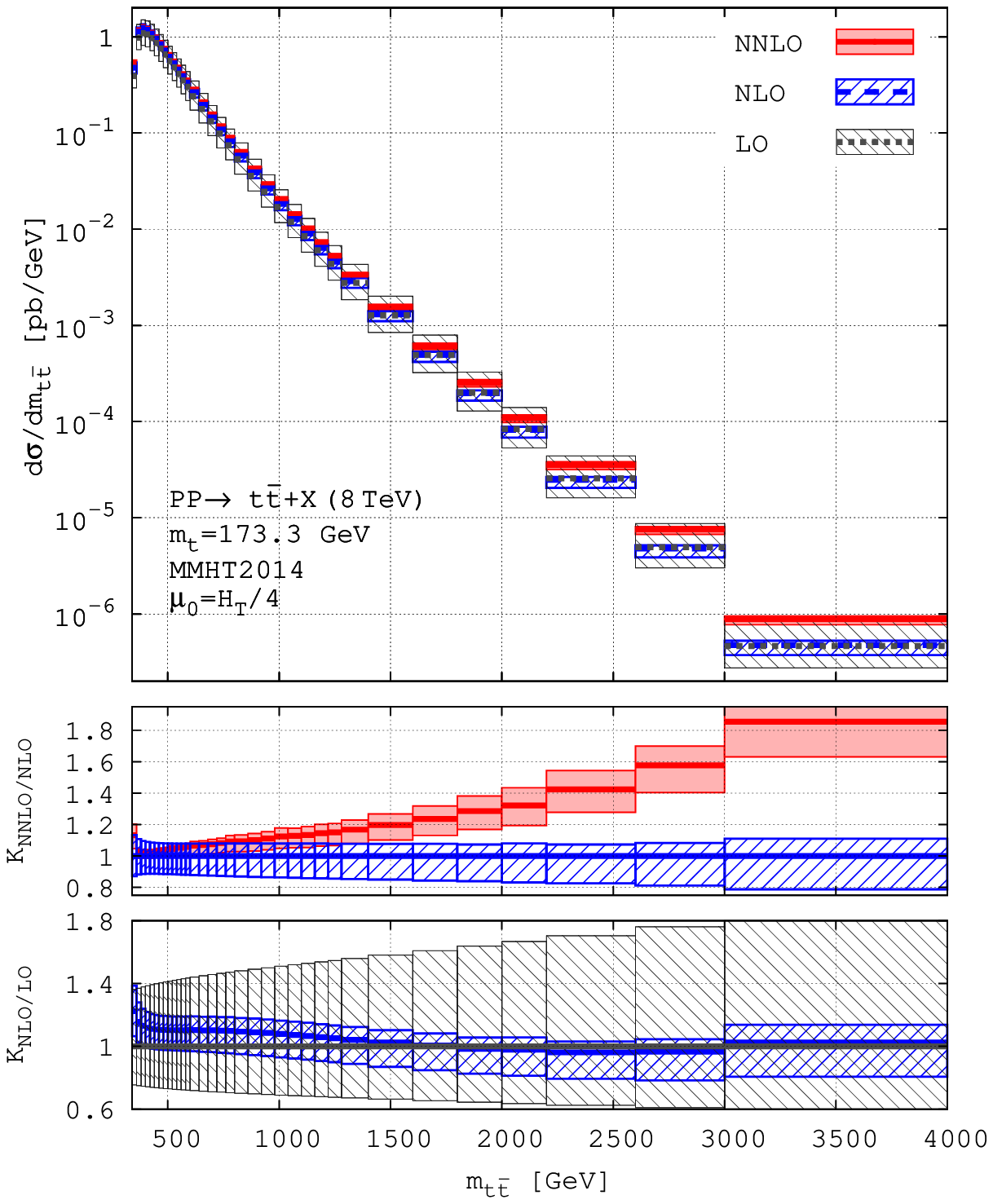}
\includegraphics[width=0.50\textwidth]{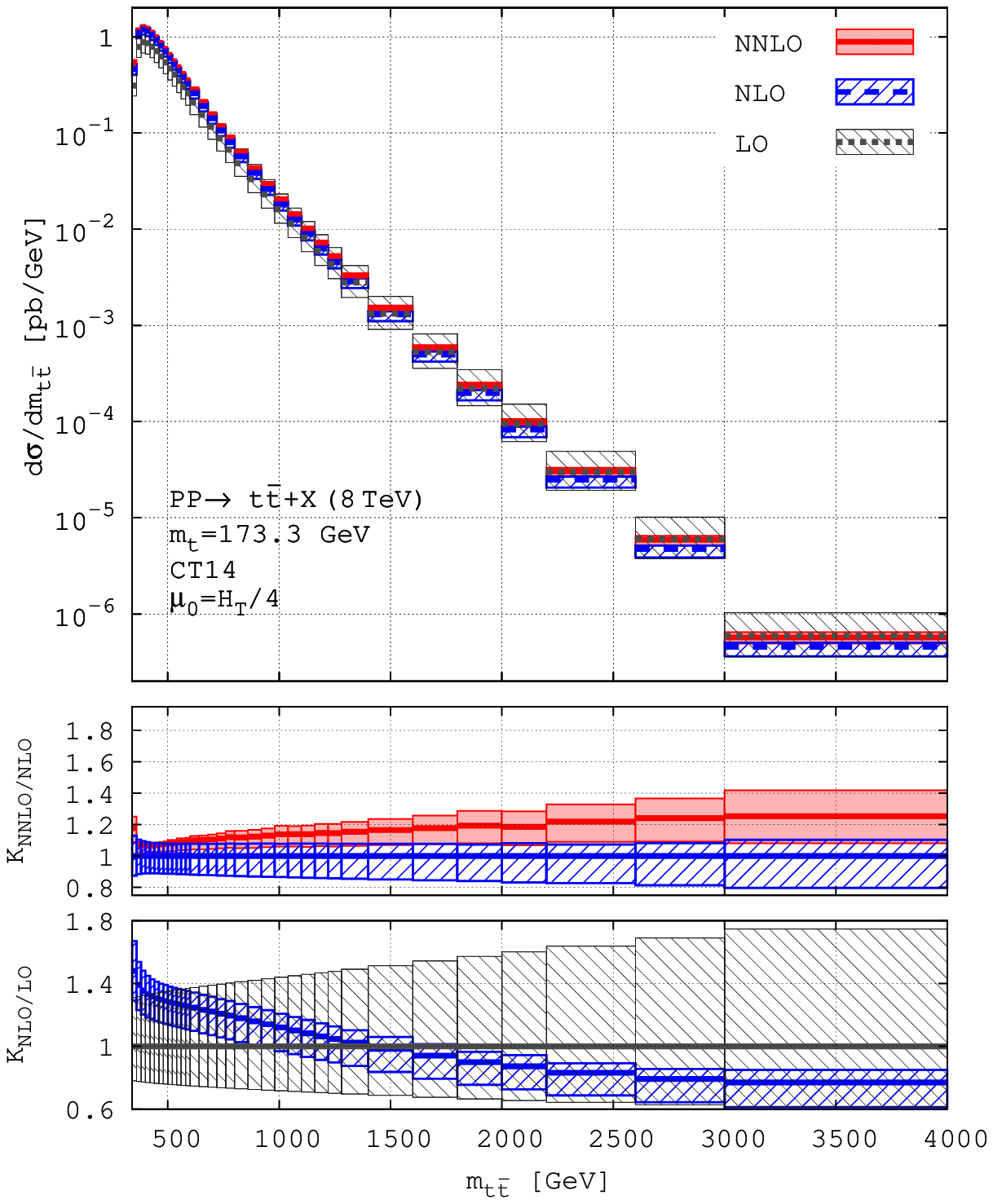}
\includegraphics[width=0.50\textwidth]{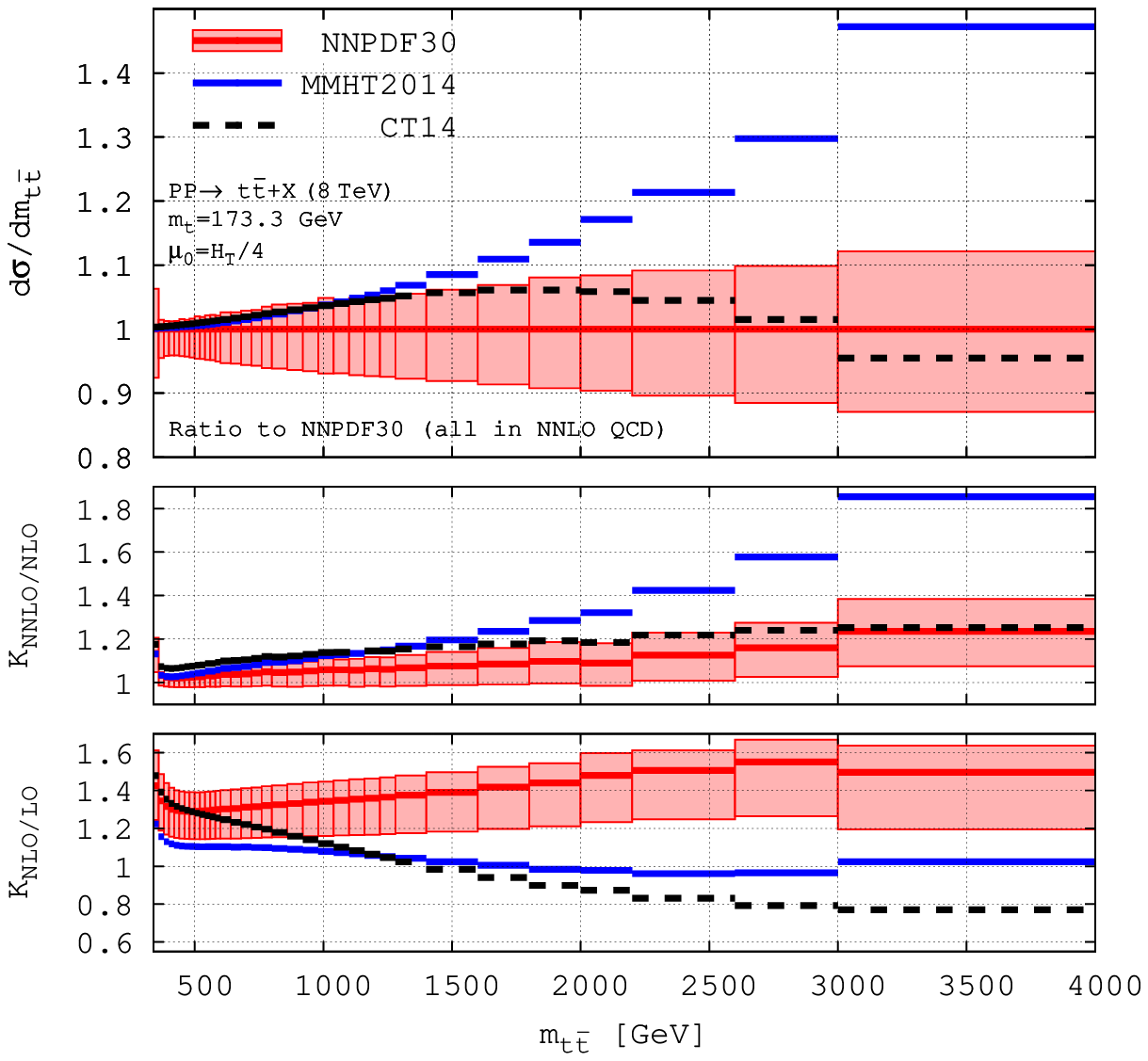}
\caption{\label{fig:diff-Mtt-8TeV-abs} As in fig.~\ref{fig:diff-PTavt-8TeV-abs}  but for the $\Mtt$ distribution.}
\end{figure}
\begin{figure}[t]
\includegraphics[width=0.50\textwidth]{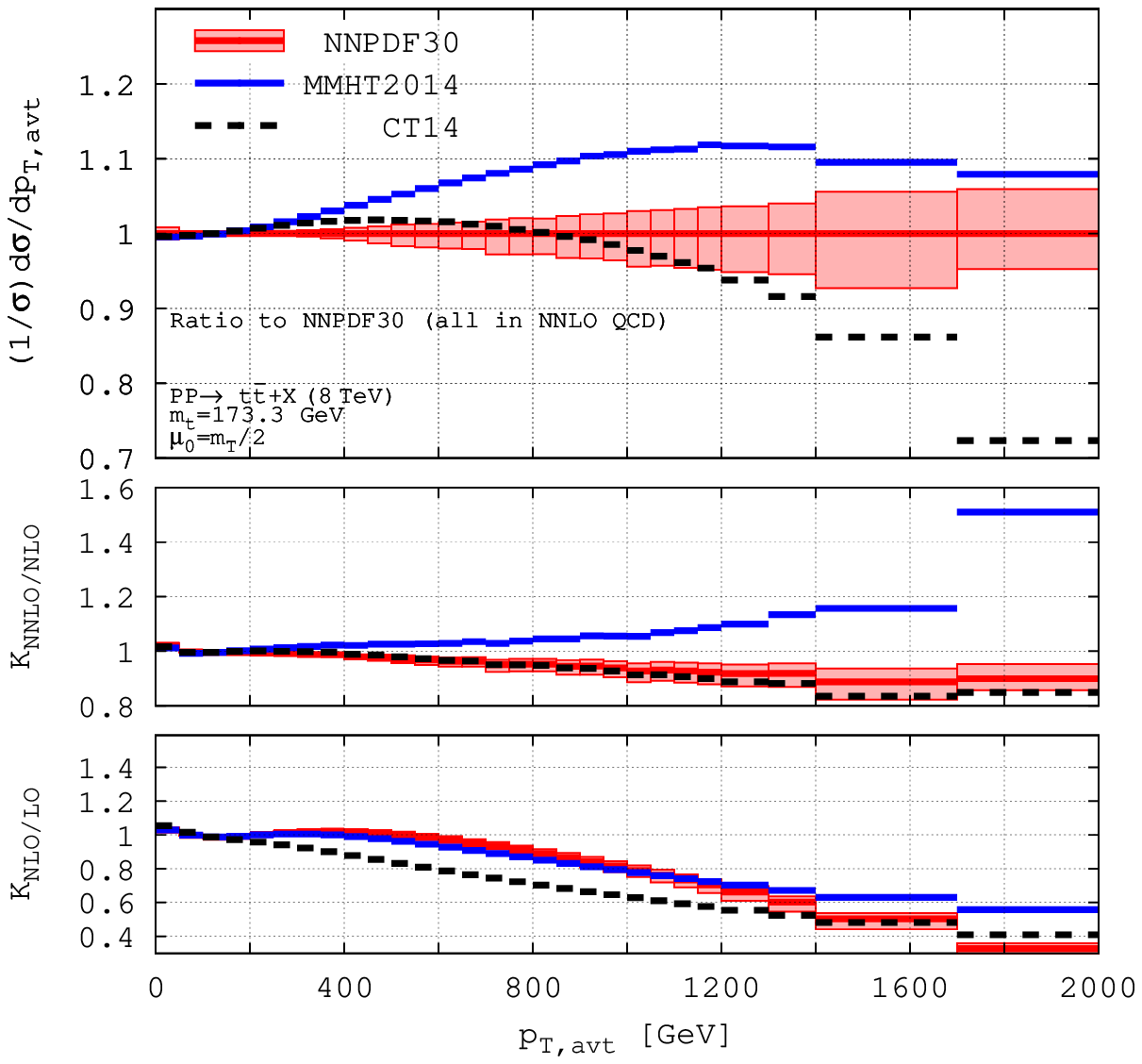}
\includegraphics[width=0.50\textwidth]{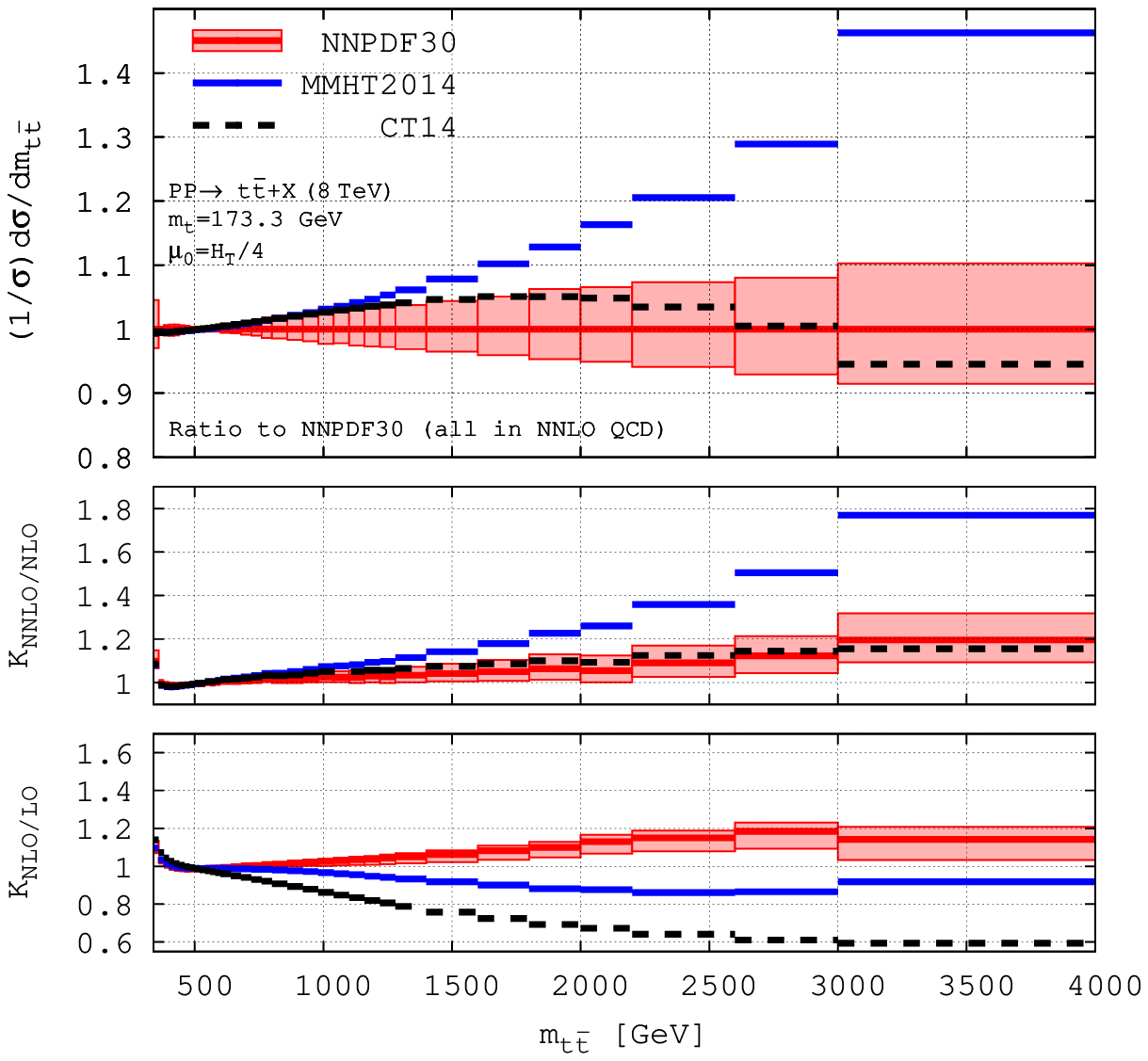}
\caption{\label{fig:diff-PTandMtt-8TeV-norm} As in figs.~\ref{fig:diff-PTavt-8TeV-abs},\ref{fig:diff-Mtt-8TeV-abs} but for the ratios of the normalised $\PTavt$ (left) and $\Mtt$ (right) distributions.}
\end{figure}

A major concern in a scale study like ours is if the conclusions drawn above apply independently of the pdf set. In figs.~\ref{fig:diff-PTavt-8TeV-abs},\ref{fig:diff-Mtt-8TeV-abs} we show the unnormalised $\PTavt$ and $\Mtt$ differential distributions based on the following three pdf sets: NNPDF 3.0, CT14 and MMHT2014. To facilitate the comparison between the three predictions, we also show the ratios of both unnormalised (figs.~\ref{fig:diff-PTavt-8TeV-abs},\ref{fig:diff-Mtt-8TeV-abs}) and normalised (fig.~\ref{fig:diff-PTandMtt-8TeV-norm}) distributions with respect to NNPDF3.0.

\begin{figure}[t]
\includegraphics[width=0.50\textwidth]{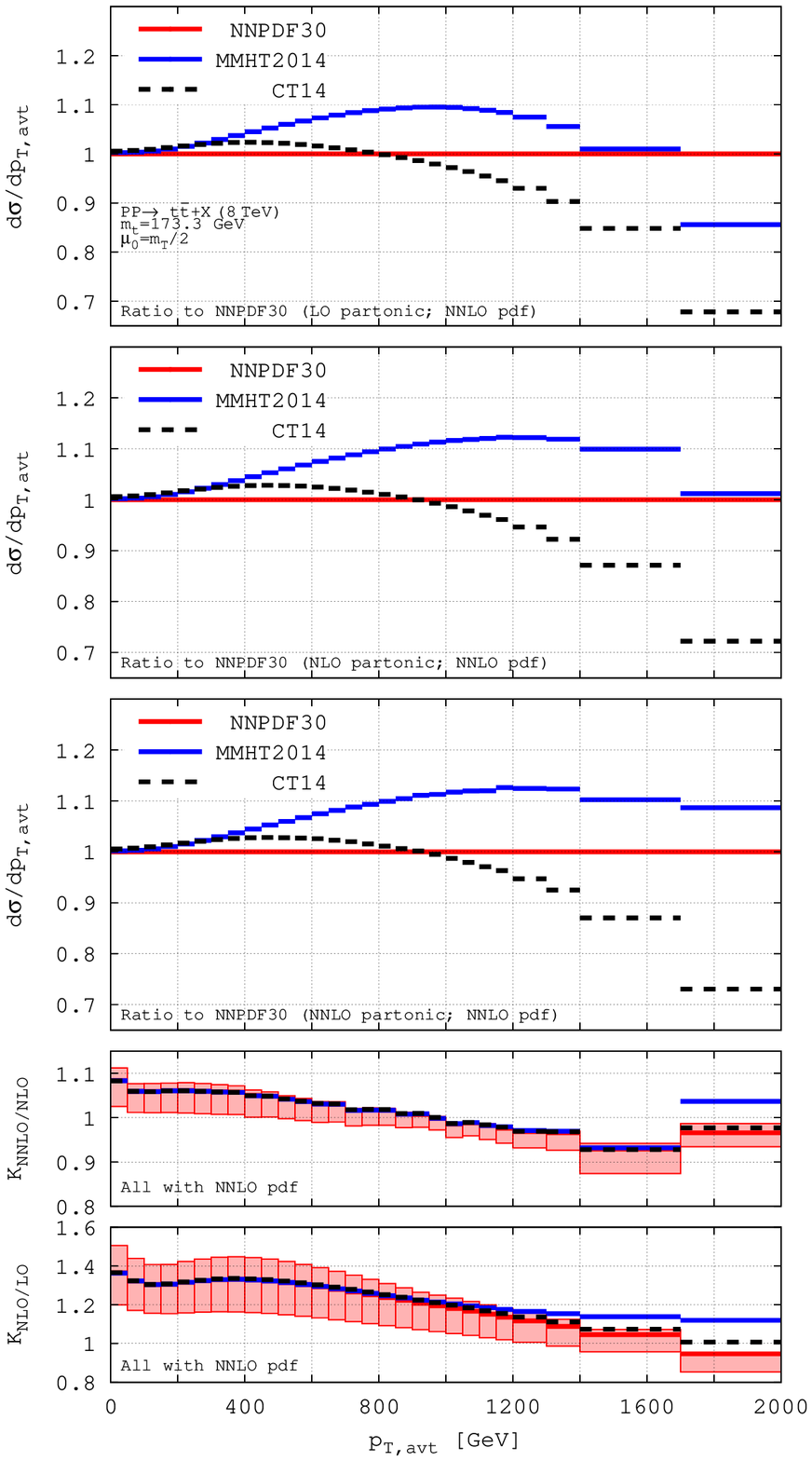}
\includegraphics[width=0.50\textwidth]{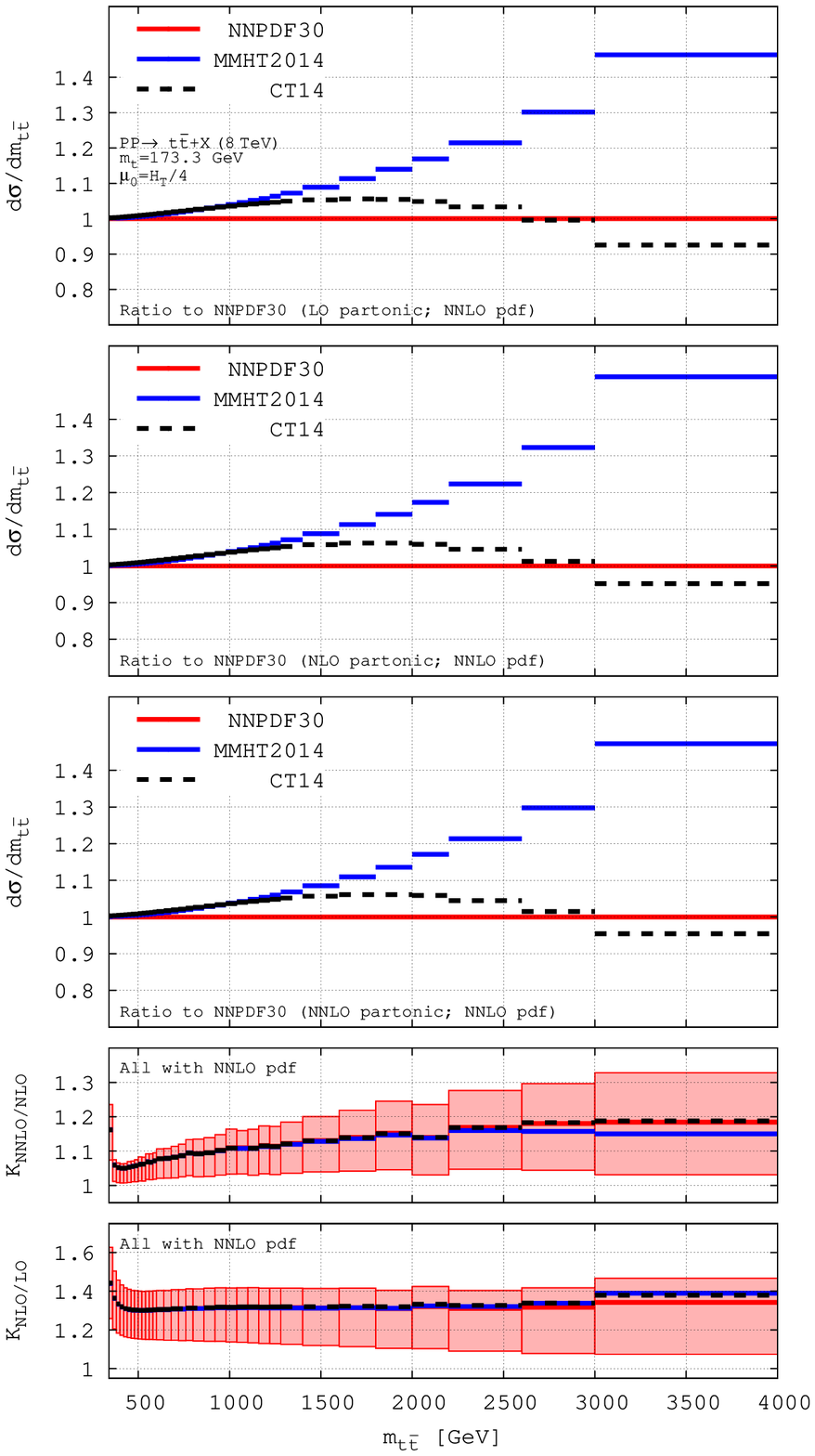}
\caption{\label{fig:diff-8TeV-all-nnlo-pdf} Absolute $\PTavt$ (left) and $\Mtt$ (right) distributions. All curves are computed with NNLO pdf; partonic cross-sections are at LO, NLO or NNLO.  Error bands are from scale variation only.}
\end{figure}

It is immediately clear that the differential distributions are significantly impacted by the choice of pdf. Furthermore, the K-factors of these three sets behave very differently. In the following we will show that these differences are due to the pdf sets themselves and are not related to the choice of dynamic scale. To that end in fig.~\ref{fig:diff-8TeV-all-nnlo-pdf} we show the $\PTavt$ and $\Mtt$ distributions always computed with NNLO pdf set while varying the order of the perturbative cross-section (from LO to NNLO). The rationale for doing this is that in a ratio where the same pdf is used both in numerator and denominator, the dependence of the pdf is reduced or even completely drops out, i.e. the ratio is effectively dependent only on the partonic cross-sections. Similarly, in a ratio where the same partonic cross-sections are used in both the numerator and denominator (but different pdf's) the dependence of the partonic cross-section is effectively removed and the ratio becomes a function of the pdf's only. In fig.~\ref{fig:diff-8TeV-all-nnlo-pdf} we observe that such cancellations indeed take place: the top three plots show near-independence with respect to the choice of the perturbative cross-section (from LO through NNLO) while the bottom two plots show the near-independence of K-factors with respect to the choice of pdf set. Fig.~\ref{fig:diff-8TeV-all-nnlo-pdf} thus confirms that the large differences between differential distributions and K-factors apparent from figs.~\ref{fig:diff-PTavt-8TeV-abs},\ref{fig:diff-Mtt-8TeV-abs},\ref{fig:diff-PTandMtt-8TeV-norm} are of pdf origin. 

\begin{figure}[t]
\includegraphics[width=0.32\textwidth]{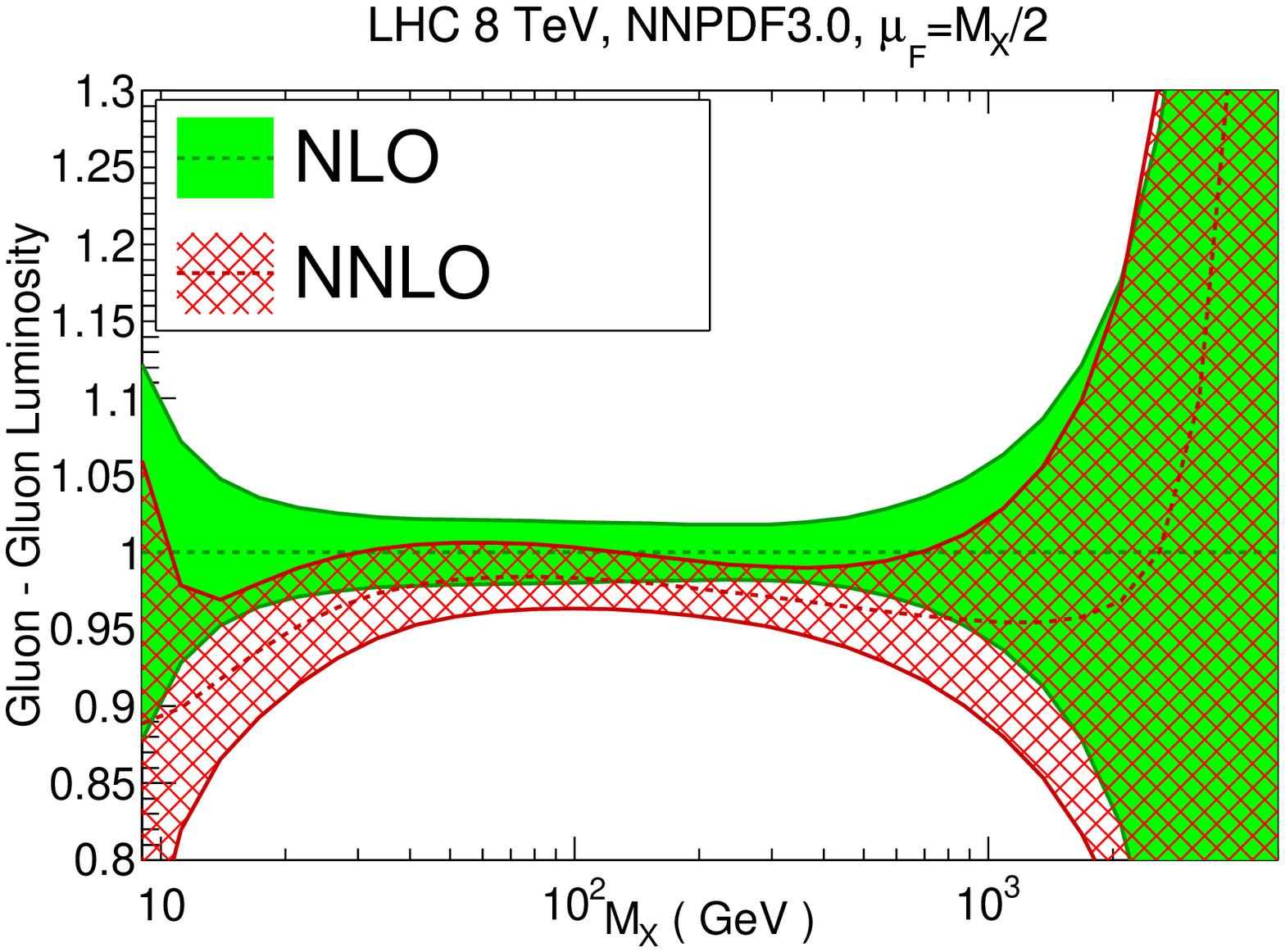}
\includegraphics[width=0.32\textwidth]{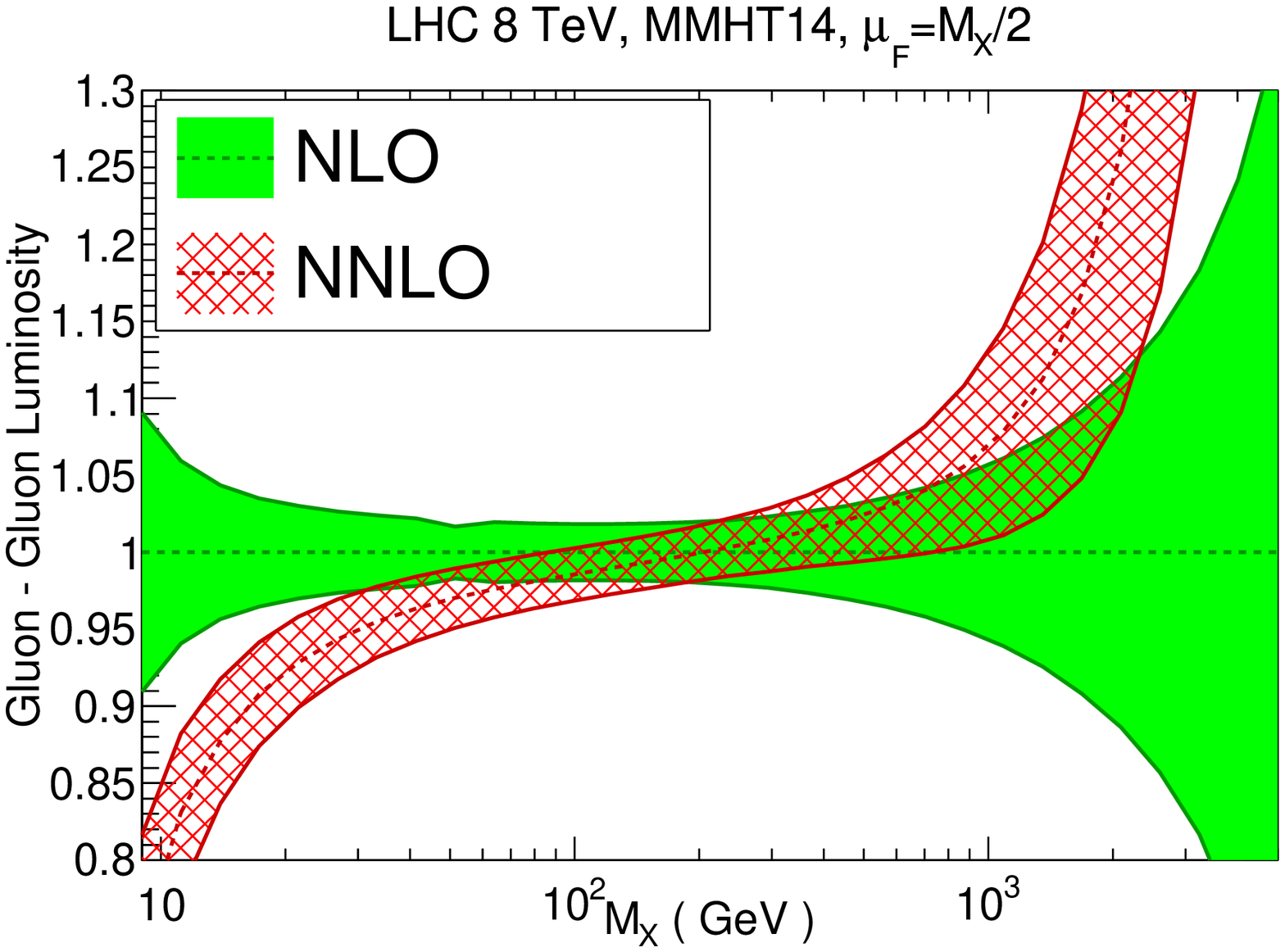}
\includegraphics[width=0.32\textwidth]{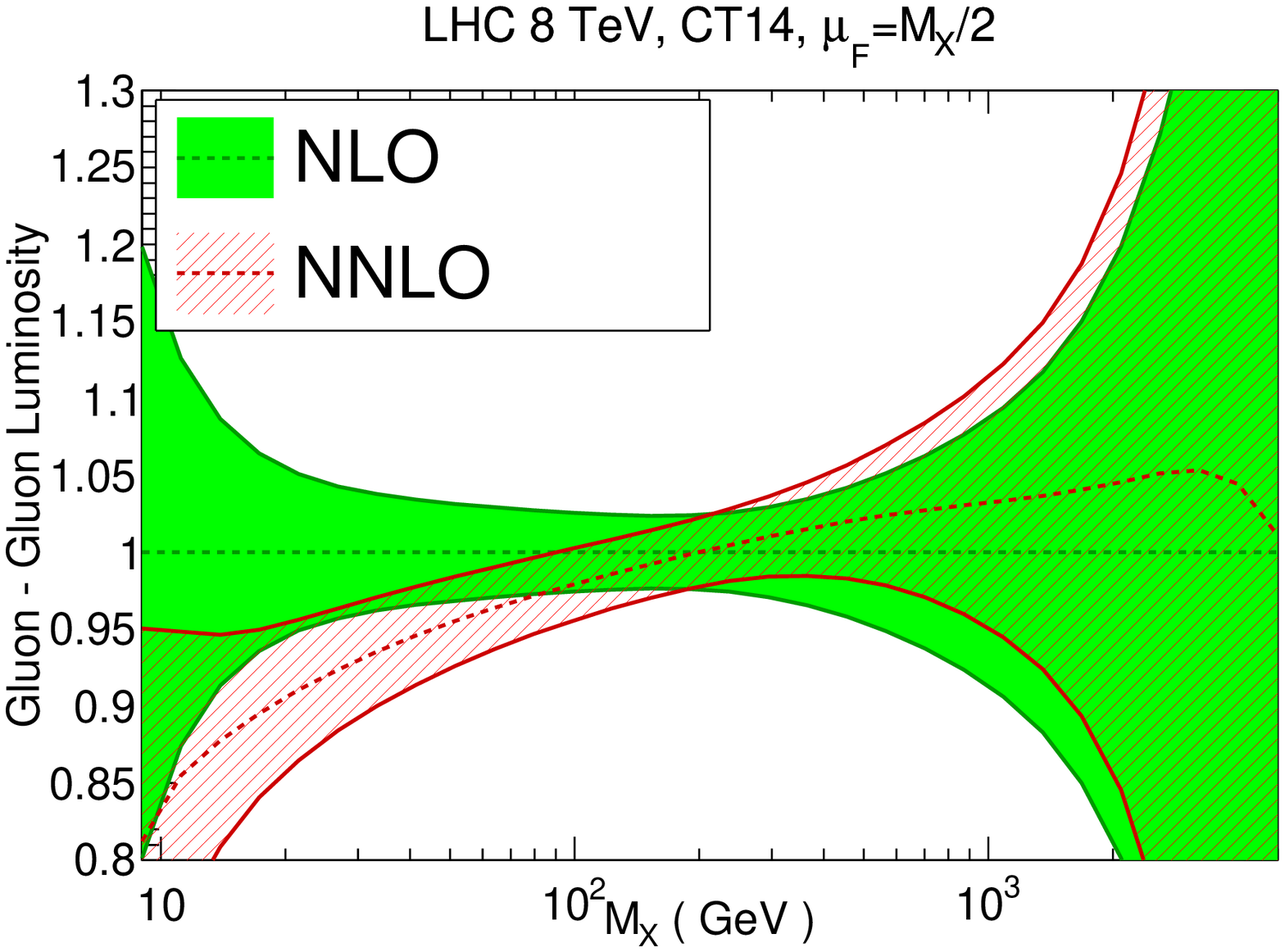}
\caption{\label{fig:gg-lumi-8TeV} LHC 8 TeV gg-luminosities for NNPDF3.0 (left), MMHT2014 (centre) and CT14 (right) as a function of the mass $M_X$ of a fictitious final state $gg\to X$. For each plot, the PDF luminosities have been normalised to the central value of the NLO result. The factorisation scale is $M_X/2$.}
\end{figure}

To further demonstrate this, in fig.~\ref{fig:gg-lumi-8TeV} we show the $gg$-luminosities for the three pdf sets.
\footnote{The plots in fig.~\ref{fig:gg-lumi-8TeV} are prepared with the help of the APFEL library \cite{Bertone:2013vaa}; we thank Juan Rojo for kindly providing us with these plots.}
We notice that above around 1 TeV the NLO and NNLO luminosities of the MMHT2014 set are incompatible with each other within the pdf error. At any rate it is evident that the growing pdf error plays a major role and that the predicted differential distributions at large values of $\PTavt$ and $\Mtt$ are likely impacted by significant uncertainty due to the imperfect knowledge of pdf. It is clear that with the large amount of top data expected during Run II of the LHC, top-quark data has very strong potential for constraining pdfs. In this work we only highlight this problem and verify that the pdf uncertainty does not affects our optimal scale-choice. Detailed analysis of pdf and how they can be improved with top data should be the subject of a dedicated study.

\begin{figure}[t]
\includegraphics[width=0.50\textwidth]{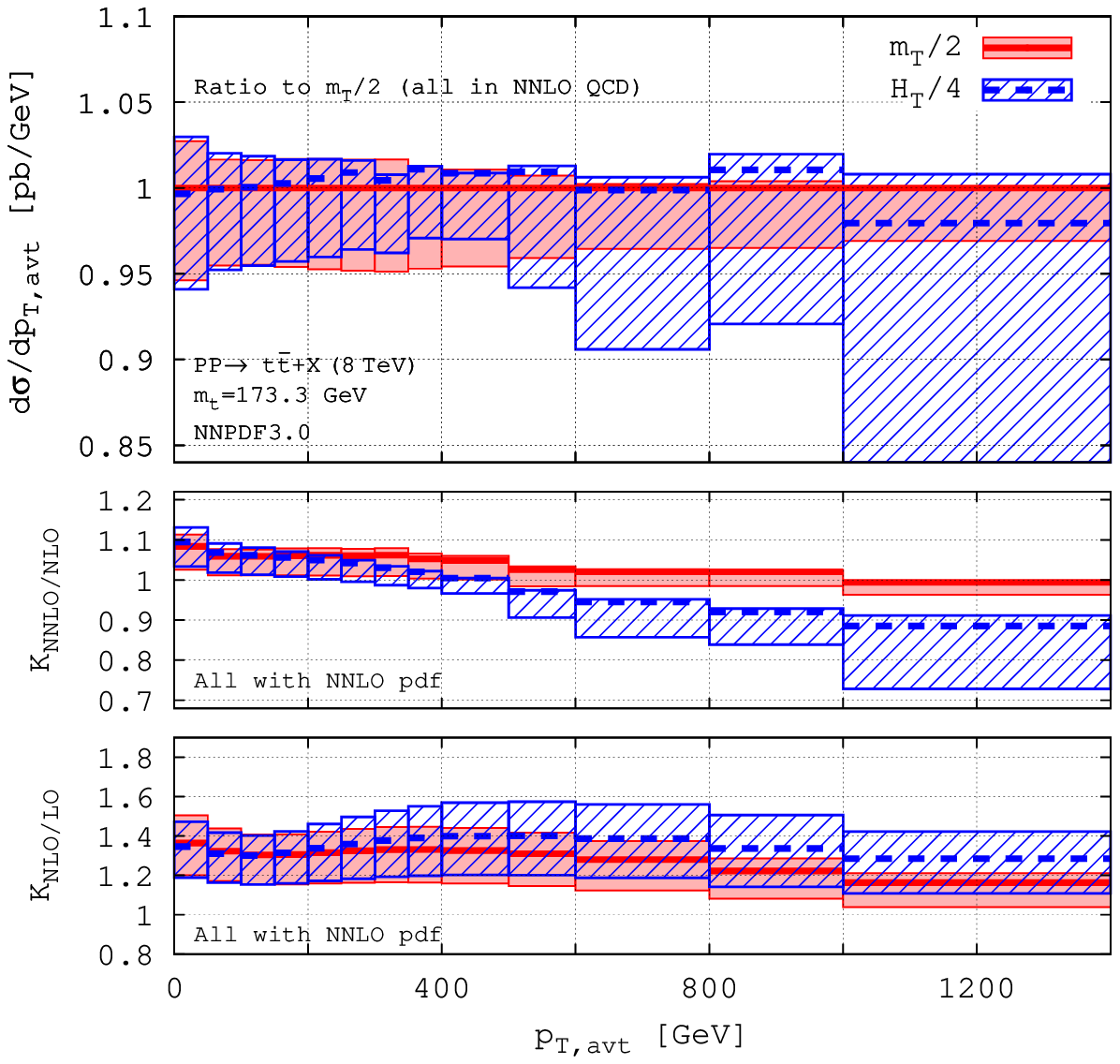}
\includegraphics[width=0.50\textwidth]{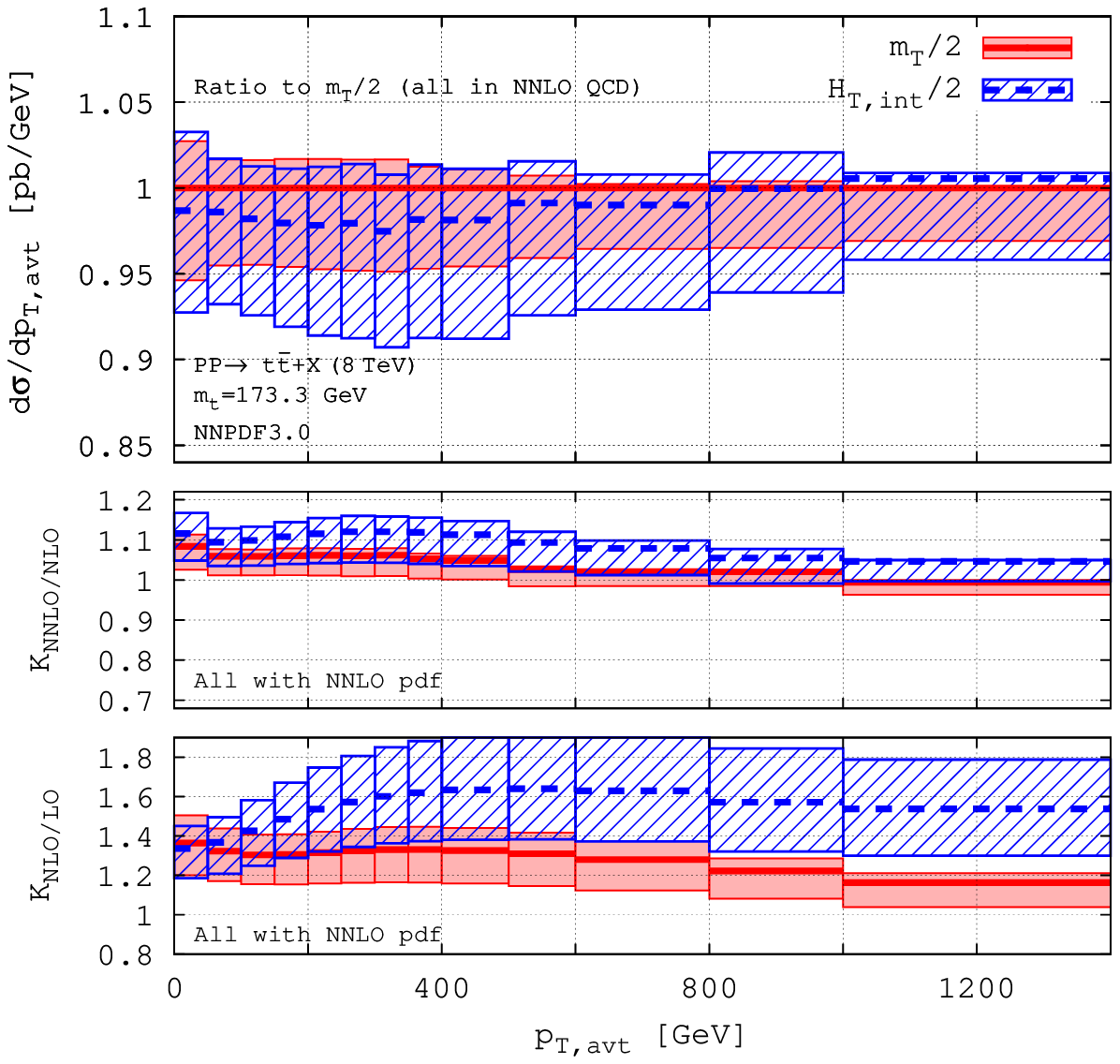}
\vskip 2mm
\includegraphics[width=0.50\textwidth]{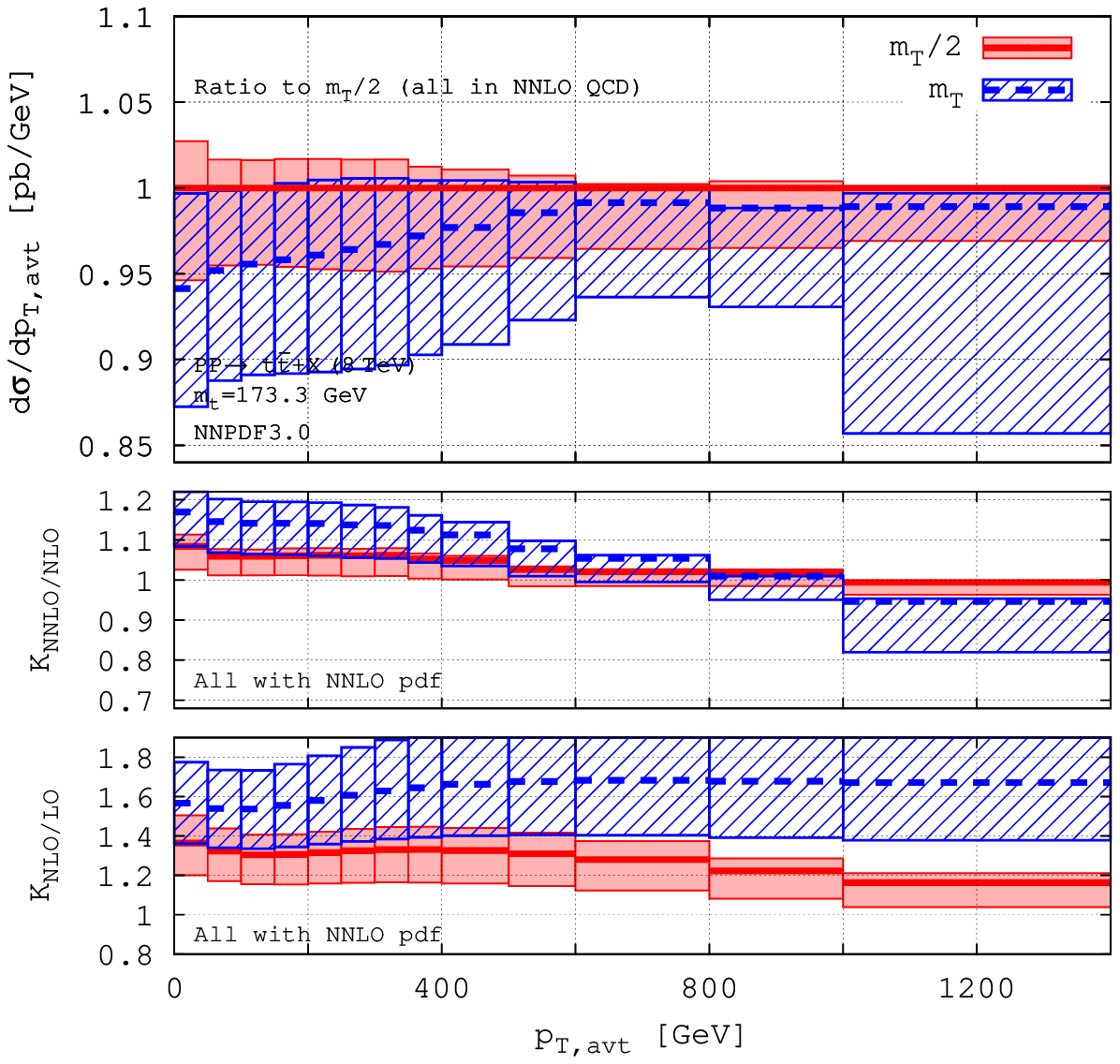}
\includegraphics[width=0.50\textwidth]{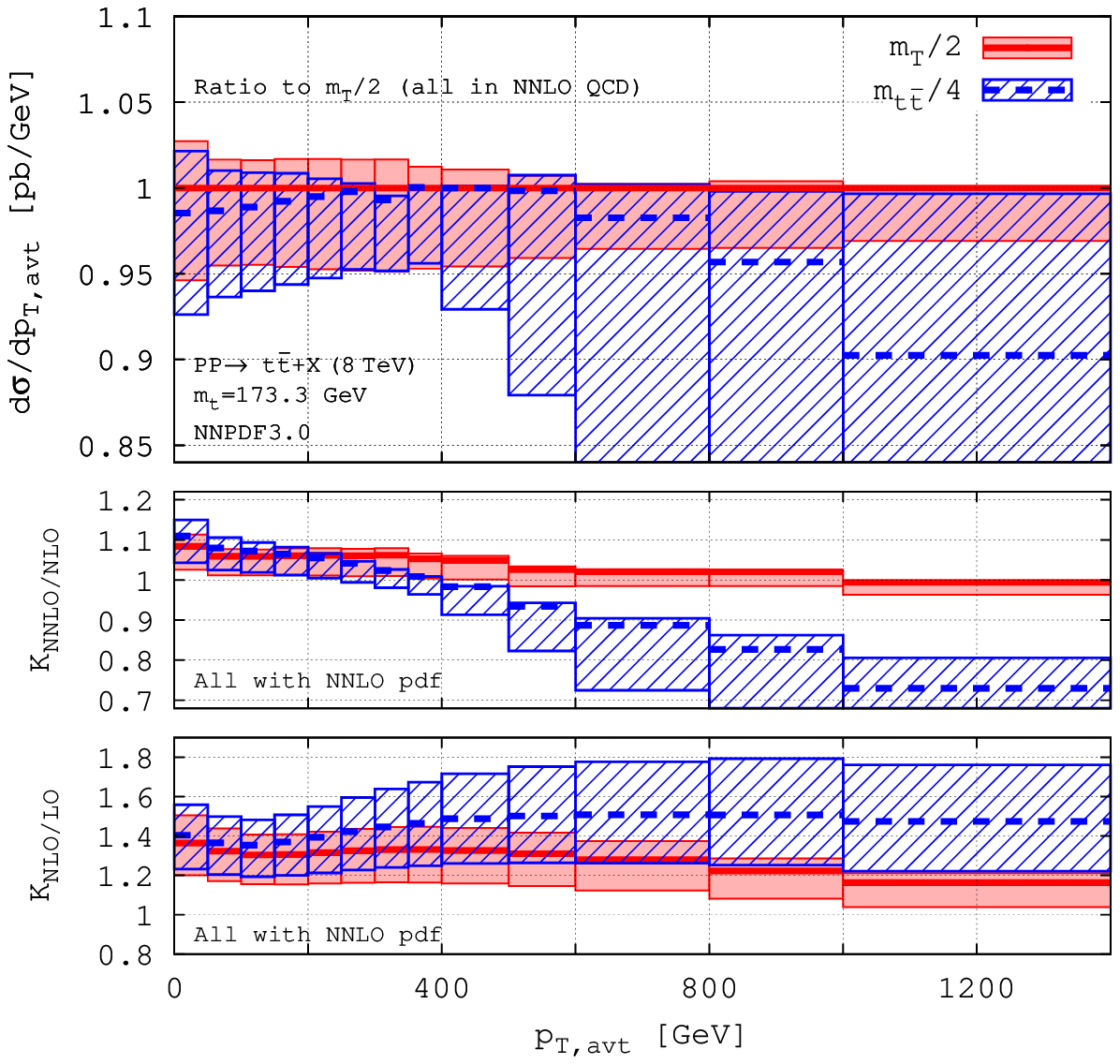}
\caption{\label{fig:diff-PTav-ratios-all_nnlopdf} As in fig.~\ref{fig:diff-PTav-ratios}, but all partonic cross-sections (LO, NLO and NNLO) are computed with NNLO pdf.}
\end{figure}

\begin{figure}[h]
\includegraphics[width=0.50\textwidth]{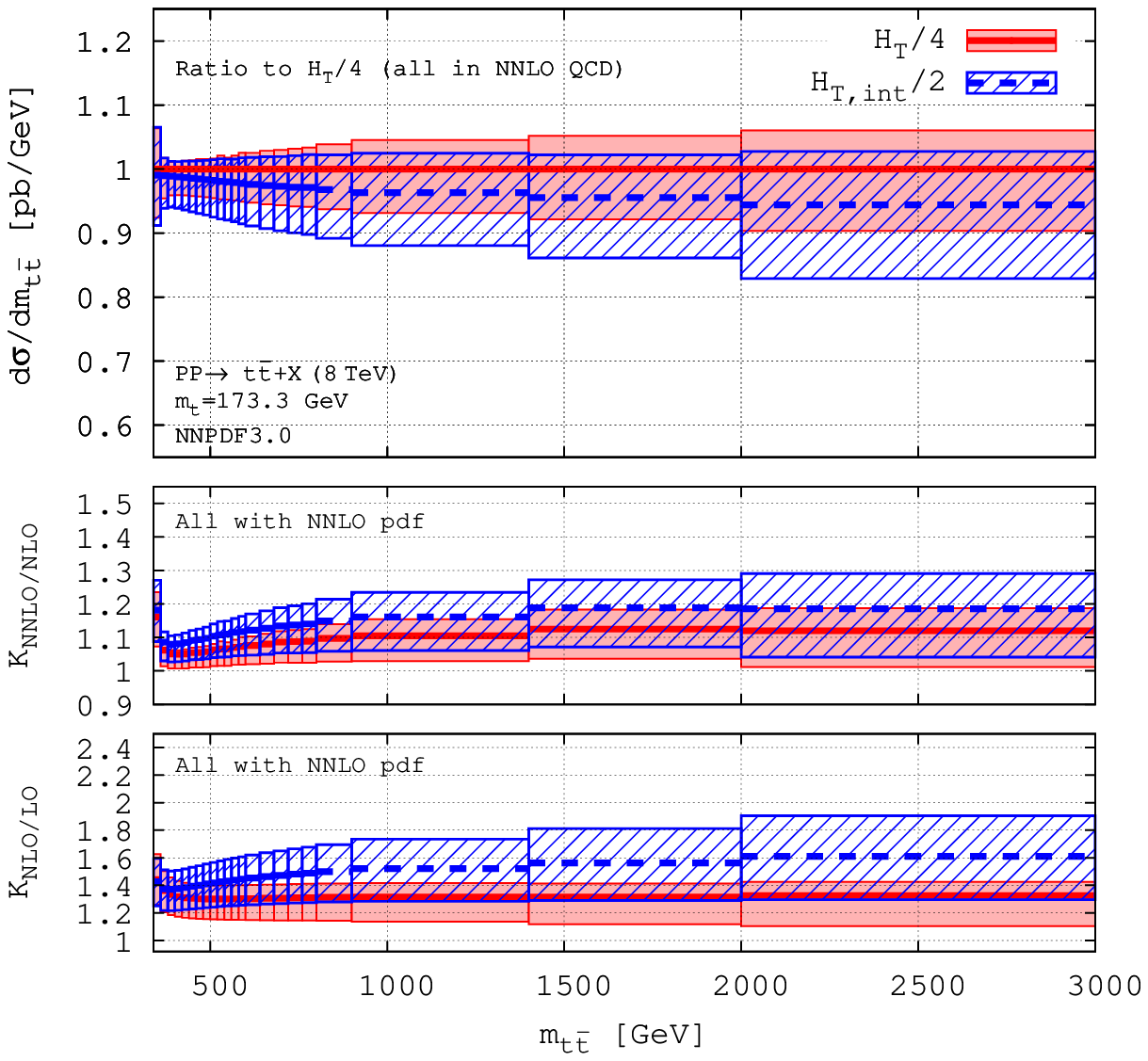}
\includegraphics[width=0.50\textwidth]{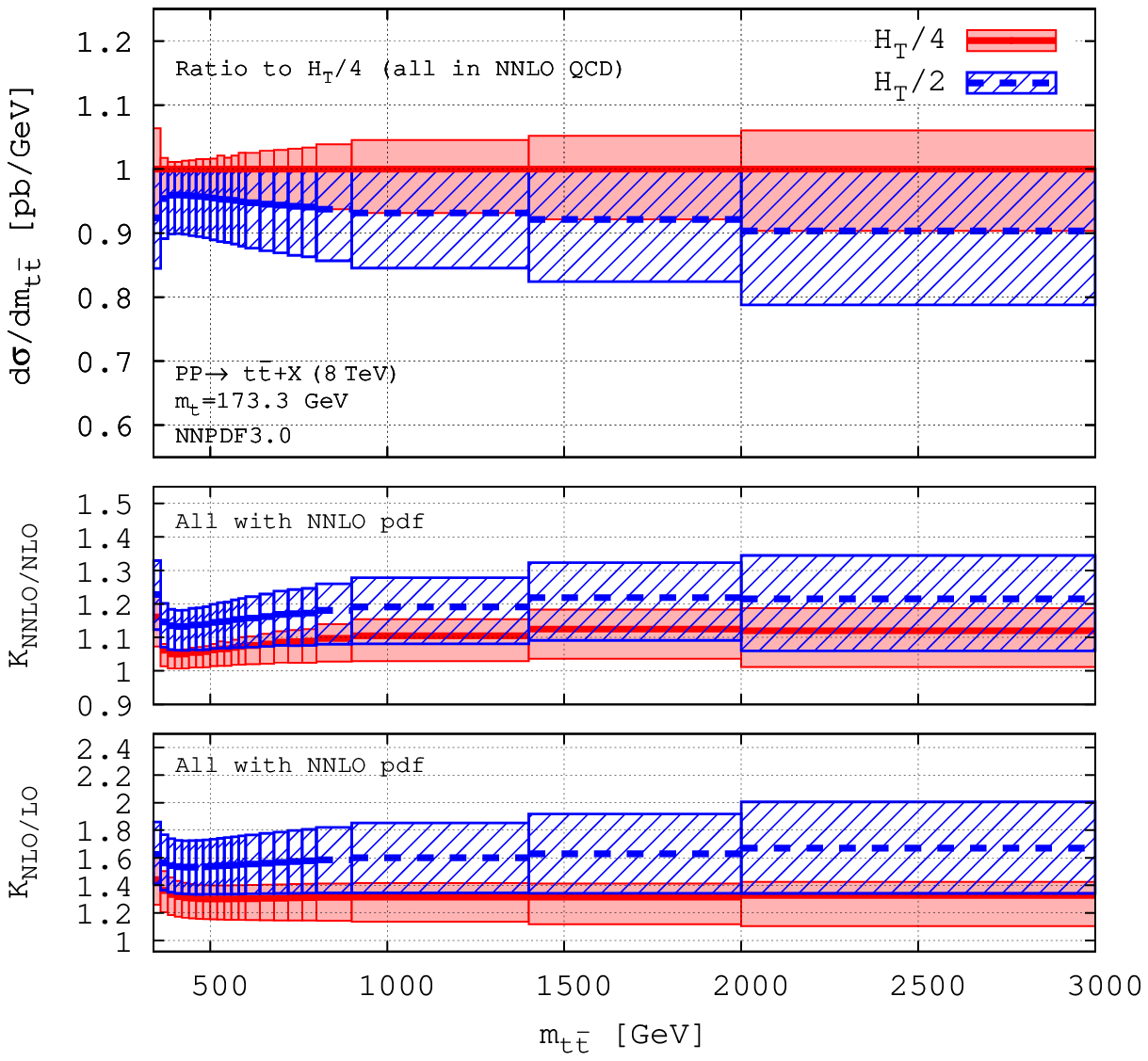}
\vskip 2mm
\includegraphics[width=0.50\textwidth]{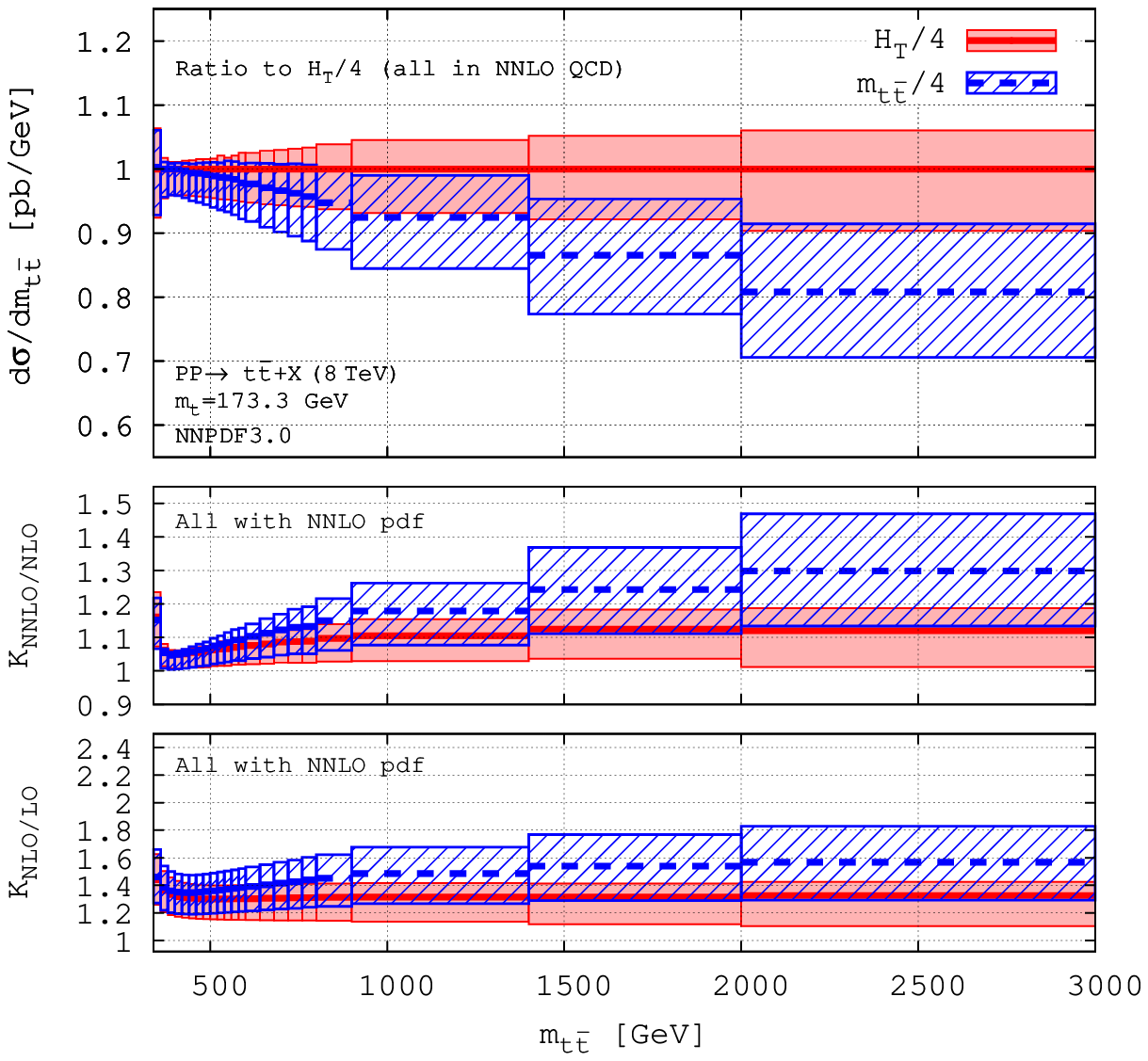}
\includegraphics[width=0.50\textwidth]{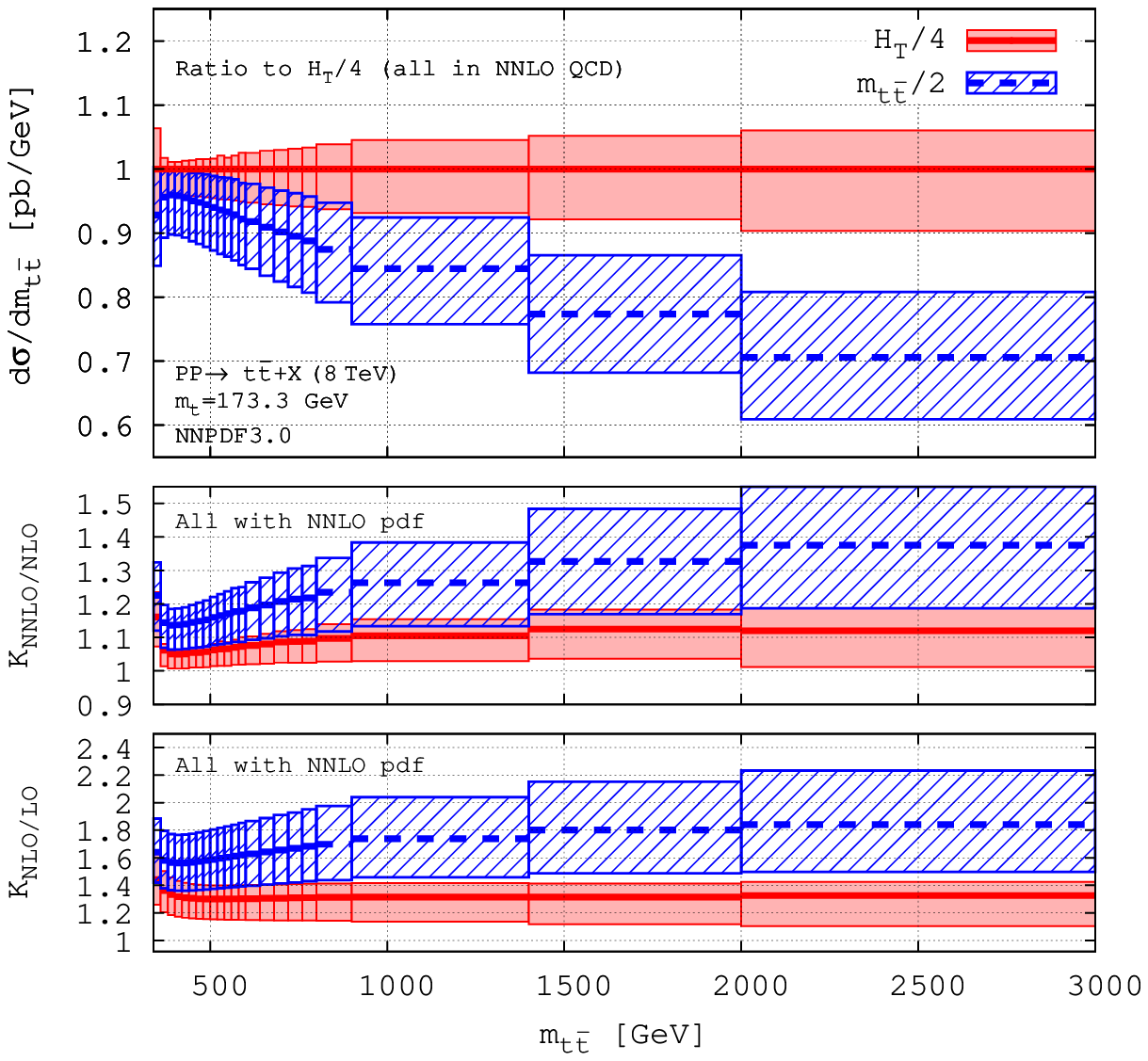}
\caption{\label{fig:diff-Mtt-ratios-all_nnlopdf} As in fig.~\ref{fig:diff-Mtt-ratios}, but all partonic cross-sections (LO, NLO and NNLO) are computed with NNLO pdf.}
\end{figure}

Finally, before closing this section, we present another proof that the conclusion derived in section \ref{sec:choosescale} regarding the choice of ``best" scale $\mu_0$ is not impacted by the choice of pdf set. Given the difference in predictions between different pdf sets such a conclusion is non-trivial and is an important test of the robustness of our chosen dynamic scales (\ref{eq:bestscale}). To that end, in figs.~\ref{fig:diff-PTav-ratios-all_nnlopdf},\ref{fig:diff-Mtt-ratios-all_nnlopdf} we show plots analogous to the ones in figs.~\ref{fig:diff-PTav-ratios},\ref{fig:diff-Mtt-ratios} but with all curves evaluated with the same NNLO pdf set (i.e. LO, NLO and NNLO partonic cross-sections are all convoluted with the same NNLO pdf). Based on the conclusions above, the K-factors for each scale should be pdf independent. We notice that all K-factors are very similar to the ones in figs.~\ref{fig:diff-PTav-ratios},\ref{fig:diff-Mtt-ratios} and most importantly, the K-factors for the ``best" scale choices eq.~(\ref{eq:bestscale}) are consistently the smallest ones, and the ones closest to unity, among all dynamic scales considered by us.

\section{Phenomenological applications}\label{sec:pheno}

As stated in the introduction, the ultimate goal of seeking a robust dynamic scale for top-pair production is to describe top production in the broadest kinematic ranges that will be accessible at the LHC. Indeed, as shown in the previous sections, the ``best" scales from eq.~(\ref{eq:bestscale}) satisfy all our criteria for a ``good" dynamic scale. In this work we calculate the NNLO QCD corrections to all stable top quark observables that have so-far been measured at the LHC. We have predictions for LHC at 8 TeV and 13 TeV. Specifically, we compute the following distributions:
$\PTavt,~\Yavt,~\Mtt,~\PTtt,~\Ytt\,,$
at LO, NLO and NNLO QCD and with three different pdf sets: NNPDF3.0, MMHT2014 and CT14. 
\footnote{The $\PTtt$ distribution is, strictly speaking, of NLO accuracy and can be easily obtained from the process $pp\to t\t j$. For this reason we do not provide explicit results for the $\PTtt$ distribution here.}

All results are available for download in electronic format with the Arxiv submission of this paper. For this reason, and due to the very large number of distributions, we do not specify here the bins and ranges of the various distributions. We would only like to remark that in order to achieve high-quality multi-TeV predictions (for example, our 13 TeV prediction for $\PTavt$ extends to 3 TeV while the one for $\Mtt$ up to 6 TeV) we have taken special care in order to populate with sufficient number of events tails of distributions that span many orders of magnitude. In doing so we have used the narrowest bins possible that allow us to keep the Monte Carlo integration error within about 1\% in almost all bins. The bins chosen do not correspond to a particular experimental analysis. They are, however, narrow enough so they might be combined to fit the usually much wider experimental bins. Another option is to fit the bin distribution with a smooth curve and then rebin that fit to any desired bin. The high quality of our result, paired with its extended range and narrow bins, should make these results useful for any future LHC experimental or theoretical analysis. 

In order to allow for the calculation of differential distributions that are normalised over any sub-range of the maximal ranges computed in this work, we make available the results for all seven $\mu_{F,R}$ scale combinations. To obtain scale variations in absolutely normalised distributions one has to simply find the min/max in each bin. For the normalised distributions, one has to first normalise each one of the seven curves within the desired range and then search for the min/max value in every bin. 
\begin{figure}[t]
\includegraphics[width=0.50\textwidth]{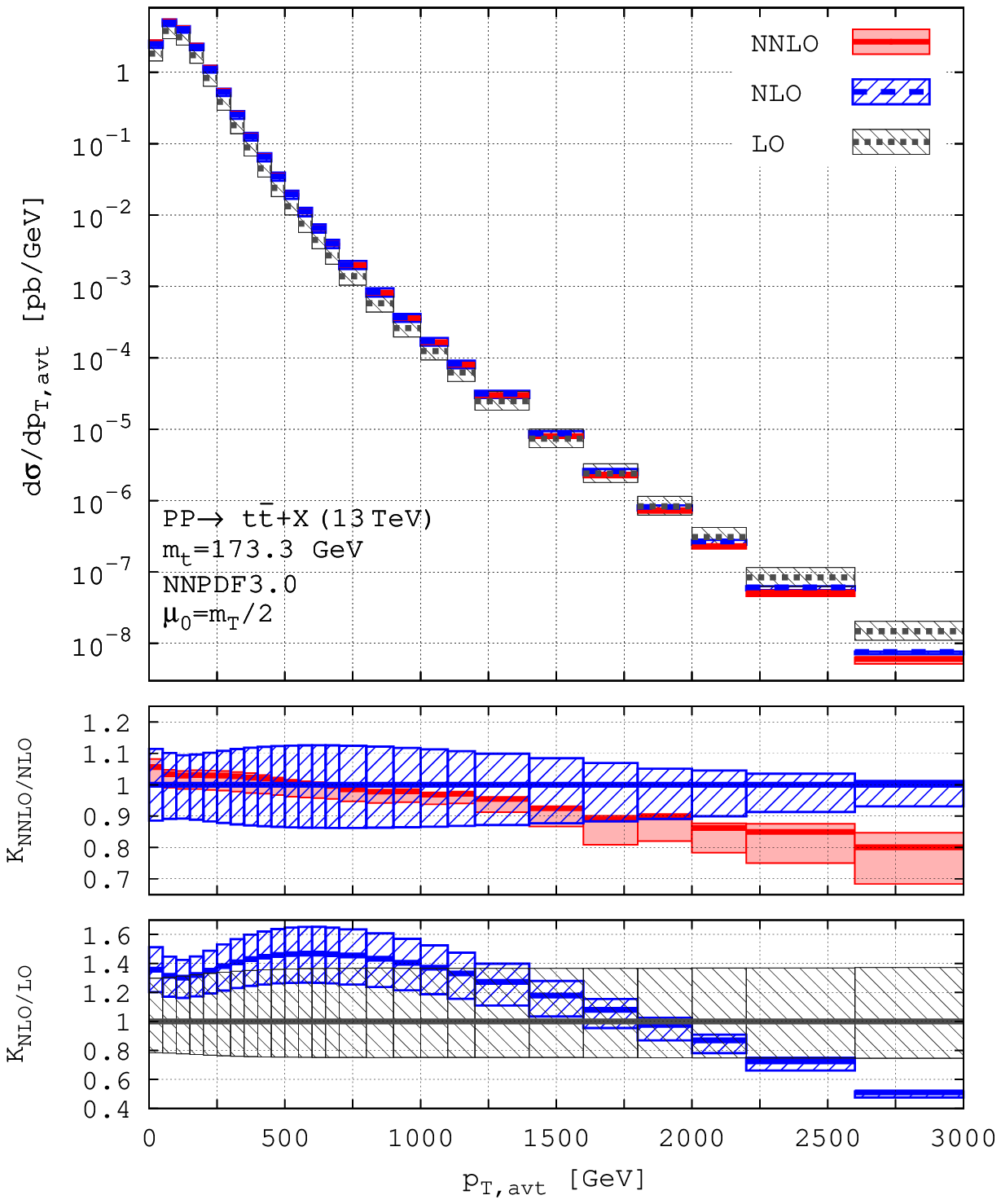}
\includegraphics[width=0.50\textwidth]{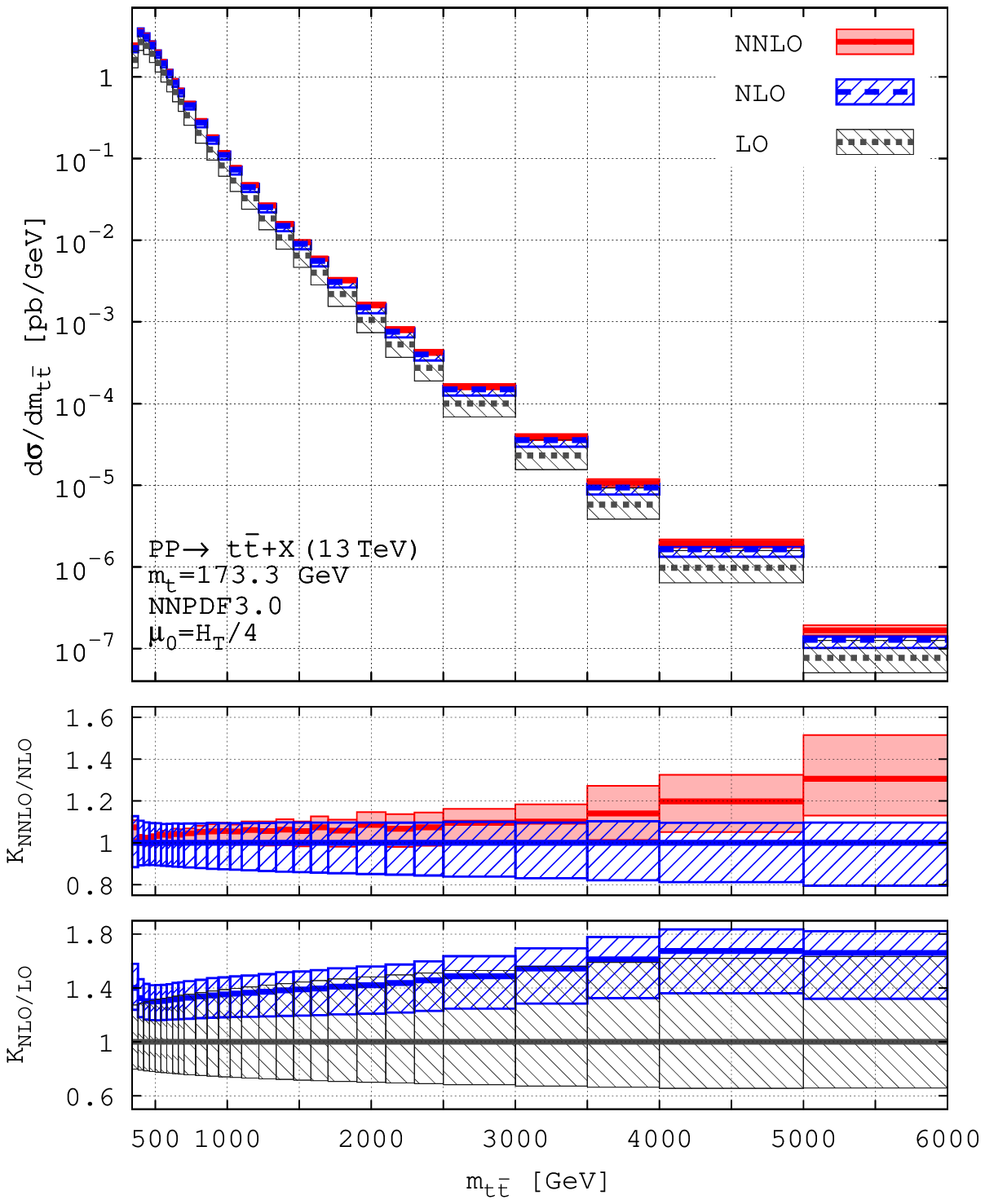}
\caption{\label{fig:diff-13-abs-mtt-pt} The $\PTavt$ (left) and $\Mtt$ (right) distributions for LHC 13 TeV. Error bands are from scale variation only.}
\end{figure}

In the following we show some representative results for LHC 13 TeV. In fig.~\ref{fig:diff-13-abs-mtt-pt} we plot the $\PTavt$ and $\Mtt$ distributions with absolute normalisation. Both are computed with NNPDF3.0 and with the optimal dynamic scales (\ref{eq:bestscale}). The distributions have behaviour similar to the case of 8 TeV shown in figs.~\ref{fig:diff-PTavt-8TeV-abs},\ref{fig:diff-Mtt-8TeV-abs}. The quality of the computation is high, with the aim of having Monte Carlo error typically within 1\% in each bin. 

The scale variation for the $\PTavt$ distribution is such that the central value is typically contained within the lower order scale variation band. At 8 TeV this is the case in the full kinematic range. At 13 TeV the NLO central scale is outside the LO error band in the interval $250\GeV - 1000\GeV$; the NNLO central value is, however, well within the NLO scale variation in this range. For very large $p_T$ both the NLO and NNLO central values at 13 TeV are outside the lower order scale bands. In this regard it is worth pointing out that the scale variation of the NLO correction, unlike the LO and NNLO ones, seems to be accidentally small at large $p_T$ and this may be the reason for such a behaviour. Furthermore, the resummation of collinear logs $\sim\ln(p_T/m_t)$ may also be playing a role in this kinematic range. 
\begin{figure}[t]
\includegraphics[width=0.50\textwidth]{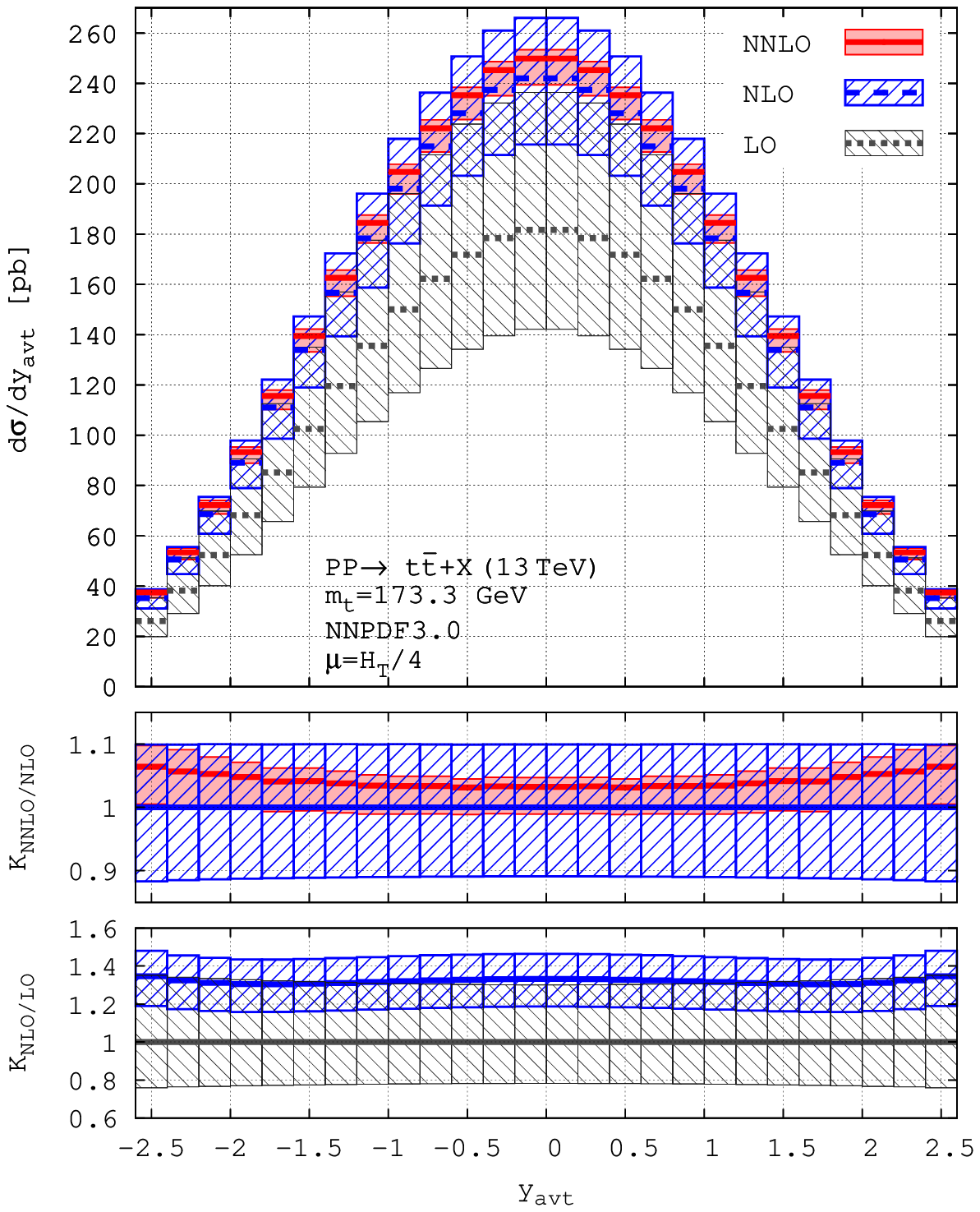}
\includegraphics[width=0.50\textwidth]{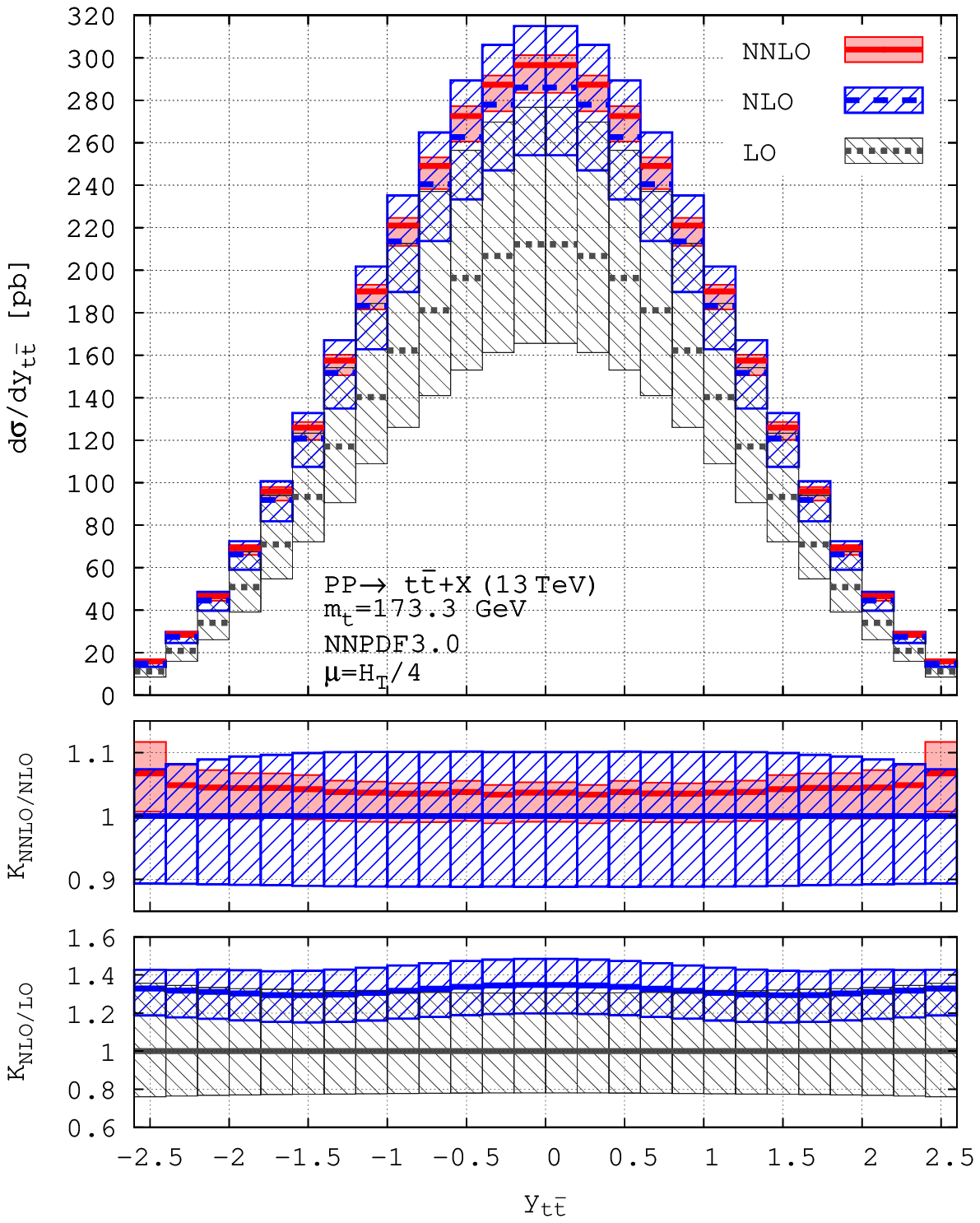}
\caption{\label{fig:diff-13-abs-yt-ytt} The $\Yavt$ (left) and $\Ytt$ (right) distributions for LHC 13 TeV. Error bands are from scale variation only.}
\end{figure}

The $\Mtt$ distribution at 13 TeV is rather well-behaved, similarly to the case of 8 TeV. Above $\Mtt\approx 3.5\TeV$ the NNLO correction tends to be outside the NLO scale variation range. This effect is comparable in size to the scale variation and so is not too significant. It would be interesting to revisit this upon supplementing the fixed order calculations with threshold and collinear resummation. The NNLO K-factor is rather mild for low $\Mtt$, although not as flat as it is for a fixed scale (see ref.~\cite{Czakon:2015owf}). The characteristic rise at absolute threshold noted in ref.~\cite{Czakon:2015owf} is also clearly visible.

\begin{figure}[t]
\includegraphics[width=0.50\textwidth]{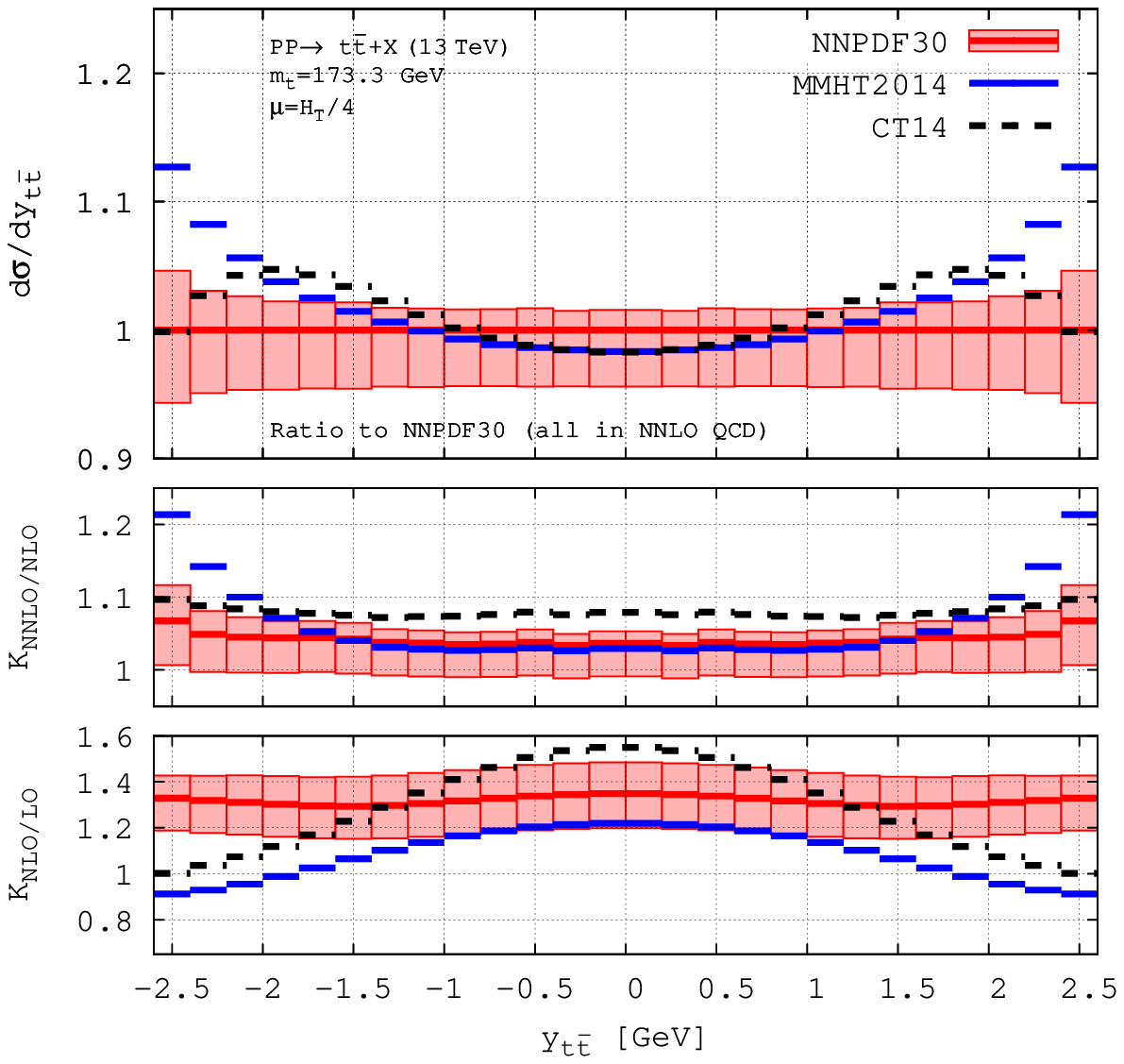}
\includegraphics[width=0.50\textwidth]{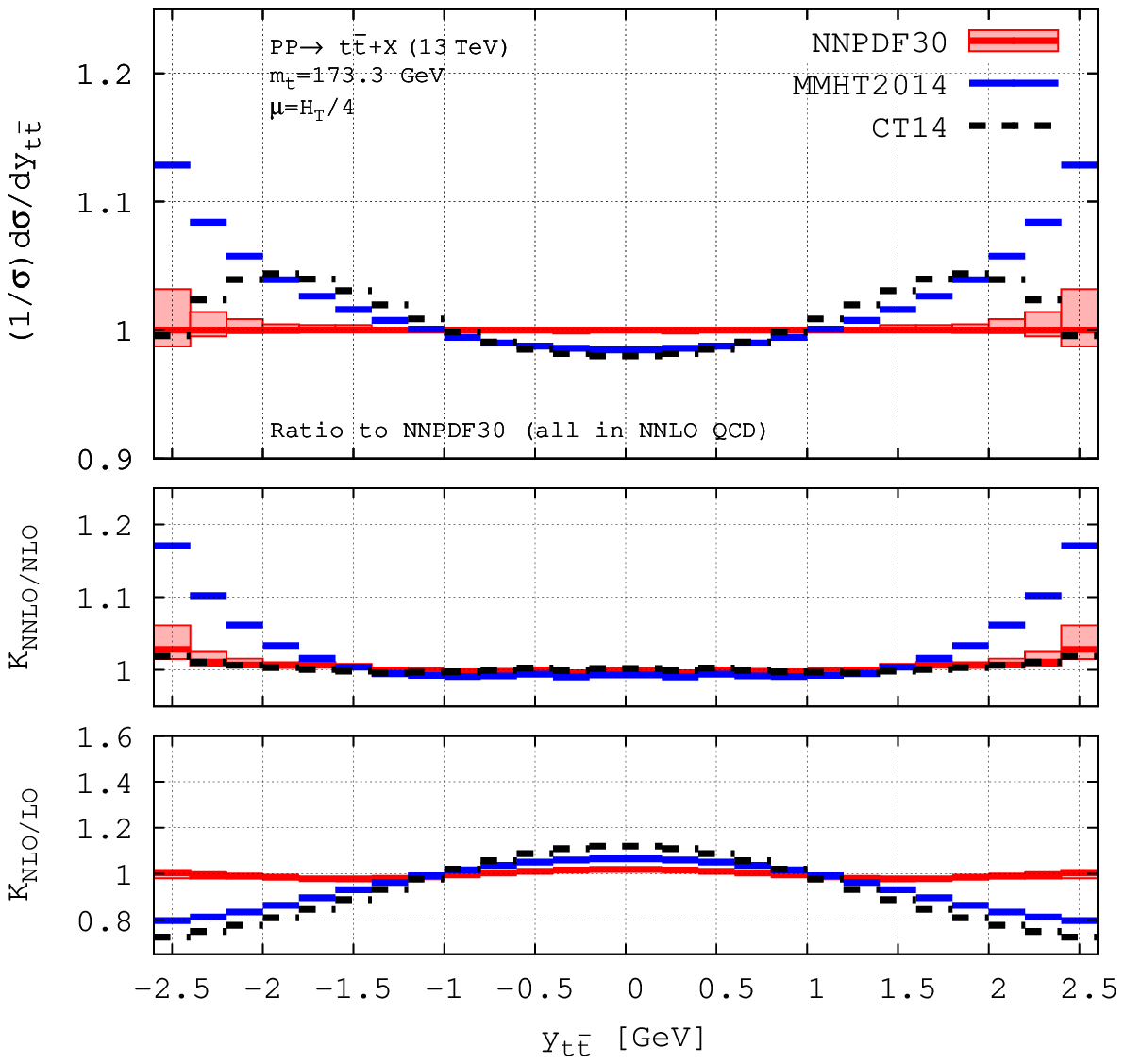}
\caption{\label{fig:diff-13-ratio-yt} The unnormalised (left) and normalised (right) $\Ytt$ distribution for LHC 13 TeV for three pdf sets. The distributions are normalised to the NNPDF3.0 central value. Error bands are from scale variation only.}
\end{figure}

In fig.~\ref{fig:diff-13-abs-yt-ytt} we show the absolutely normalised $\Yavt$ and $\Ytt$ distributions. Both are computed with NNPDF3.0 and with the optimal dynamic scale (\ref{eq:bestscale}). We notice good perturbative convergence as well as the tendency for the NLO and NNLO results to be within the scale error bands of the lower orders for both distributions. The MC errors are very small and the calculations of both spectra are of very high quality. In view of the importance of the $\Ytt$ distributions for fits of parton distribution functions in fig.~\ref{fig:diff-13-ratio-yt} we show this distribution computed with all three pdf sets considered in this work. For both the unnormalised and normalised distributions we show the ratios with respect to the central value computed with NNPDF3.0. A large spread among the various pdf sets is evident. It is moreover particularly significant in the normalised $\Ytt$ distribution where the differences due to different pdf is on-par with the scale error. Clearly, the $\Ytt$ distribution suffers from significant pdf error and could, in turn, be used as a strong constraint on pdfs from high-precision LHC data.

We conclude this section with the following two comments. First, in this work we have not computed the pdf errors for any pdf set. As we conclude in the previous sections, however, pdf related uncertainties become the dominant source of error long before one reaches the end points of the computed ranges. To gain insight into the size of the pdf error we have compared predictions based on three pdf sets. It appears that at present the constraining factor in doing TeV analyses is the knowledge of pdfs. For this reason the result of the present work should be used with some care. Future precision progress will critically depend on the availability of improved pdf sets. In order to facilitate the use of our calculations with any future pdf set, we will release in the near future our results also as tables in the {\tt fastNLO} library format \cite{Kluge:2006xs,Britzger:2012bs}. 

Second, we would like to emphasise that besides pdf errors, the results we present here will also be affected by the resummation of collinear logs and possibly by EW effects. Those contributions will require dedicated future studies. In any case the NNLO QCD result computed in this work offers the base for such future additions.

\section{Conclusions}\label{sec:conclusions}

The main result of this work is the extension of the recently computed NNLO QCD differential distributions for stable top quark pair production at the LHC beyond the small $p_T$/$\Mtt$ regime studied so far at LHC Run I. The results derived here make it possible to describe stable top quark production into the multi-TeV regime which will be explored in detail during LHC Run II. We have presented high-quality predictions for most top-quark distributions for both LHC 8 TeV and 13 TeV. Our results are in the form of binned distributions and are computed with three different pdf sets. All results are available for download in electronic form with the Arxiv submission of this work. The relatively small bin sizes for our results, coupled with their small Monte Carlo errors, would allow one to easily produce high-quality analytic fits to all distributions. We expect that such fits could subsequently be used for further rebinning to a different bin size, at the expense of tolerable errors. This way our results could be extended to accommodate diverse bin configurations; in order to also allow for a (fast) change of parton distribution sets we will release in the near future our results as {\tt fastNLO} library tables. This way, our results should satisfy most of the requirements for stable top quark distributions of both theorists and LHC collaborations over the span of LHC Run II.

At the technical level, the new ingredient that makes it possible to extend our previous NNLO QCD results to the widest ranges achievable at the LHC is a new {\it dynamic} renormalisation and factorisation scale $\mu_0$. We derive such a scale based on the principle of fastest perturbative convergence, i.e. we require the scale be such that, both at NLO and NNLO, it introduces the smallest possible K-factors across the full kinematic range. Since the small $p_T$ behaviour of such a scale is strongly correlated with the well-understood total top-pair cross-section, we also find it desirable to have good numerical agreement with the value of the NNLO+NNLL cross-section. 

The following scales satisfy our requirements best: $\mu_0=m_T/2$ to be used for the description of the $p_T$ distribution of top/antitop quarks and $\mu_0=H_T/4$ for all other distributions. These functional forms, along with other functional forms that we found to be less suitable, have been used in the past in NLO QCD calculations; the main new feature we uncover is that the scale $\mu_0$ need to be a factor of 2 smaller compared to the typical form in past studies. We demonstrate that such functional forms for $\mu_0$ lead to fast perturbative convergence, small-to-moderate scale errors and return NNLO total cross-section which differs from the NNLO+NNLL $\sigma_{\rm tot}(m_t)$ value at the sub-percent level.

A convincing derivation of a ``good" dynamic scale is possible because of the full control over both NLO and NNLO corrections. Furthermore the reduced error of the NNLO-accurate cross-section makes it much easier to distinguish between various dynamic scale candidates. For example, we find that $\Mtt$-based dynamic scales are disfavoured, a result which may have implications in matching the NNLO results with NNLL soft-gluon resummation. We have also noted that the behaviour of the total $t\t$ cross-section through NNLO+NNLL is very similar to the Higgs production cross-section through resummed N$^3$LO. 

We estimate that the error due to missing higher orders is typically within 5\%, at least for kinematic ranges of current phenomenological interest. Such typical-size estimate, however, should only be used as a rough guide for the scale error of differential distributions in NNLO QCD and one should keep in mind that the actual error varies across kinematic ranges and across distributions. Specifically, the top/antitop and top-pair rapidities seem to be under very good control in the full kinematic ranges considered here. The $\PTavt$ distribution seems to also be reliably predicted for $\PTavt$ as large as 2 TeV. The $\Mtt$ distribution's scale variation is within 5\% for masses of up to 2 TeV, but is steadily increasing towards larger scales. For example, for $\Mtt=4$ TeV, the scale error is as large as 10\%. Moreover, the overlap between various perturbative orders is not as good for very large $\PTavt$ and $\Mtt$.

Very importantly, by comparing predictions with three different pdf sets, we show that for $p_T$ and $\Mtt$ that are just into the TeV range, as well as for medium and large values of $\Ytt$, the uncertainty due to the imperfect knowledge of pdfs very fast becomes the dominant source of error. Therefore, our results should be used with care over extended ranges with current pdf sets and one should be mindful of the implied pdf error (which is not plotted in any of the figures or included in the supplied electronic files). In fact, it seems to us, truly precise top-quark predictions in the TeV range will only be possible once a new generation of pdf sets becomes available and it seems likely that such pdf sets will utilise, to some degree, LHC top quark data. We should also emphasise that the direct phenomenological relevance of our results in the TeV range is additionally subject to the following so-far unaccounted effects: resummation of large collinear logs $\ln^n(p_T/m_t)$, fixed-versus-variable flavour number scheme ambiguity for top production as well as inclusion of EW corrections. The range of phenomenological relevance for these effects, however, has yet to be carefully investigated.

In conclusion, we mention a number of other lessons that can be drawn from our work. First, our approach to finding an appropriate dynamical scale is quite generic and it may benefit other LHC processes that are now - or will soon be - known at NNLO. In particular, we notice that our best scales have a feature that may well be process-independent: they tend to reflect the observable already encoded in the LO kinematics. Second, in the past \cite{Langenfeld:2009wd,Dowling:2013baa} the use of the $\overline{\rm MS}$ scheme for the top-quark mass has been advocated for, among others, improved convergence of the perturbative series. Our work shows that in order to achieve good convergence no special choice for $m_t$ is needed. Third, our experience shows that the principle of fastest perturbative convergence works quite well. This may be contrasted, for example, with the principle of minimal sensitivity that has been used in the past in the context of NLO studies.

\begin{acknowledgments}
M.C. thanks the CERN Theoretical Physics Department and Emmanuel College Cambridge for hospitality during the completion of this work. A.M. Thanks the KITP at UCSB for hospitality during the completion of this work. The work of M.C. was supported in part by grants of the DFG and BMBF. The work of D.H. and A.M. is supported by the UK Science and Technology Facilities Council [grants ST/L002760/1 and ST/K004883/1]. The work of A.M. was supported in part by the National Science Foundation Grant NSF PHY11-25915.
\end{acknowledgments}


\begin{thebibliography}{99}

%\cite{Czakon:2015owf}
\bibitem{Czakon:2015owf} 
  M.~Czakon, D.~Heymes and A.~Mitov,
  %``High-precision differential predictions for top-quark pairs at the LHC,''
  Phys.\ Rev.\ Lett.\  {\bf 116}, no. 8, 082003 (2016)
  [arXiv:1511.00549 [hep-ph]].

%\cite{Czakon:2014xsa}
\bibitem{Czakon:2014xsa} 
  M.~Czakon, P.~Fiedler and A.~Mitov,
  %``Resolving the Tevatron Top Quark Forward-Backward Asymmetry Puzzle: Fully Differential Next-to-Next-to-Leading-Order Calculation,''
  Phys.\ Rev.\ Lett.\  {\bf 115}, no. 5, 052001 (2015)
  [arXiv:1411.3007 [hep-ph]].

%\cite{Czakon:2016ckf}
\bibitem{Czakon:2016ckf} 
  M.~Czakon, P.~Fiedler, D.~Heymes and A.~Mitov,
  %``NNLO QCD predictions for fully-differential top-quark pair production at the Tevatron,''
  JHEP {\bf 1605}, 034 (2016)
  [arXiv:1601.05375 [hep-ph]].
  %%CITATION = doi:10.1007/JHEP05(2016)034;%%

%\cite{Baernreuther:2012ws}
\bibitem{Baernreuther:2012ws} 
  P.~Bärnreuther, M.~Czakon and A.~Mitov,
  %``Percent Level Precision Physics at the Tevatron: First Genuine NNLO QCD Corrections to $q \bar{q} \to t \bar{t} + X$,''
  Phys.\ Rev.\ Lett.\  {\bf 109}, 132001 (2012)
  [arXiv:1204.5201 [hep-ph]].

%\cite{Czakon:2012zr}
\bibitem{Czakon:2012zr} 
  M.~Czakon and A.~Mitov,
  %``NNLO corrections to top-pair production at hadron colliders: the all-fermionic scattering channels,''
  JHEP {\bf 1212}, 054 (2012)
  [arXiv:1207.0236 [hep-ph]].
    
%\cite{Czakon:2012pz}
\bibitem{Czakon:2012pz} 
  M.~Czakon and A.~Mitov,
  %``NNLO corrections to top pair production at hadron colliders: the quark-gluon reaction,''
  JHEP {\bf 1301}, 080 (2013)
  [arXiv:1210.6832 [hep-ph]].
    
%\cite{Czakon:2013goa}
\bibitem{Czakon:2013goa} 
  M.~Czakon, P.~Fiedler and A.~Mitov,
  %``Total Top-Quark Pair-Production Cross Section at Hadron Colliders Through $O(?\frac{4}{S})$,''
  Phys.\ Rev.\ Lett.\  {\bf 110}, 252004 (2013)
  [arXiv:1303.6254 [hep-ph]].

%\cite{Beenakker:1993yr}
\bibitem{Beenakker:1993yr} 
  W.~Beenakker, A.~Denner, W.~Hollik, R.~Mertig, T.~Sack and D.~Wackeroth,
  %``Electroweak one loop contributions to top pair production in hadron colliders,''
  Nucl.\ Phys.\ B {\bf 411}, 343 (1994).
  %%CITATION = doi:10.1016/0550-3213(94)90454-5;%%
  
%\cite{Kuhn:2005it}
\bibitem{Kuhn:2005it} 
  J.~H.~Kuhn, A.~Scharf and P.~Uwer,
  %``Electroweak corrections to top-quark pair production in quark-antiquark annihilation,''
  Eur.\ Phys.\ J.\ C {\bf 45}, 139 (2006)
  [hep-ph/0508092].
  %%CITATION = doi:10.1140/epjc/s2005-02423-6;%%

%\cite{Bernreuther:2005is}
\bibitem{Bernreuther:2005is} 
  W.~Bernreuther, M.~Fuecker and Z.~G.~Si,
  %``Mixed QCD and weak corrections to top quark pair production at hadron colliders,''
  Phys.\ Lett.\ B {\bf 633}, 54 (2006)
  [hep-ph/0508091].
  %%CITATION = doi:10.1016/j.physletb.2005.11.056;%%

%\cite{Kuhn:2006vh}
\bibitem{Kuhn:2006vh} 
  J.~H.~Kuhn, A.~Scharf and P.~Uwer,
  %``Electroweak effects in top-quark pair production at hadron colliders,''
  Eur.\ Phys.\ J.\ C {\bf 51}, 37 (2007)
  [hep-ph/0610335].
  %%CITATION = doi:10.1140/epjc/s10052-007-0275-x;%%

%\cite{Bernreuther:2006vg}
\bibitem{Bernreuther:2006vg} 
  W.~Bernreuther, M.~Fuecker and Z.~G.~Si,
  %``Weak interaction corrections to hadronic top quark pair production,''
  Phys.\ Rev.\ D {\bf 74}, 113005 (2006)
  [hep-ph/0610334].
  %%CITATION = doi:10.1103/PhysRevD.74.113005;%%
  
%\cite{Bernreuther:2008md}
\bibitem{Bernreuther:2008md} 
  W.~Bernreuther, M.~Fucker and Z.~G.~Si,
  %``Weak interaction corrections to hadronic top quark pair production: Contributions from quark-gluon and b anti-b induced reactions,''
  Phys.\ Rev.\ D {\bf 78}, 017503 (2008)
  [arXiv:0804.1237 [hep-ph]].
  %%CITATION = doi:10.1103/PhysRevD.78.017503;%%

%\cite{Manohar:2012rs}
\bibitem{Manohar:2012rs} 
  A.~V.~Manohar and M.~Trott,
  %``Electroweak Sudakov Corrections and the Top Quark Forward-Backward Asymmetry,''
  Phys.\ Lett.\ B {\bf 711}, 313 (2012)
  [arXiv:1201.3926 [hep-ph]].
  %%CITATION = doi:10.1016/j.physletb.2012.04.013;%%
    
%\cite{Kuhn:2013zoa}
\bibitem{Kuhn:2013zoa} 
  J.~H.~Kühn, A.~Scharf and P.~Uwer,
  %``Weak Interactions in Top-Quark Pair Production at Hadron Colliders: An Update,''
  Phys.\ Rev.\ D {\bf 91}, no. 1, 014020 (2015)
  [arXiv:1305.5773 [hep-ph]].
  %%CITATION = doi:10.1103/PhysRevD.91.014020;%%
  
%\cite{Campbell:2015vua}
\bibitem{Campbell:2015vua} 
  J.~M.~Campbell, D.~Wackeroth and J.~Zhou,
  %``Electroweak Corrections at the LHC with MCFM,''
  PoS DIS {\bf 2015}, 130 (2015)
  [arXiv:1508.06247 [hep-ph]].
  %%CITATION = ARXIV:1508.06247;%%
  
%\cite{Hollik:2007sw}
\bibitem{Hollik:2007sw} 
  W.~Hollik and M.~Kollar,
  %``NLO QED contributions to top-pair production at hadron collider,''
  Phys.\ Rev.\ D {\bf 77}, 014008 (2008)
  [arXiv:0708.1697 [hep-ph]].
  %%CITATION = doi:10.1103/PhysRevD.77.014008;%%
  
%\cite{Bernreuther:2010ny}
\bibitem{Bernreuther:2010ny} 
  W.~Bernreuther and Z.~G.~Si,
  %``Distributions and correlations for top quark pair production and decay at the Tevatron and LHC.,''
  Nucl.\ Phys.\ B {\bf 837}, 90 (2010)
  [arXiv:1003.3926 [hep-ph]].
  %%CITATION = doi:10.1016/j.nuclphysb.2010.05.001;%%

%\cite{Pagani:2016caq}
\bibitem{Pagani:2016caq} 
  D.~Pagani, I.~Tsinikos and M.~Zaro,
  %``The impact of the photon PDF and electroweak corrections on $t \bar t$ distributions,''
  arXiv:1606.01915 [hep-ph].
  %%CITATION = ARXIV:1606.01915;%%

%\cite{Denner:2010jp}
\bibitem{Denner:2010jp} 
  A.~Denner, S.~Dittmaier, S.~Kallweit and S.~Pozzorini,
  %``NLO QCD corrections to WWbb production at hadron colliders,''
  Phys.\ Rev.\ Lett.\  {\bf 106}, 052001 (2011)
  [arXiv:1012.3975 [hep-ph]].
  
%\cite{Bevilacqua:2010qb}
\bibitem{Bevilacqua:2010qb} 
  G.~Bevilacqua, M.~Czakon, A.~van Hameren, C.~G.~Papadopoulos and M.~Worek,
  %``Complete off-shell effects in top quark pair hadroproduction with leptonic decay at next-to-leading order,''
  JHEP {\bf 1102}, 083 (2011)
  [arXiv:1012.4230 [hep-ph]].

%\cite{Denner:2012yc}
\bibitem{Denner:2012yc} 
  A.~Denner, S.~Dittmaier, S.~Kallweit and S.~Pozzorini,
  %``NLO QCD corrections to off-shell top-antitop production with leptonic decays at hadron colliders,''
  JHEP {\bf 1210}, 110 (2012)
  [arXiv:1207.5018 [hep-ph]].

%\cite{Papanastasiou:2013dta}
\bibitem{Papanastasiou:2013dta} 
  A.~S.~Papanastasiou, R.~Frederix, S.~Frixione, V.~Hirschi and F.~Maltoni,
  %``Single-top $t$-channel production with off-shell and non-resonant effects,''
  Phys.\ Lett.\ B {\bf 726}, 223 (2013)
  [arXiv:1305.7088 [hep-ph]].
  
%\cite{Frederix:2013gra}
\bibitem{Frederix:2013gra} 
  R.~Frederix,
  %``Top Quark Induced Backgrounds to Higgs Production in the $WW^{(*)}\to ll\nu\nu$ Decay Channel at Next-to-Leading-Order in QCD,''
  Phys.\ Rev.\ Lett.\  {\bf 112}, no. 8, 082002 (2014)
  [arXiv:1311.4893 [hep-ph]].

%\cite{Cascioli:2013wga}
\bibitem{Cascioli:2013wga} 
  F.~Cascioli, S.~Kallweit, P.~Maierhöfer and S.~Pozzorini,
  %``A unified NLO description of top-pair and associated Wt production,''
  Eur.\ Phys.\ J.\ C {\bf 74}, no. 3, 2783 (2014)
  [arXiv:1312.0546 [hep-ph]].
    
%\cite{Bevilacqua:2015qha}
\bibitem{Bevilacqua:2015qha} 
  G.~Bevilacqua, H.~B.~Hartanto, M.~Kraus and M.~Worek,
  %``Top Quark Pair Production in Association with a Jet with Next-to-Leading-Order QCD Off-Shell Effects at the Large Hadron Collider,''
  Phys.\ Rev.\ Lett.\  {\bf 116}, no. 5, 052003 (2016)
  [arXiv:1509.09242 [hep-ph]].

%\cite{Frederix:2016rdc}
\bibitem{Frederix:2016rdc} 
  R.~Frederix, S.~Frixione, A.~S.~Papanastasiou, S.~Prestel and P.~Torrielli,
  %``Off-shell single-top production at NLO matched to parton showers,''
  arXiv:1603.01178 [hep-ph].  

%\cite{Mitov:2012gt}
\bibitem{Mitov:2012gt} 
  A.~Mitov and G.~Sterman,
  %``Final state interactions in single- and multi-particle inclusive cross sections for hadronic collisions,''
  Phys.\ Rev.\ D {\bf 86}, 114038 (2012)
  [arXiv:1209.5798 [hep-ph]].

%\cite{Cacciari:2008zb}
\bibitem{Cacciari:2008zb} 
  M.~Cacciari, S.~Frixione, M.~L.~Mangano, P.~Nason and G.~Ridolfi,
  %``Updated predictions for the total production cross sections of top and of heavier quark pairs at the Tevatron and at the LHC,''
  JHEP {\bf 0809}, 127 (2008)
  [arXiv:0804.2800 [hep-ph]].

%\cite{Nason:1989zy}
\bibitem{Nason:1989zy} 
  P.~Nason, S.~Dawson and R.~K.~Ellis,
  %``The One Particle Inclusive Differential Cross-Section for Heavy Quark Production in Hadronic Collisions,''
  Nucl.\ Phys.\ B {\bf 327}, 49 (1989)
  Erratum: [Nucl.\ Phys.\ B {\bf 335}, 260 (1990)].
  %%CITATION = doi:10.1016/0550-3213(90)90180-L, 10.1016/0550-3213(89)90286-1;%%

%\cite{Beenakker:1990maa}
\bibitem{Beenakker:1990maa} 
  W.~Beenakker, W.~L.~van Neerven, R.~Meng, G.~A.~Schuler and J.~Smith,
  %``QCD corrections to heavy quark production in hadron hadron collisions,''
  Nucl.\ Phys.\ B {\bf 351}, 507 (1991).
  %%CITATION = doi:10.1016/S0550-3213(05)80032-X;%%  

%\cite{Forte:2010ta}
\bibitem{Forte:2010ta} 
  S.~Forte, E.~Laenen, P.~Nason and J.~Rojo,
  %``Heavy quarks in deep-inelastic scattering,''
  Nucl.\ Phys.\ B {\bf 834}, 116 (2010)
  [arXiv:1001.2312 [hep-ph]].
  %%CITATION = doi:10.1016/j.nuclphysb.2010.03.014;%%
  
%\cite{Ball:2011mu}
\bibitem{Ball:2011mu} 
  R.~D.~Ball {\it et al.},
  %``Impact of Heavy Quark Masses on Parton Distributions and LHC Phenomenology,''
  Nucl.\ Phys.\ B {\bf 849}, 296 (2011)
  [arXiv:1101.1300 [hep-ph]].
  %%CITATION = doi:10.1016/j.nuclphysb.2011.03.021;%%  

%\cite{Han:2014nja}
\bibitem{Han:2014nja} 
  T.~Han, J.~Sayre and S.~Westhoff,
  %``Top-Quark Initiated Processes at High-Energy Hadron Colliders,''
  JHEP {\bf 1504}, 145 (2015)
  [arXiv:1411.2588 [hep-ph]].
  %%CITATION = doi:10.1007/JHEP04(2015)145;%%  
  
  %\cite{Czakon:2013xaa}
\bibitem{Czakon:2013xaa} 
  M.~Czakon, P.~Fiedler, A.~Mitov and J.~Rojo,
  %``Further exploration of top pair hadroproduction at NNLO,''
  arXiv:1305.3892 [hep-ph].
  %%CITATION = ARXIV:1305.3892;%%
  
%\cite{Mangano:1991jk}
\bibitem{Mangano:1991jk} 
  M.~L.~Mangano, P.~Nason and G.~Ridolfi,
  %``Heavy quark correlations in hadron collisions at next-to-leading order,''
  Nucl.\ Phys.\ B {\bf 373}, 295 (1992).
  %%CITATION = doi:10.1016/0550-3213(92)90435-E;%%

%\cite{Frixione:1995fj}
\bibitem{Frixione:1995fj} 
  S.~Frixione, M.~L.~Mangano, P.~Nason and G.~Ridolfi,
  %``Top quark distributions in hadronic collisions,''
  Phys.\ Lett.\ B {\bf 351}, 555 (1995)
  [hep-ph/9503213].
  %%CITATION = doi:10.1016/0370-2693(95)00430-S;%%

%\cite{Ahrens:2010zv}
\bibitem{Ahrens:2010zv} 
  V.~Ahrens, A.~Ferroglia, M.~Neubert, B.~D.~Pecjak and L.~L.~Yang,
  %``Renormalization-Group Improved Predictions for Top-Quark Pair Production at Hadron Colliders,''
  JHEP {\bf 1009}, 097 (2010)
  [arXiv:1003.5827 [hep-ph]].
  %%CITATION = doi:10.1007/JHEP09(2010)097;%%
  
%\cite{Ferroglia:2013zwa}
\bibitem{Ferroglia:2013zwa} 
  A.~Ferroglia, B.~D.~Pecjak and L.~L.~Yang,
  %``Top-quark pair production at high invariant mass: an NNLO soft plus virtual approximation,''
  JHEP {\bf 1309}, 032 (2013)
  [arXiv:1306.1537 [hep-ph]].
  %%CITATION = doi:10.1007/JHEP09(2013)032;%%

%\cite{Ferroglia:2015ivv}
\bibitem{Ferroglia:2015ivv} 
  A.~Ferroglia, B.~D.~Pecjak, D.~J.~Scott and L.~L.~Yang,
  %``QCD resummations for boosted top production,''
  PoS TOP {\bf 2015}, 052 (2016)
  [arXiv:1512.02535 [hep-ph]].
  %%CITATION = ARXIV:1512.02535;%%

%\cite{Pecjak:2016nee}
\bibitem{Pecjak:2016nee} 
  B.~D.~Pecjak, D.~J.~Scott, X.~Wang and L.~L.~Yang,
  %``Resummed differential cross sections for top-quark pairs at the LHC,''
  Phys.\ Rev.\ Lett.\  {\bf 116}, no. 20, 202001 (2016)
  [arXiv:1601.07020 [hep-ph]].
  %%CITATION = doi:10.1103/PhysRevLett.116.202001;%%  
  
%\cite{Berger:2010zx}
\bibitem{Berger:2010zx} 
  C.~F.~Berger {\it et al.},
  %``Precise Predictions for W + 4 Jet Production at the Large Hadron Collider,''
  Phys.\ Rev.\ Lett.\  {\bf 106}, 092001 (2011)
  [arXiv:1009.2338 [hep-ph]].
  %%CITATION = doi:10.1103/PhysRevLett.106.092001;%%

%\cite{Boughezal:2016yfp}
\bibitem{Boughezal:2016yfp} 
  R.~Boughezal, X.~Liu and F.~Petriello,
  %``A comparison of NNLO QCD predictions with 7 TeV ATLAS and CMS data for $V$+jet processes,''
  arXiv:1602.05612 [hep-ph].
  %%CITATION = ARXIV:1602.05612;%%

%\cite{Melnikov:2009wh}
\bibitem{Melnikov:2009wh} 
  K.~Melnikov and G.~Zanderighi,
  %``W+3 jet production at the LHC as a signal or background,''
  Phys.\ Rev.\ D {\bf 81}, 074025 (2010)
  [arXiv:0910.3671 [hep-ph]].
  %%CITATION = doi:10.1103/PhysRevD.81.074025;%%

%\cite{Alwall:2007fs}
\bibitem{Alwall:2007fs} 
  J.~Alwall {\it et al.},
  %``Comparative study of various algorithms for the merging of parton showers and matrix elements in hadronic collisions,''
  Eur.\ Phys.\ J.\ C {\bf 53}, 473 (2008)
  [arXiv:0706.2569 [hep-ph]].
  %%CITATION = doi:10.1140/epjc/s10052-007-0490-5;%%

%\cite{Catani:2001cc}
\bibitem{Catani:2001cc} 
  S.~Catani, F.~Krauss, R.~Kuhn and B.~R.~Webber,
  %``QCD matrix elements + parton showers,''
  JHEP {\bf 0111}, 063 (2001)
  [hep-ph/0109231].
  %%CITATION = doi:10.1088/1126-6708/2001/11/063;%%

%\cite{Bauer:2009km}
\bibitem{Bauer:2009km} 
  C.~W.~Bauer and B.~O.~Lange,
  %``Scale setting and resummation of logarithms in pp ---> V + jets,''
  arXiv:0905.4739 [hep-ph].
  %%CITATION = ARXIV:0905.4739;%%

%\cite{Chatrchyan:2012bja}
\bibitem{Chatrchyan:2012bja} 
  S.~Chatrchyan {\it et al.} [CMS Collaboration],
  %``Measurements of differential jet cross sections in proton-proton collisions at $\sqrt{s}=7$ TeV with the CMS detector,''
  Phys.\ Rev.\ D {\bf 87}, no. 11, 112002 (2013)
  [Phys.\ Rev.\ D {\bf 87}, no. 11, 119902 (2013)]
  [arXiv:1212.6660 [hep-ex]].
  %%CITATION = doi:10.1103/PhysRevD.87.112002, 10.1103/PhysRevD.87.119902;%%

%\cite{Aad:2014vwa}
\bibitem{Aad:2014vwa} 
  G.~Aad {\it et al.} [ATLAS Collaboration],
  %``Measurement of the inclusive jet cross-section in proton-proton collisions at $ \sqrt{s}=7 $ TeV using 4.5 fb$^{?1}$ of data with the ATLAS detector,''
  JHEP {\bf 1502}, 153 (2015)
  [JHEP {\bf 1509}, 141 (2015)]
  [arXiv:1410.8857 [hep-ex]].
  %%CITATION = doi:10.1007/JHEP02(2015)153, 10.1007/JHEP09(2015)141;%%

%\cite{Carrazza:2014hra}
\bibitem{Carrazza:2014hra} 
  S.~Carrazza and J.~Pires,
  %``Perturbative QCD description of jet data from LHC Run-I and Tevatron Run-II,''
  JHEP {\bf 1410}, 145 (2014)
  [arXiv:1407.7031 [hep-ph]].
  %%CITATION = doi:10.1007/JHEP10(2014)145;%%
  
%\cite{Aad:2013tea}
\bibitem{Aad:2013tea} 
  G.~Aad {\it et al.} [ATLAS Collaboration],
  %``Measurement of dijet cross sections in $pp$ collisions at 7 TeV centre-of-mass energy using the ATLAS detector,''
  JHEP {\bf 1405}, 059 (2014)
  [arXiv:1312.3524 [hep-ex]].
  %%CITATION = doi:10.1007/JHEP05(2014)059;%%

%\cite{Francavilla:2015yxa}
\bibitem{Francavilla:2015yxa} 
  P.~Francavilla,
  %``Measurements of inclusive jet and dijet cross sections at the Large Hadron Collider,''
  Int.\ J.\ Mod.\ Phys.\ A {\bf 30}, no. 31, 1546003 (2015)
  [arXiv:1510.01943 [hep-ex]].
  %%CITATION = doi:10.1142/S0217751X15460033;%%

%\cite{Grunberg:1980ja}
\bibitem{Grunberg:1980ja} 
  G.~Grunberg,
  %``Renormalization Group Improved Perturbative QCD,''
  Phys.\ Lett.\  {\bf 95B}, 70 (1980)
  Erratum: [Phys.\ Lett.\  {\bf 110B}, 501 (1982)].
  %%CITATION = doi:10.1016/0370-2693(80)90402-5;%%

%\cite{Grunberg:1982fw}
\bibitem{Grunberg:1982fw} 
  G.~Grunberg,
  %``Renormalization Scheme Independent QCD and QED: The Method of Effective Charges,''
  Phys.\ Rev.\ D {\bf 29}, 2315 (1984).
  %%CITATION = doi:10.1103/PhysRevD.29.2315;%%

%\cite{Grunberg:1989xf}
\bibitem{Grunberg:1989xf} 
  G.~Grunberg,
  %``On Some Ambiguities in the Method of Effective Charges,''
  Phys.\ Rev.\ D {\bf 40}, 680 (1989).
  %%CITATION = doi:10.1103/PhysRevD.40.680;%%

%\cite{Stevenson:1981vj}
\bibitem{Stevenson:1981vj} 
  P.~M.~Stevenson,
  %``Optimized Perturbation Theory,''
  Phys.\ Rev.\ D {\bf 23}, 2916 (1981).
  %%CITATION = doi:10.1103/PhysRevD.23.2916;%%

%\cite{Kubo:1982gd}
\bibitem{Kubo:1982gd} 
  J.~Kubo and S.~Sakakibara,
  %``Equivalence of the Fastest Apparent Convergence Criterion and the Principle of Minimal Sensitivity in Perturbative Quantum Chromodynamics,''
  Phys.\ Rev.\ D {\bf 26}, 3656 (1982).
  %%CITATION = doi:10.1103/PhysRevD.26.3656;%%
  
%\cite{Stevenson:1986cu}
\bibitem{Stevenson:1986cu} 
  P.~M.~Stevenson and H.~D.~Politzer,
  %``Optimized Perturbation Theory Applied To Factorization Scheme Dependence,''
  Nucl.\ Phys.\ B {\bf 277}, 758 (1986).
  %%CITATION = doi:10.1016/0550-3213(86)90467-0;%%

%\cite{Maxwell:2000mm}
\bibitem{Maxwell:2000mm} 
  C.~J.~Maxwell and A.~Mirjalili,
  %``Complete renormalization group improvement: Avoiding factorization and renormalization scale dependence in QCD predictions,''
  Nucl.\ Phys.\ B {\bf 577}, 209 (2000)
  [hep-ph/0002204].

%\cite{Maltoni:2007tc}
\bibitem{Maltoni:2007tc} 
  F.~Maltoni, T.~McElmurry, R.~Putman and S.~Willenbrock,
  %``Choosing the Factorization Scale in Perturbative QCD,''
  hep-ph/0703156 [HEP-PH].
  %%CITATION = HEP-PH/0703156;%%

%\cite{Boos:2003yi}
\bibitem{Boos:2003yi} 
  E.~Boos and T.~Plehn,
  %``Higgs boson production induced by bottom quarks,''
  Phys.\ Rev.\ D {\bf 69}, 094005 (2004)
  [hep-ph/0304034].
  %%CITATION = doi:10.1103/PhysRevD.69.094005;%%
  
%\cite{Maltoni:2003pn}
\bibitem{Maltoni:2003pn} 
  F.~Maltoni, Z.~Sullivan and S.~Willenbrock,
  %``Higgs-boson production via bottom-quark fusion,''
  Phys.\ Rev.\ D {\bf 67}, 093005 (2003)
  [hep-ph/0301033].
  %%CITATION = doi:10.1103/PhysRevD.67.093005;%%

%\cite{Brodsky:1982gc}
\bibitem{Brodsky:1982gc} 
  S.~J.~Brodsky, G.~P.~Lepage and P.~B.~Mackenzie,
  %``On the Elimination of Scale Ambiguities in Perturbative Quantum Chromodynamics,''
  Phys.\ Rev.\ D {\bf 28}, 228 (1983).
  %%CITATION = doi:10.1103/PhysRevD.28.228;%%

%\cite{Brodsky:2011ig}
\bibitem{Brodsky:2011ig} 
  S.~J.~Brodsky and L.~Di Giustino,
  %``Setting the Renormalization Scale in QCD: The Principle of Maximum Conformality,''
  Phys.\ Rev.\ D {\bf 86}, 085026 (2012)
  [arXiv:1107.0338 [hep-ph]].
  %%CITATION = doi:10.1103/PhysRevD.86.085026;%%
  
%\cite{Brodsky:2011ta}
\bibitem{Brodsky:2011ta} 
  S.~J.~Brodsky and X.~G.~Wu,
  %``Scale Setting Using the Extended Renormalization Group and the Principle of Maximum Conformality: the QCD Coupling Constant at Four Loops,''
  Phys.\ Rev.\ D {\bf 85}, 034038 (2012)
  [Phys.\ Rev.\ D {\bf 86}, 079903 (2012)]
  [arXiv:1111.6175 [hep-ph]].
  %%CITATION = doi:10.1103/PhysRevD.85.034038, 10.1103/PhysRevD.86.079903;%%

%\cite{Brodsky:2012rj}
\bibitem{Brodsky:2012rj} 
  S.~J.~Brodsky and X.~G.~Wu,
  %``Eliminating the Renormalization Scale Ambiguity for Top-Pair Production Using the Principle of Maximum Conformality,''
  Phys.\ Rev.\ Lett.\  {\bf 109}, 042002 (2012)
  [arXiv:1203.5312 [hep-ph]].
  %%CITATION = doi:10.1103/PhysRevLett.109.042002;%%

%\cite{Mojaza:2012mf}
\bibitem{Mojaza:2012mf} 
  M.~Mojaza, S.~J.~Brodsky and X.~G.~Wu,
  %``Systematic All-Orders Method to Eliminate Renormalization-Scale and Scheme Ambiguities in Perturbative QCD,''
  Phys.\ Rev.\ Lett.\  {\bf 110}, 192001 (2013)
  [arXiv:1212.0049 [hep-ph]].
  %%CITATION = doi:10.1103/PhysRevLett.110.192001;%%
  
%\cite{Brodsky:2013vpa}
\bibitem{Brodsky:2013vpa} 
  S.~J.~Brodsky, M.~Mojaza and X.~G.~Wu,
  %``Systematic Scale-Setting to All Orders: The Principle of Maximum Conformality and Commensurate Scale Relations,''
  Phys.\ Rev.\ D {\bf 89}, 014027 (2014)
  [arXiv:1304.4631 [hep-ph]].
  %%CITATION = doi:10.1103/PhysRevD.89.014027;%%  

%\cite{Ma:2015dxa}
\bibitem{Ma:2015dxa} 
  H.~H.~Ma, X.~G.~Wu, Y.~Ma, S.~J.~Brodsky and M.~Mojaza,
  %``Setting the renormalization scale in perturbative QCD: Comparisons of the principle of maximum conformality with the sequential extended Brodsky-Lepage-Mackenzie approach,''
  Phys.\ Rev.\ D {\bf 91}, no. 9, 094028 (2015)
  [arXiv:1504.01260 [hep-ph]].
  %%CITATION = doi:10.1103/PhysRevD.91.094028;%%

%\cite{Cacciari:2011ze}
\bibitem{Cacciari:2011ze} 
  M.~Cacciari and N.~Houdeau,
  %``Meaningful characterisation of perturbative theoretical uncertainties,''
  JHEP {\bf 1109}, 039 (2011)
  [arXiv:1105.5152 [hep-ph]].
  %%CITATION = doi:10.1007/JHEP09(2011)039;%%
  
%\cite{Bagnaschi:2014wea}
\bibitem{Bagnaschi:2014wea} 
  E.~Bagnaschi, M.~Cacciari, A.~Guffanti and L.~Jenniches,
  %``An extensive survey of the estimation of uncertainties from missing higher orders in perturbative calculations,''
  JHEP {\bf 1502}, 133 (2015)
  [arXiv:1409.5036 [hep-ph]].
  %%CITATION = doi:10.1007/JHEP02(2015)133;%%

%\cite{David:2013gaa}
\bibitem{David:2013gaa} 
  A.~David and G.~Passarino,
  %``How well can we guess theoretical uncertainties?,''
  Phys.\ Lett.\ B {\bf 726}, 266 (2013)
  [arXiv:1307.1843].
  %%CITATION = doi:10.1016/j.physletb.2013.08.025;%%  

%\cite{Buckley:2014ana}
\bibitem{Buckley:2014ana} 
  A.~Buckley, J.~Ferrando, S.~Lloyd, K.~Nordström, B.~Page, M.~Rüfenacht, M.~Schönherr and G.~Watt,
  %``LHAPDF6: parton density access in the LHC precision era,''
  Eur.\ Phys.\ J.\ C {\bf 75}, 132 (2015)
  [arXiv:1412.7420 [hep-ph]].
  %%CITATION = doi:10.1140/epjc/s10052-015-3318-8;%%
  
%\cite{Martin:2009iq}
\bibitem{Martin:2009iq} 
  A.~D.~Martin, W.~J.~Stirling, R.~S.~Thorne and G.~Watt,
  %``Parton distributions for the LHC,''
  Eur.\ Phys.\ J.\ C {\bf 63}, 189 (2009)
  [arXiv:0901.0002 [hep-ph]].
  %%CITATION = doi:10.1140/epjc/s10052-009-1072-5;%%

%\cite{Ball:2014uwa}
\bibitem{Ball:2014uwa} 
  R.~D.~Ball {\it et al.} [NNPDF Collaboration],
  %``Parton distributions for the LHC Run II,''
  JHEP {\bf 1504}, 040 (2015)
  [arXiv:1410.8849 [hep-ph]].
  %%CITATION = doi:10.1007/JHEP04(2015)040;%%  

%\cite{Czakon:2011xx}
\bibitem{Czakon:2011xx} 
  M.~Czakon and A.~Mitov,
  %``Top++: A Program for the Calculation of the Top-Pair Cross-Section at Hadron Colliders,''
  Comput.\ Phys.\ Commun.\  {\bf 185}, 2930 (2014)
  [arXiv:1112.5675 [hep-ph]].
  %%CITATION = doi:10.1016/j.cpc.2014.06.021;%%

%\cite{Anastasiou:2016cez}
\bibitem{Anastasiou:2016cez} 
  C.~Anastasiou, C.~Duhr, F.~Dulat, E.~Furlan, T.~Gehrmann, F.~Herzog, A.~Lazopoulos and B.~Mistlberger,
  %``High precision determination of the gluon fusion Higgs boson cross-section at the LHC,''
  JHEP {\bf 1605}, 058 (2016)
  [arXiv:1602.00695 [hep-ph]].
  %%CITATION = doi:10.1007/JHEP05(2016)058;%%

%\cite{Dulat:2015mca}
\bibitem{Dulat:2015mca} 
  S.~Dulat {\it et al.},
  %``New parton distribution functions from a global analysis of quantum chromodynamics,''
  Phys.\ Rev.\ D {\bf 93}, no. 3, 033006 (2016)
  [arXiv:1506.07443 [hep-ph]].
  %%CITATION = doi:10.1103/PhysRevD.93.033006;%%

%\cite{Harland-Lang:2014zoa}
\bibitem{Harland-Lang:2014zoa} 
  L.~A.~Harland-Lang, A.~D.~Martin, P.~Motylinski and R.~S.~Thorne,
  %``Parton distributions in the LHC era: MMHT 2014 PDFs,''
  Eur.\ Phys.\ J.\ C {\bf 75}, no. 5, 204 (2015)
  [arXiv:1412.3989 [hep-ph]].
  %%CITATION = doi:10.1140/epjc/s10052-015-3397-6;%%

%\cite{Bertone:2013vaa}
\bibitem{Bertone:2013vaa} 
  V.~Bertone, S.~Carrazza and J.~Rojo,
  %``APFEL: A PDF Evolution Library with QED corrections,''
  Comput.\ Phys.\ Commun.\  {\bf 185}, 1647 (2014)
  [arXiv:1310.1394 [hep-ph]].
  %%CITATION = doi:10.1016/j.cpc.2014.03.007;%%
  
%\cite{Kluge:2006xs}
\bibitem{Kluge:2006xs} 
  T.~Kluge, K.~Rabbertz and M.~Wobisch,
  %``FastNLO: Fast pQCD calculations for PDF fits,''
  hep-ph/0609285.
  %%CITATION = HEP-PH/0609285;%%

%\cite{Britzger:2012bs}
\bibitem{Britzger:2012bs} 
  D.~Britzger {\it et al.} [fastNLO Collaboration],
  %``New features in version 2 of the fastNLO project,''
  arXiv:1208.3641 [hep-ph].
  %%CITATION = doi:10.3204/DESY-PROC-2012-02/165;%%

%\cite{Langenfeld:2009wd}
\bibitem{Langenfeld:2009wd} 
  U.~Langenfeld, S.~Moch and P.~Uwer,
  %``Measuring the running top-quark mass,''
  Phys.\ Rev.\ D {\bf 80}, 054009 (2009)
  [arXiv:0906.5273 [hep-ph]].
  %%CITATION = doi:10.1103/PhysRevD.80.054009;%%
  
%\cite{Dowling:2013baa}
\bibitem{Dowling:2013baa} 
  M.~Dowling and S.~O.~Moch,
  %``Differential distributions for top-quark hadro-production with a running mass,''
  Eur.\ Phys.\ J.\ C {\bf 74}, no. 11, 3167 (2014)
  [arXiv:1305.6422 [hep-ph]].
  %%CITATION = doi:10.1140/epjc/s10052-014-3167-x;%%

\end{thebibliography}
\end{document}